\documentclass[aps,twocolumn,groupedaddress,superscriptaddress]{revtex4-1}

\usepackage{graphicx,amsmath,amssymb}
\usepackage{soul}
\begin{document}
\begin{titlepage}
\title{Length scale dependence of the Stokes-Einstein and Adam-Gibbs relations in model glass formers}

\author{Anshul D. S. Parmar}
\affiliation{Jawaharlal Nehru Center for Advanced Scientific Research, Jakkur Campus, Bengaluru 560064, India}
\affiliation{TIFR Center for Interdisciplinary Sciences, 21 Brundavan Colony, Narsingi, Hyderabad  500075, India}
\author{Shiladitya Sengupta}
\affiliation{Dept. of Fundamental Engineering, Institute of Industrial Science, The University of Tokyo, Komaba 4-6-1, Meguro-ku, Tokyo 153-8505, Japan.}
\author{Srikanth Sastry}
\affiliation{Jawaharlal Nehru Center for Advanced Scientific Research, Jakkur Campus, Bengaluru 560064, India}

\begin{abstract}{
The Adam-Gibbs (AG) relation connects the dynamics of a glass-forming liquid to its the thermodynamics {\it via.} the configurational entropy, and is of fundamental importance in descriptions of glassy behaviour.  The breakdown of the Stokes-Einstein (SEB) relation between the diffusion coefficient and the viscosity (or structural relaxation times) in glass formers raises the question as to which dynamical quantity the AG relation describes. By performing molecular dynamics simulations, we show that the AG relation is valid over the widest temperature range for the diffusion coefficient and not for the viscosity or  relaxation times. Studying relaxation times defined at a given wavelength, we find that  
SEB and the deviation from the AG relation occur below a temperature at which the correlation length of dynamical heterogeneity equals the wavelength probed.}
\end{abstract} 
 \maketitle
\end{titlepage}

It is now clear from extensive research over the last two decades
that, as a liquid is gradually (super)cooled, its dynamics becomes
spatially heterogeneous (``dynamical heterogeneity'')
\cite{Silescu,EdRev,DHbook} and the collective nature of the
underlying relaxation processes can be quantified by various growing
correlation length scales \cite{KDS-rev,KDS-rev2}. Dynamics can be
described by different measures - the translational diffusion
coefficient ($D$), the shear viscosity ($\eta$) or the
$\alpha$-relaxation time ($\tau_{\alpha}$). At high temperatures,
where particle motions are diffusive, all these time scales are
mutually coupled. $D$ and $\eta$ are related {\it via} the Stokes-Einstein (SE) relation \cite{HM,LL} : $D = {mk_B \over c\pi}{T \over R\eta}$, ($m,R$ are respectively mass and radius of a diffusing particle, $T$ is the temperature of the liquid and the constant $c$ depends on stick or slip boundary condition). At high temperatures, owing to exponential decay of self-intermediate scattering function $F_s(k,t) = \exp(-Dk^2 t)$ at all probe wave vectors $k$, $D$ also gets coupled to the relaxation time $\tau (k)$ measured from $F_s(k,t)$ : $Dk^2\tau(k) = constant$. Further, $\tau_{\alpha}$ is often used as a proxy for $\eta$ (or $\eta / T$) in the SE relation to save computational cost. At low temperatures in dense, viscous liquids, $D$ becomes much bigger than the value estimated from $\tau_{\alpha}$ ($\eta$) using the SE relation. This phenomenon is known as the breakdown of the SE relation (SEB) \cite{EdRev,EdigerGr,Kumar-etal,Becker-etal,Sengupta2013,SEBKA-4,SEB-Hopping}. The regime showing SEB is often found to obey a fractional SE relation: $D \propto \tau_{\alpha}^{-\xi}$, where the SE exponent $\xi \in [0,1]$ measures the extent of SEB.

\begin{figure*}[htp]
  \centering
  \includegraphics[scale=0.331]{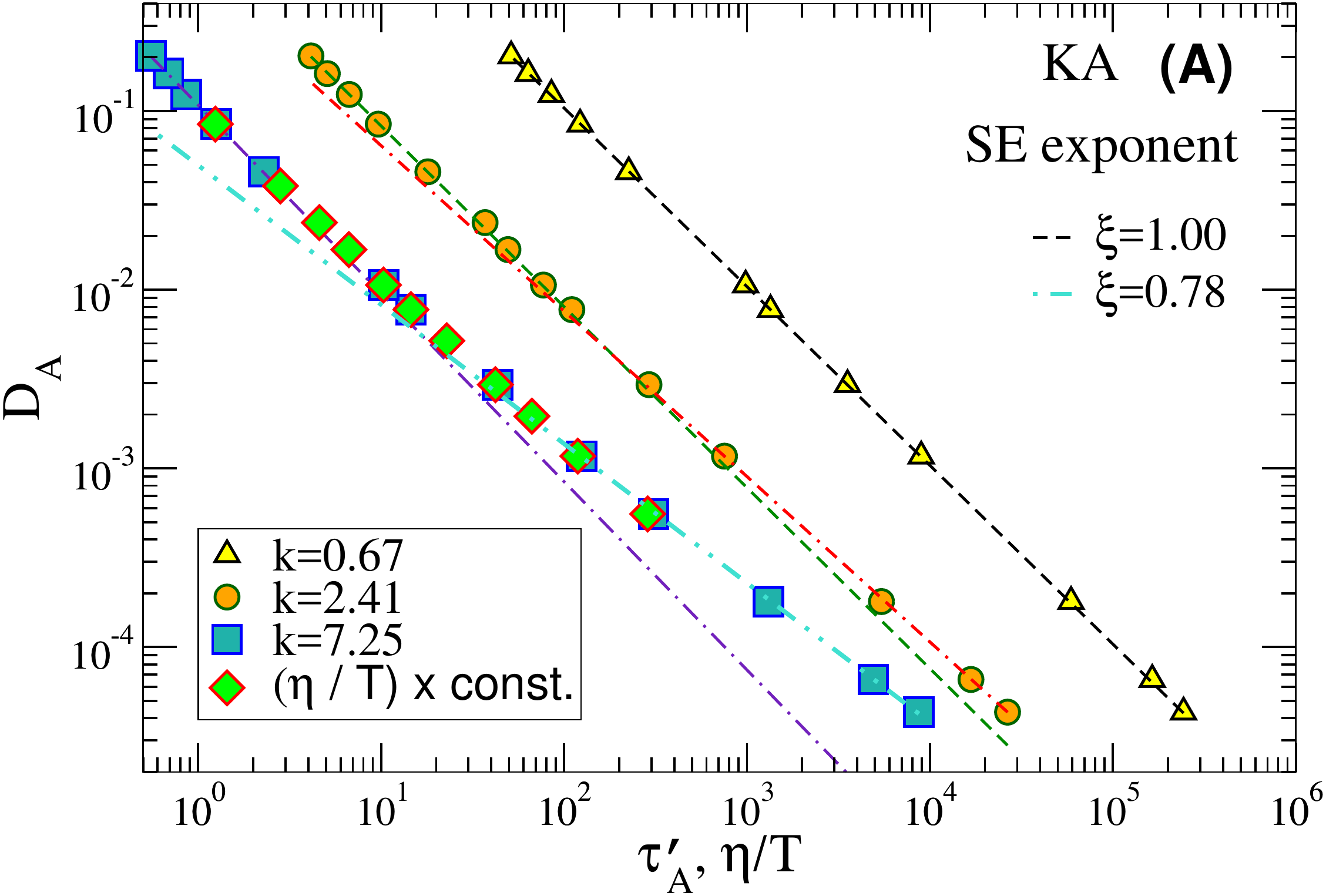}
  \includegraphics[scale=0.33]{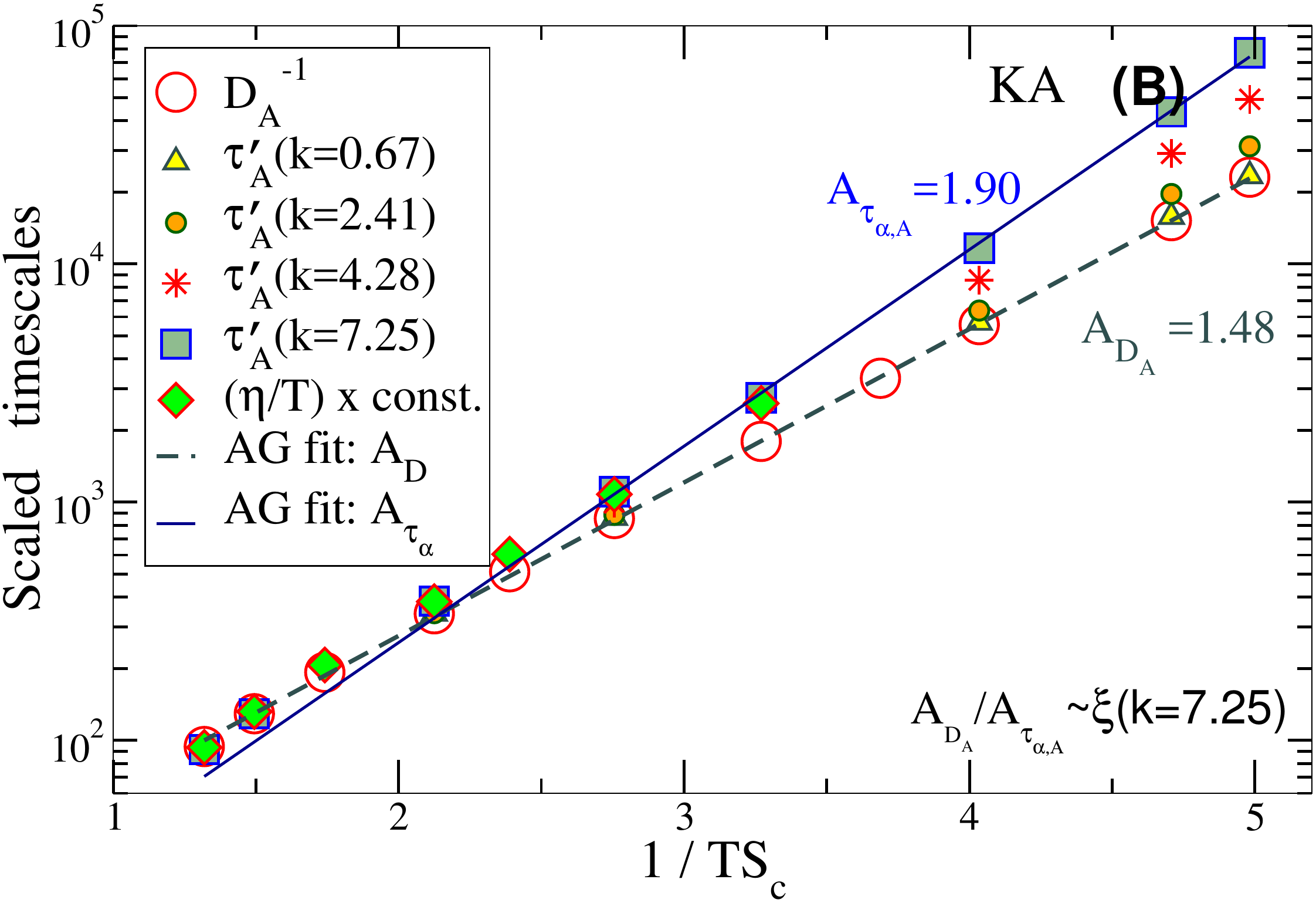}\vspace{-3.5mm}
  \caption{\emph{(A)}: Wave number ($k$) dependence of the Stokes Einstein (SE) breakdown in the KA model. $k$ dependent relaxation times are computed from {$F^A_s(k,t)$} using the definition {$F^A_s(k,\tau')=0.01$} ({ see SM for a comparison with the standard definition of $1/e$}). { The SE breakdown using the shear viscosity is also shown. Lines are power law fits to {$D_A = \mathcal{A} X^{-\xi}$}, where $X$ is either {$\tau'_A$} or $\eta/T$. At high $k$, high temperature and low temperature regimes can be clearly seen. The shear viscosity $\eta$ is coupled to {$\tau'_A(k^*)$, i.e. $\tau_{\alpha,A}$}}. 
\emph{(B)}: Testing the AG relation for { $D_A$}, $\eta/T$ and $\tau_A'(k)$ for different $k$ values for the KA model. { The {$\tau_A'(k)$} and $\eta/T$ have been scaled so that they coincide with {$D_A^{-1}$} at one high temperature ($1/TS_c \approx 1.49$). Lines are fits to the AG relation $X(T) = X_o \exp (A_X/TS_c)$, where {$X = \tau_A'(k),\,\eta/T\text{ or}\, D_A$}. The activation energy for {$\tau_A'(k)$} and $\eta/T$ shows change of slope as a consequence of the SEB(\emph{i.e.} $A_D/A_{\tau'(k)} = \xi(k)$).
The AG relation holds for the {$D_A$} in the considered temperature range but there is a systematic deviation in {$\tau'_A(k,T)$} from the AG relation due to the presence of SEB}. 
However, the deviation observed in the KA model is not large enough to obtain a reliable length scale from the $k$ dependence of such deviation.}
\label{fig:SEplotKA}
\end{figure*}

\begin{figure}[htp]
\centering
\includegraphics[scale=0.24]{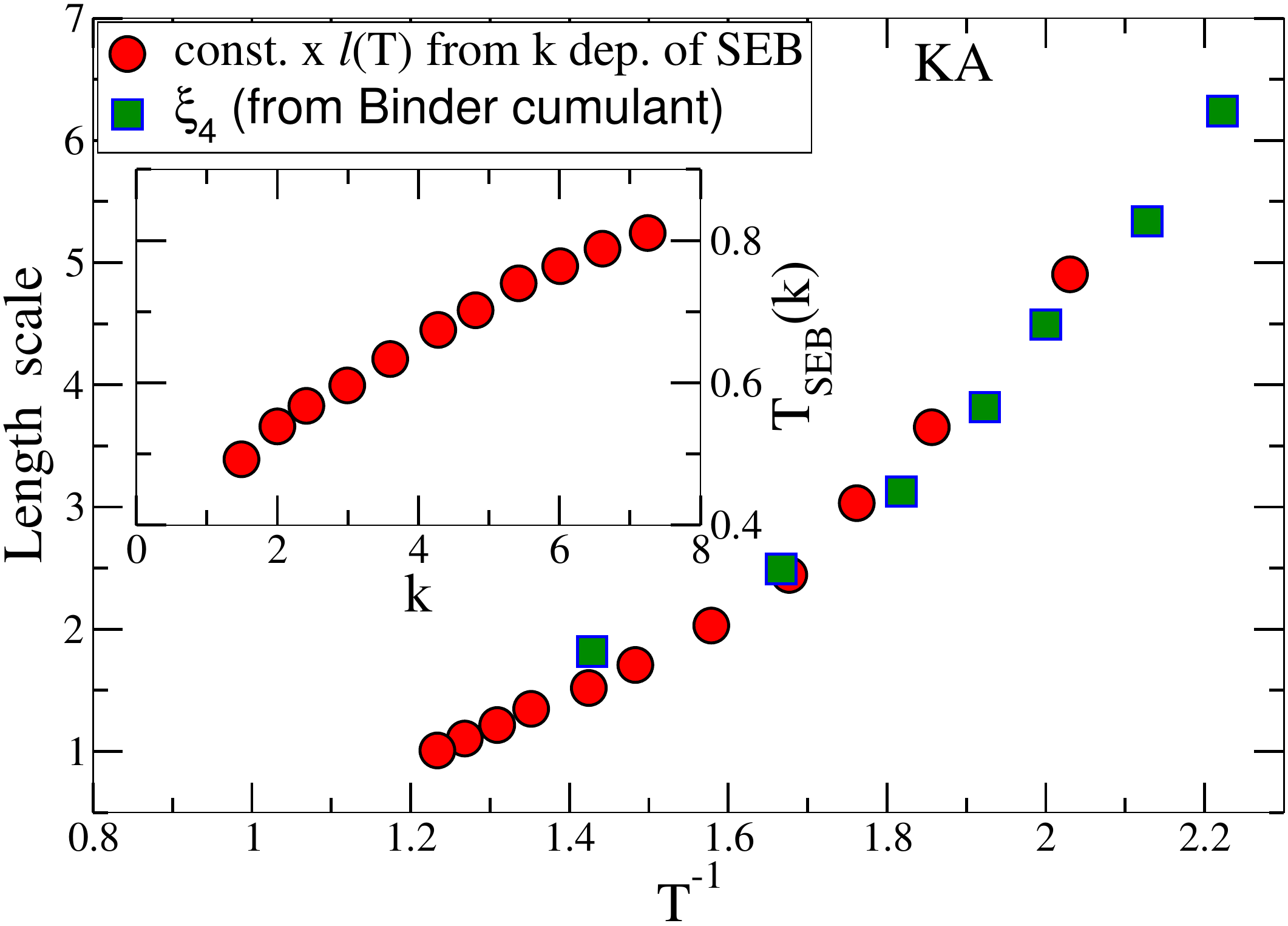} \vspace{-3.5mm}
\caption{The inset shows $k$ dependence of the SE breakdown temperature $(T_{SEB})$. See text for detailed procedure. The main panel shows a direct comparison of the length scale $l(T)$ obtained from the $k$ dependence of the SE breakdown temperature with the dynamic heterogeneity length scale $ \xi_4(T)$ obtained from finite size scaling of Binder cumulant $T$ shows that they are proportional in KA model. DH length scale data obtained with permission from Ref. \cite{KDS-PNAS}.}
\label{fig:KA2}
\end{figure}
{The question naturally arises as to what causes the SE
breakdown.  A commonly held point of view in the literature}
\cite{EdRev,Silescu,LL,EdigerGr} interprets the {SEB} as a
consequence of the dynamical heterogeneity (DH) developing in the
liquid upon cooling.  DH simply means that there are populations of
slow and fast particles which form transient clusters, making the dynamics spatially heterogeneous. The existence of DH leads to the expectation of a distribution of diffusion coefficients and relaxation times \cite{SEBKA-4,SEBKA-5} corresponding to populations of different mobility. The observed $D$ is dominated by the fast population, while the observed $\tau$ (or $\eta$) is governed mainly by the slow population, leading to the decoupling and the SEB. The observed decoupling (SEB) is an \emph{average} effect in such a picture, which however, does not clarify the role played by a heterogeneity length scale.
\begin{figure*}[htp]
\centering
\includegraphics[scale=0.285]{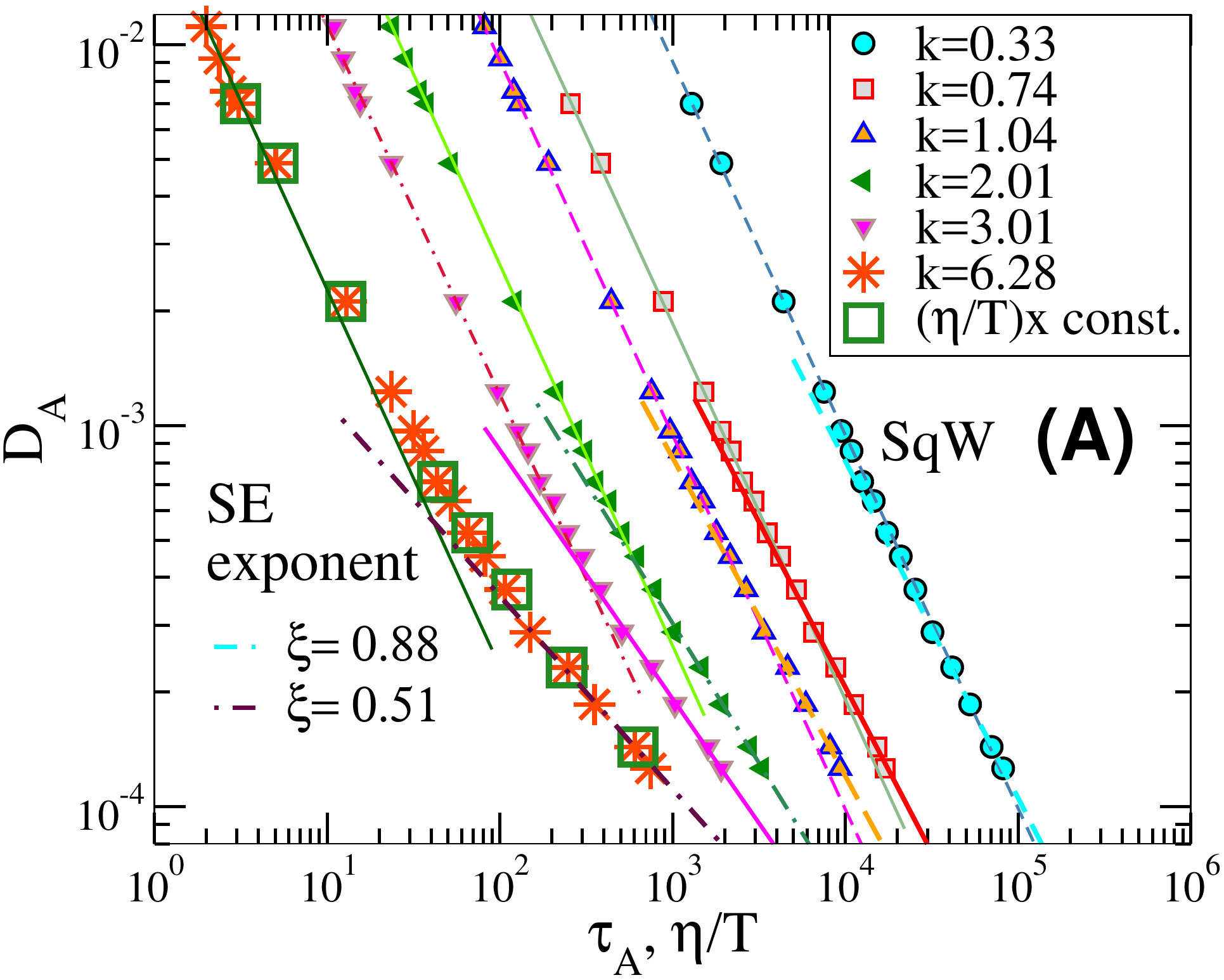}
\includegraphics[scale=0.27]{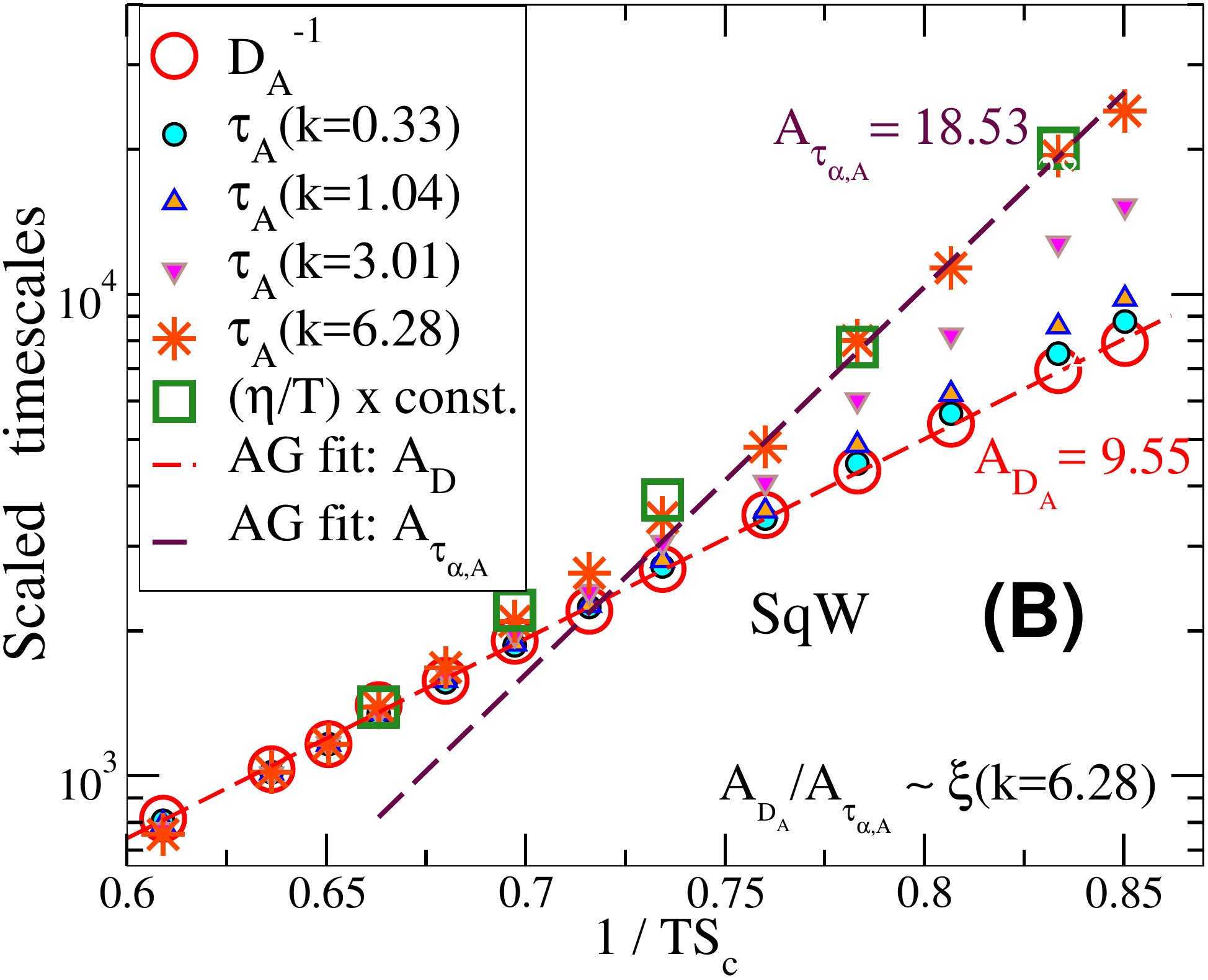}\vspace{2mm}
\includegraphics[scale=0.29]{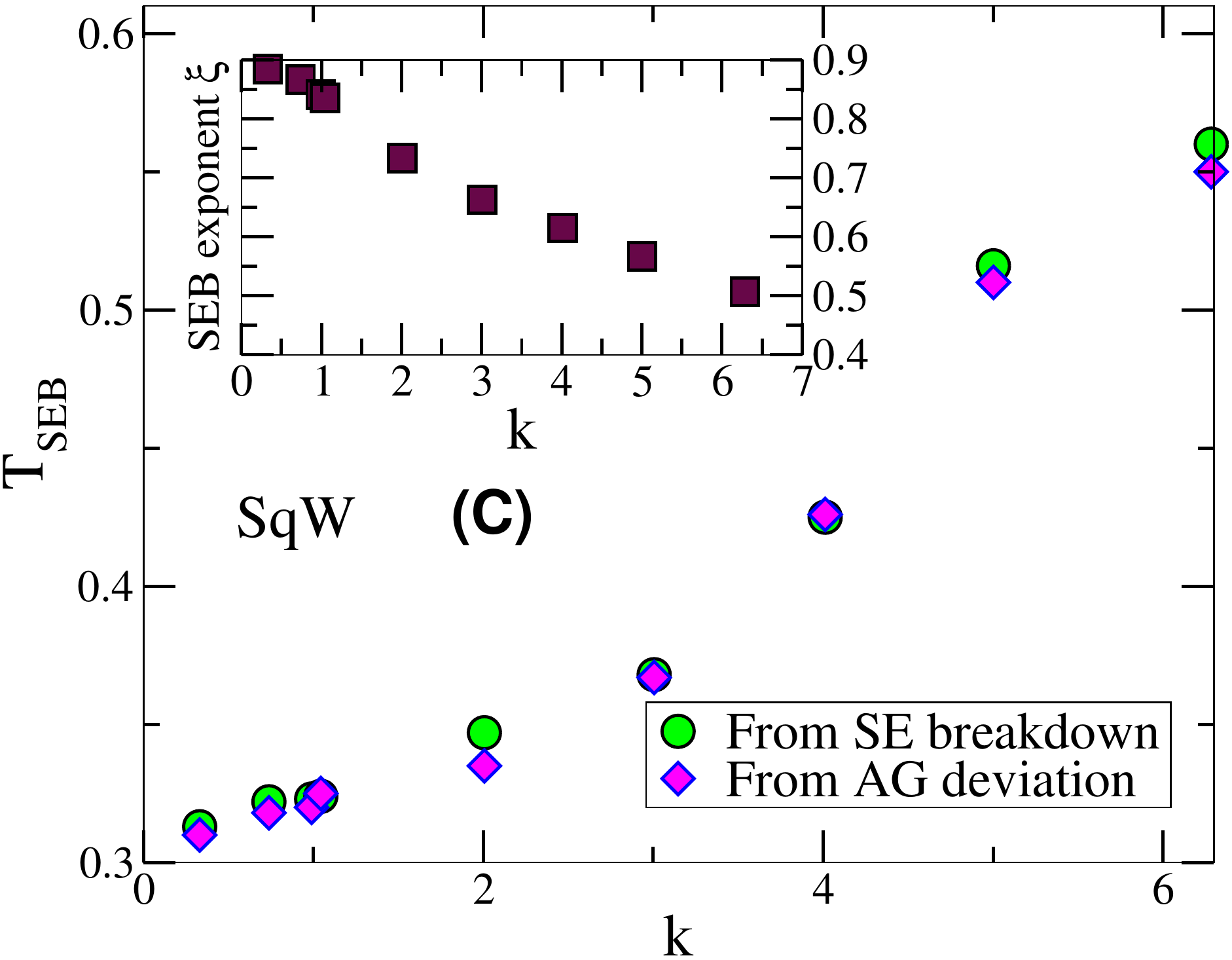}
\caption{ \emph{(A)}: Wave number ($k$) dependence of the Stokes Einstein breakdown in the SqW model. $k^{*}= 2 \pi$ is the first peak of the static structure factor $S(k)$ for A type of particles. As $k$ increases, gradually a fractional SE regime appears at low T. {The viscosity plotted against diffusivity shows that $\tau_{\alpha,A} \propto \eta/T$ is a good description of data across the breakdown temperature.
Lines are power law fits to $D_A = AX^{-\xi}$ , where $X$ is either $\tau_A$ or $\eta/T$}.
 \emph{(B)}: {Testing the AG relation for {$D_A$}, $\eta/T$ and relaxation times $\tau(k)$ for different k values for the SqW model. {$\tau_A(k)$}  and $\eta/T$ have been scaled so that they coincide with $D_A^{-1}$ at one high temperature ( $1/TS_c \approx 0.65$). Lines are fits to the AG relation $X(T) = X_o \exp (A_X/TS_c)$, where {$X = \tau_A(k),\, \eta/T \text\,{or}\, D_A$}. The AG relation is valid for {$D_A$} across the SEB temperature. The relaxation times and the viscosity show two regimes with different slopes for the activation energy $A_{\tau, \eta/T}$. In the presence of SEB, the ratio of activation energies for viscosity and {$\tau_A(k)$} to that of {$D_A$} is given by the fractional SE exponent $\xi$. 
$T_{SEB}(k)$ is obtained as the temperature at which {$\tau_A(k)$} shows deviations from the AG relation}. 
\emph{(C)}: {Inset- $k$ dependence of the SEB exponent shows that the breakdown is very strong in the SqW model}. Main: $k$ dependence of $T_{SEB}$ for the SqW model, obtained from breakdown of both the SE and the AG relations. For every $T$ there is a normal SE regime above a cross-over length scale $l(T)$ and breakdown of the SE relation  below that length scale.}
\label{fig:SEplotSqW}
\end{figure*}
\begin{figure}[htp]
\centering
\includegraphics[scale=0.22]{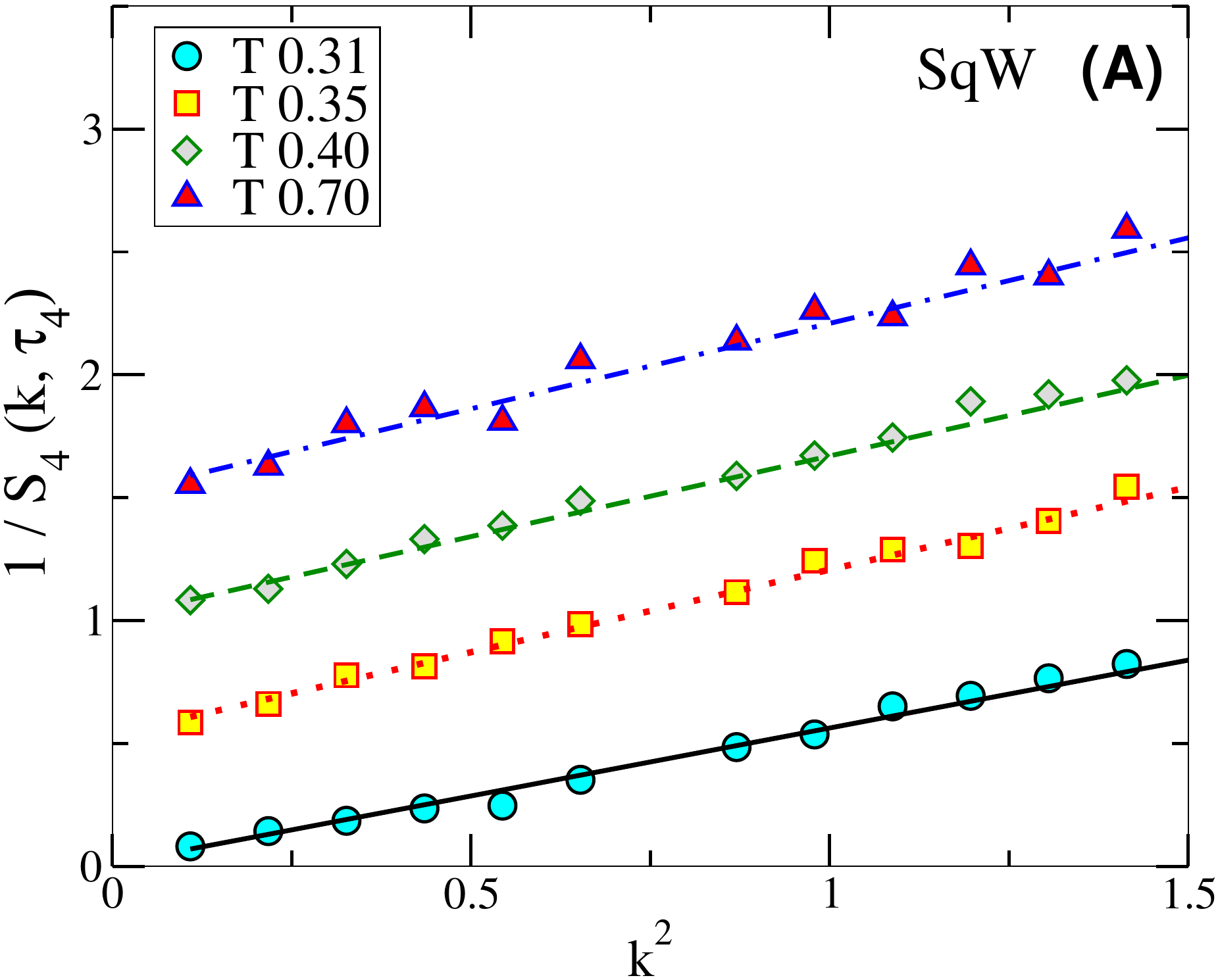}\hspace{2mm}
\includegraphics[scale=0.22]{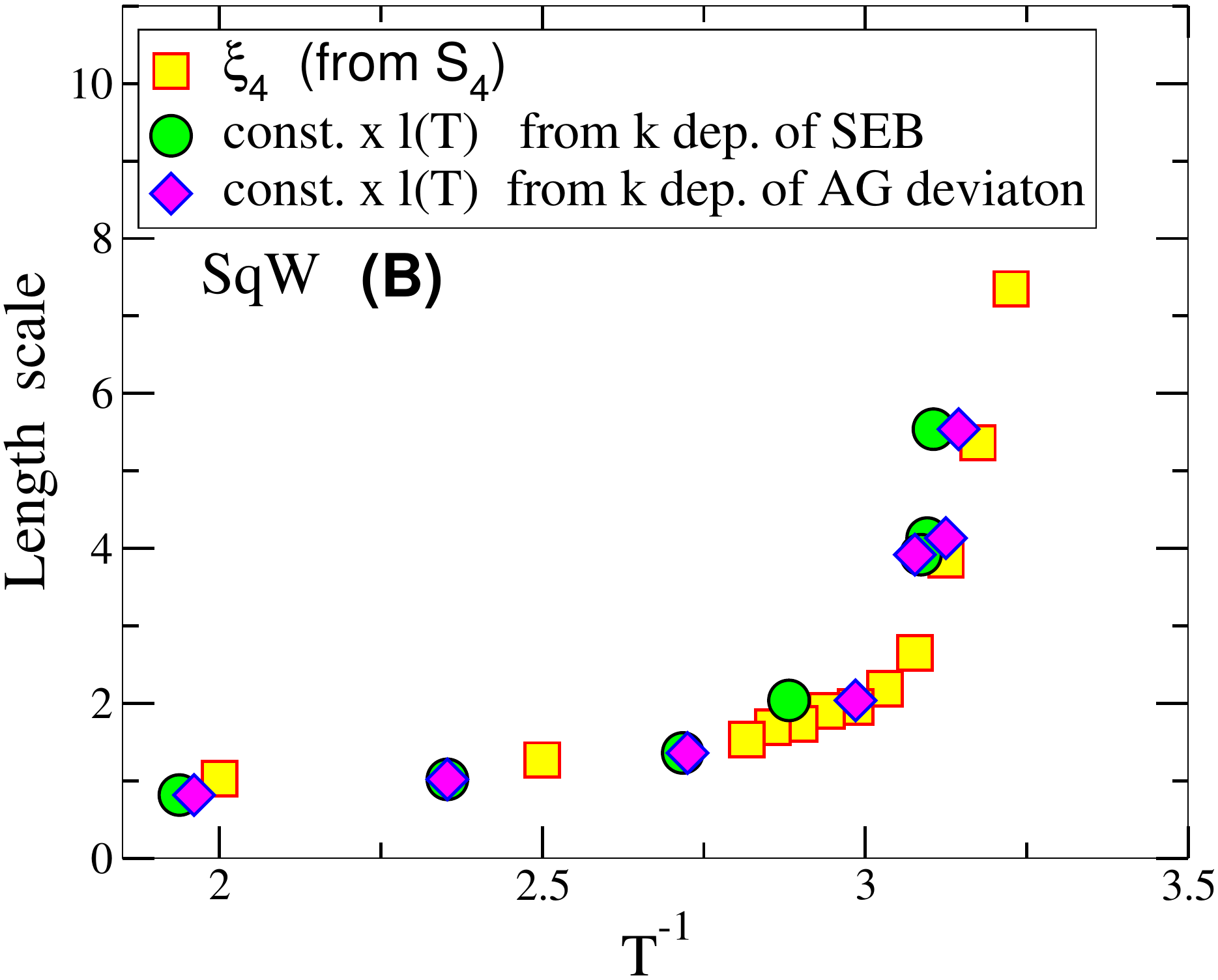}\vspace{-2mm}
\caption{\emph{(A)}: $k$ dependence of the 4-point structure factor $S_4(k,\tau_4)$ for the SqW model. Lines are fits to Ornstein-Zernicke relation $\lim_{k\rightarrow 0} {1 \over S_4} = A (1 + k^2 \xi_4^2)$ from which the dynamical heterogeneity lengthscale $\xi_4$ is estimated. \emph{(B)}: Comparison of the DH length scale and the cross-over length scale, showing they are the same within numerical precision. }
\label{fig:lengthSqW}
\end{figure}

The breakdown of the SE relation poses a puzzle with regard to the
celebrated  Adam-Gibbs (AG) relation, which describes
\cite{AG,AG-expt1,AG-expt2} the viscous slowdown upon cooling in terms
of a parallel decrease in the configurational entropy ($S_c$)
\cite{ScKA,SWB} of the liquid. Its rationalization is considered
fundamental to understanding the glass transition problem \cite{AG-water,AG-fragility,KDS-PNAS,AG-DH,GET,Freed,AG-S2,RFOT-Rev-BB,RFOT-WolynesGr,RFOT-Rev-KT,RFOT-fragility}. The AG relation can be written as $X(T) = X_0 \exp\left({A_X \over TS_c(T)}\right)$, where $X$ is an appropriate measure of dynamics {\it i.e.} $D, \eta$ or $\tau_{\alpha}$. It has been tested in many different systems and has proven to be an extremely useful, predictive relationship between dynamics and thermodynamics below the onset temperature \cite{AG-water,AG-fragility,KDS-PNAS,AG-DH,AG-S2}, despite ambiguity about the range of temperatures over which one should expect it to hold (\cite{SEB-Hopping}). The breakdown of the SE relation raises the question of which dynamical quantity should obey the Adam-Gibbs relation. The $k$ dependence of the breakdown of the SE relation implies a $k$ dependence of the AG relation itself which, to our knowledge, has not been tested before.

{ Here we study two different glass-formers - (a) the well-known Kob-Andersen (KA) binary mixture \cite{KAref} and (b) the square well (SqW) model \cite{SqWdet,SPZ,ST,SWB} by performing NVT-MD simulations. For the KA model, we perform simulations for system size  $N=1000$, number density $\rho = 1.20$ and temperature range  $T \in [0.46, 5.00]$. {Temperature is kept constant using} the Brown and Clarke algorithm \cite{BC}. To simulate the SqW model we carry out the event driven NVT-MD simulations. The SqW model is a 50:50 binary mixture of particles having hard core repulsion at contact as well as an attractive interaction. The ratio of diameters of the two types of particles is $1.20$, with $\sigma_{AA} = 1.20\sigma_{BB}$. Further, for the AB interaction, the hard core diameter is additive, \emph{i.e.} $\sigma_{AB} = (\sigma_{AA} + \sigma_{BB})/2$. The width of the attractive shell ($\Delta_{ij}$) is defined such as $\Delta_{ij}/(\sigma_{ij} + \Delta_{ij})= 0.03$. System size is $N=  5324$ particles, number density $\rho$=  0.77, and temperature range $T \in [0.31,0.7]$. {Temperature is kept constant using} the Lowe-Anderson thermostat \cite{Thermo1,Thermo2}. At each state point, runs are executed for duration $> 100 \tau_{\alpha}$, where $\tau_{\alpha}$ is the $\alpha-$ relaxation time. 
{A length scale dependent relaxation time {$\tau_A(k)$} is computed from the self part of the intermediate scattering function for $A$ type particles, ($F^A_s (k,t)$}$\equiv {1 \over N_A} \sum_{i=1}^{N_A} \exp(i {\bf k}.({\bf r}_i (0) - {\bf r}_i (t)) $  where ${\bf{r}}_i(t)$ is the position of particle $i$)}.
 Configurational entropy is calculated by subtracting the vibrational component from the total entropy of the system, $S_c(T) = S_{total}(T)-S_{vib}(T)$ following \cite{ScKA,SWB}. For each model, we compute the $k$ dependence of the SE relation and its breakdown and extract a temperature dependent length scale from the said $k$ dependences.  We show that this length scale compares very well with a dynamical heterogeneity length scale estimated independently using standard definitions. We also show that the breakdown of the SE relation  corresponds to a deviation from the AG relation for the relaxation times at appropriate wavenumbers $k$. { For diffusion coefficients, the AG relation holds for the entire temperature range but for the relaxation times and viscosity there is a deviation from the AG behaviour at low temperatures in a manner dictated by the SE breakdown}.

The SEB in the KA model is well documented
\cite{SEBKA-4,SEBKA-5,SEBKA-1,SEBKA-2,SEBKA-3,SEBKA-6,SEBKA-7}. In the
KA model, for $k = k^{*}(\sim 7.25)$ (first peak of the static
structure factor $S(k)$), the SE relation breaks down close to the
onset temperature of slow dynamics. A fractional SE relation $D_A\propto \tau_A^{-\xi}$ is followed in the low temperature regime. Fig. \ref{fig:SEplotKA}A shows the 
{diffusion coefficient (of species A)} {\it vs.} the relaxation time data for different $k$ values. 
{ We have considered the conventional definition {$F^A_s(k,\tau_A)=1/e$} as well as a longer timescale {$F^A_s(k,\tau_A') = 0.01$} \cite{Chong} as measures of relaxation time and for subsequent analyses in the KA model, consider only the longer timescales {$\tau_A'$} (see Supplementary Material (SM)).}
{ We also measure the shear viscosity \cite{Leporini2001} ($\eta$) and find that $\eta/T \propto$ {$\tau_{\alpha,A}$}\cite{Sengupta2013}, \emph{i.e.} the viscosity decouples from diffusivity.}
At low $k$ we see no {SEB}. At high $k$, data show two distinct power law regimes (fit lines) at high and at low $T$.
To estimate the temperature of the SE breakdown $T_{SEB}(k)$, we study the product {$D_A \tau_A'(k)$} scaled by the corresponding value at a reference high T {\it vs.} the relaxation time for different $k$, deviation from $1$ will indicate {SEB} (see SM).

In Fig. \ref{fig:SEplotKA}B, we plot { $D_A$}, $\eta/T$ and {$\tau_A'(k)$}
(shifted to coincide all data sets at a chosen temperature) as a function of $(TS_c)^{-1}$. { For diffusion coefficient and relaxation time for $k \rightarrow 0$  the AG relation is valid for the entire temperature range studied. However, as k increases, a change of slope \cite{SEBKA-3} in the data can be seen. The viscosity which is decoupled from {$D_A$} but remains coupled to {$\tau_{\alpha,A}$}, also show similar behaviour. We consider this as a manifestation of the k-dependent {SEB}. However, the observed change in slope in the KA model is gradual, making it difficult to extract a reliable temperature of deviation from AG behaviour. We discuss the deviation from the AG relation in more detail in the context of the SqW model.

The SE breakdown temperature $T_{SEB}$ estimated from 
deviation from SE relation 
shows a systematic increase with increasing $k$ (inset Fig. \ref{fig:KA2}). Thus the SE breakdown follows a scenario where both the breakdown exponent and the breakdown temperature are $k$ dependent. The dependence of $T_{SEB}$ on the probe wave number $k$ can be inverted to obtain a temperature dependent length scale $l(T)$, which we compare with an independently evaluated heterogeneity length.  At any given temperature, particle displacements probed above this length scale will show diffusion coefficients and relaxation times to be coupled while probing below this length scale will reveal decoupling between timescales. Thus this length is similar to the non-Fickian to Fickian crossover lengthscale \cite{Chong,Chong-Kob,BCG,Berthier}, which however has been claimed to be distinct \cite{BCG} from a heterogeneity length.  In Fig. \ref{fig:KA2}, we show the temperature dependences of the length scale of SEB and  the length scale of dynamical heterogeneity $\xi_4(T)$ as computed in Ref. \cite{KDS-PNAS} from the finite size scaling analysis of Binder cumulant. They are directly proportional to each other. This provides a new demonstration that the breakdown of the SE relation is strongly related to the emergence of  dynamical heterogeneity with cooling.


Next, we study another glass-former, {\it viz.} the SqW model.  { As in the case of the KA model, we show diffusion coefficient and relaxation times for $A$ type particles}. Fig. \ref{fig:SEplotSqW}A, shows { $D_A$} {\it vs.} {$\tau_A(k)$}  (In the SqW model, we measure {$\tau_A(k)$} using {$F^A_s(k,t)=1/e$)}. 
Like the KA model, the SqW model shows no {SEB} for $k \rightarrow 0$. At large $k$, there is a high T regime where the SE relation holds and a low T regime obeying a fractional SE relation. However, the fractional exponent reaches much lower values, implying that the SEB in the SqW model is stronger  than in the KA model. {We also measure shear viscosity from Helfand moments \cite{Alder,haile,HM} (see SM) and find that, like the KA model, the SqW model shows $\eta/T \propto \tau_\alpha,A$, and viscosity decouples from the diffusivity}.
Note that in the SqW model, the configurational entropy can be reliably estimated \cite{ScKA,SWB} in the temperature range where the {SEB} occurs (for all the $k$).  Consequently, in the SqW model, the AG relation can be tested over a temperature range which shows both the normal SE relation as well as a $k$ dependent breakdown, making it possible to address for which quantity the AG relation is valid.
If the AG relation is valid for {$D_A$, $\tau_A(k)$} and $\eta/T$
then {these  timescales} 
should overlap up to a constant in the normal SE regime or show deviations in the SEB or fractional SE regime, when plotted against $(TS_c)^{-1}$. The change of slope between {$D_A$} $vs$ {$\tau_A(k)$} { and $\eta/T$} should occur at the breakdown temperature $T_{SEB}$.  In Fig. \ref{fig:SEplotSqW}B, we show the AG plot for {$D_A$}, {$\tau_A(k)$} and $\eta/T$ (shifted to match all data sets at a selected high temperature) {\it vs.} $(TS_c)^{-1}$. For {$D_A$}, the AG relation is valid in the full range shown. For relaxation times {$\tau_A(k)$}  { (and viscosity)}, two distinct regimes are obtained with a crossover at a $k$ dependent temperature.

{
The AG theory proposed a mechanism for structural relaxation {\it via} the concept of CRRs. Recent works have proposed cooperatively moving highly mobile particles, called ``strings", as candidates for CRR \cite{AG-DH,Freed,string2,string3}, in that the string length is found to be inversely proportional to the configurational entropy as envisaged in the AG theory. The string life time is found to be proportional to $t^*$, the time when the non-Gaussian parameter $\alpha_2$ is maximum  \cite{AG-DH}. In turn,  $t^*$ ({for type $A$})  is proportional to {$D_A^{-1}$}, and decouples from {$\tau_{\alpha,A}$} or $\eta$ in the presence of SEB (see SM).   Another strong evidence comes from the observation that diffusion coefficients for finite segments of a system's trajectory are related to the configurational entropy estimated for the same segment \cite{Nave2006}. These results strongly suggest that the time scale described by the AG relation is  proportional to {$D_A$} and not to {$\tau_{\alpha,A}$} (or $\eta$). Theoretical investigations along the lines of  \cite{AG-DH,GET,Freed,Nave2006} therefore offer a promising way of investigating further and rationalising our results.}

%
%

At high $k$, as the temperature decreases, there is a marked change of
slope at the crossover temperature, which decreases as $k\rightarrow
0$. The $k$ dependence of this temperature ($T_{SEB}$) is shown in Fig. \ref{fig:SEplotSqW}C. For comparison, we also show the $T_{SEB}$ obtained directly (from the temperature dependence of the product {$D_A\tau_A(k)$} scaled by the corresponding value at a reference high temperature). There is a reasonably good agreement between the two estimates, { implying that the breakdown of the AG relation for relaxation and viscosity is a manifestation of the {SEB}}. We compute the length scale of dynamical heterogeneity $\xi_4(T)$ \cite{KDS-PNAS}  from the four-point dynamic structure factor $S_4(k,\tau_4)$ ($\tau_4$ is the time at which $S_4(k=0,t)$ is maximum) (Fig. \ref{fig:lengthSqW}A) and in Fig. \ref{fig:lengthSqW}B, compare it to the length scale of {SEB} $l(T)$ obtained from the $k$ dependence of $T_{SEB}$. The data show that they are proportional.
{The strong SEB and large length scales suggest enhanced DH in the SqW model.The SqW model was originally a model for attractive colloids, and systematic studies suggest that the DH is more pronounced in attractive glass-formers \cite{JCP07,DHexpt,DHsimu}. At low temperatures and at the characteristic time scale, the peak values of dynamical susceptibility $\chi_4(t)$ and the non-Gaussian parameter {$\alpha^A_2(t)$} are bigger in the SqW model, showing that the DH is stronger in the SqW model than in the KA model (see SM).} 
{Considering $A$ and $B$ particle diffusion coefficients \cite{SEBKA-6}, we find that they obey a fractional power law relationship over a wide temperature range, and thus the Adam-Gibbs relation hold for both consistently \cite{Sengupta2013} (see SM for details).}


In summary, we have studied the wave number $k$ dependent relaxation
times in two model liquids. We have shown that the breakdown of the SE
relation occurs at $k$ dependent temperatures $T_{SEB}(k)$ that
decrease with decreasing $k$. This allows us to identify a SEB length
scale based on $T_{SEB}(k)$. We compare the length scale with an
independently estimated dynamical heterogeneity length and demonstrate
that they are the same up to an undetermined multiplicative constant.
The fractional SE exponent varies continuously from a value close to
$1$ at the lowest $k$ to  values as low as $0.51$ for one of the
models (SqW). For diffusion coefficients, the Adam-Gibbs relation is
valid throughout the temperature range we describe. For relaxation
times and viscosity, at higher temperature range where the SE relation is valid, the AG relation with the same activation free energy is valid;  at temperatures below the crossover at the SEB temperature, a different AG relation is valid, given by the fractional SE exponent. In presence of the SEB, the AG relation cannot be valid for both diffusion and relaxation times. However the question of which quantity is better described by the AG relation hasn't been addressed in the literature before. Why diffusion coefficient, rather than relaxation times or viscosity, should be better described by the AG relation is an interesting question that needs to be addressed further in future works.


\bigskip

\clearpage

\part*{}   

\setcounter{figure}{0}
\renewcommand{\thefigure}{S\arabic{figure}}%

\begin{center}
\textbf{Length scale dependence of the Stokes-Einstein and Adam-Gibbs relations in model glass formers (Supplementary Material)}
\\
{Anshul D. S. Parmar, Shiladitya Sengupta and Srikanth Sastry}
\date{\today}
\end{center}
Here we provide additional information regarding the following aspects of analysis of studied model glasses: Stokes-Einstein breakdown and viscosity data for 
($i$)  KA model,
($ii$) SqW model,
($iii$) Dynamical heterogeneity,
($iv$) Coupling of $t^*$ and diffusion time scales and 
($v$) Stokes-Einstein and Adam-Gibbs relations for diffusion coefficients and relaxation times of  $A$ and $B$ components in the KA and SqW binary mixtures.
\section{KA model}

\begin{figure}[h!]
  \centering
   \includegraphics[scale=0.35]{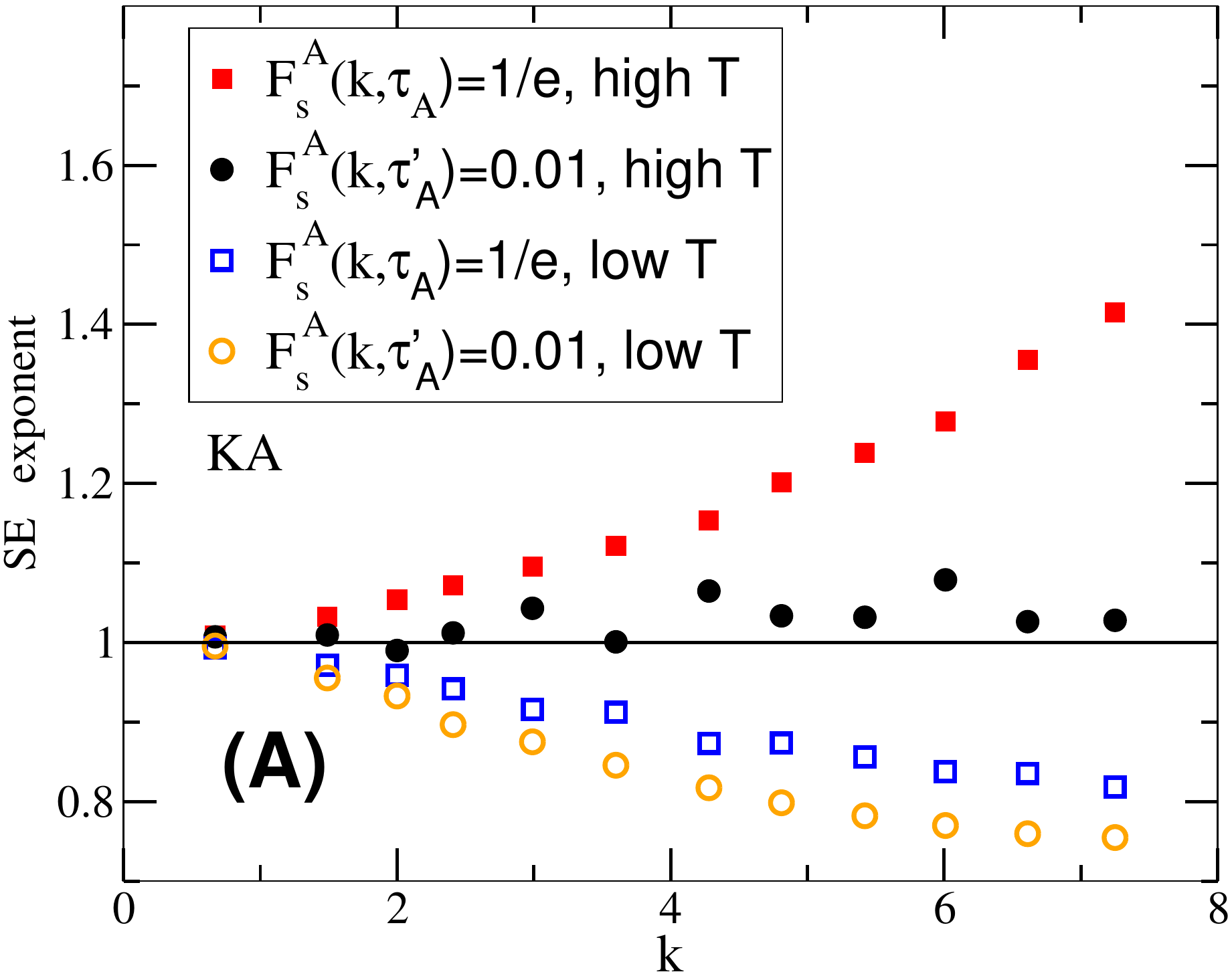}
  \includegraphics[scale=0.255]{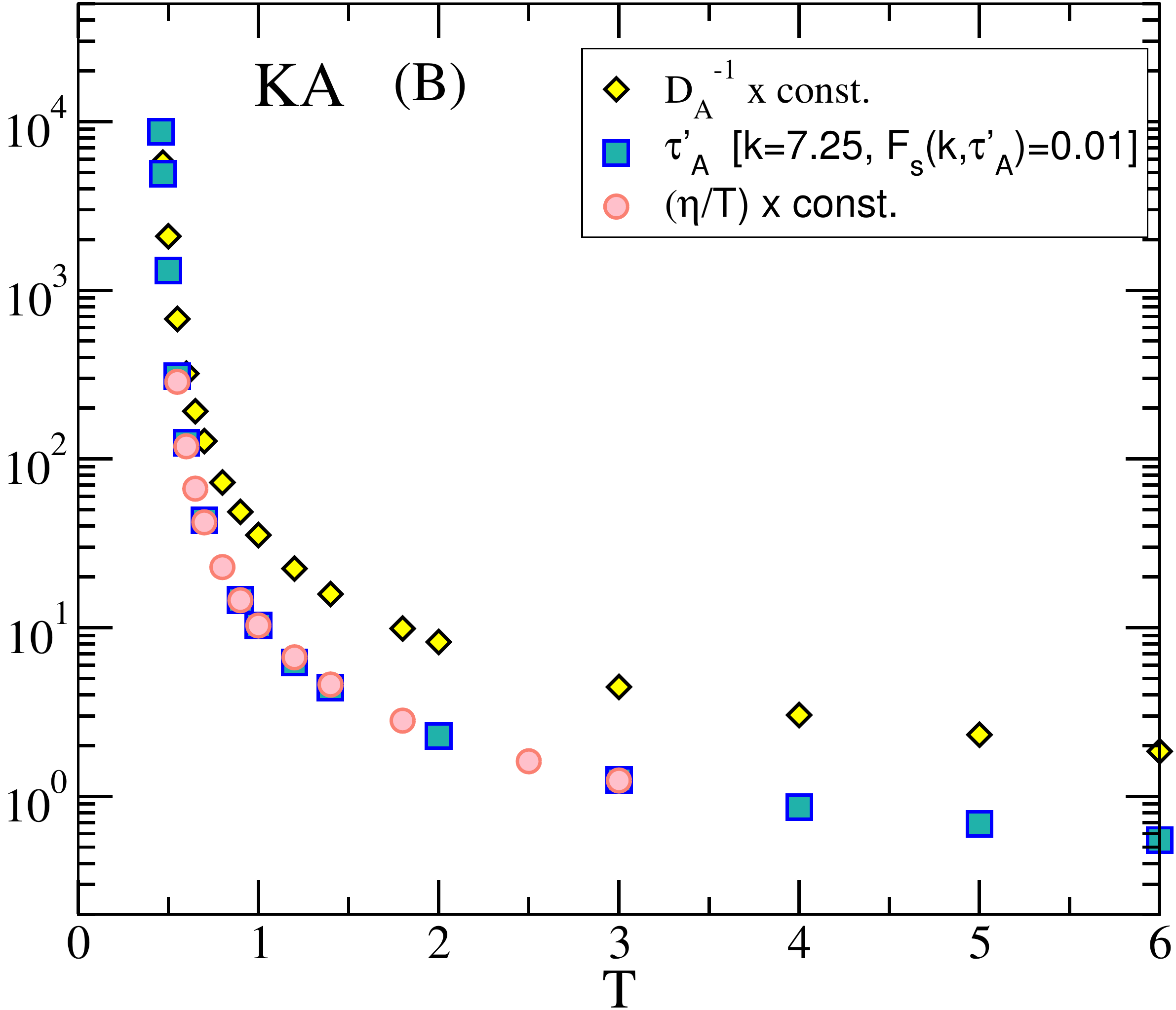} 
  \caption{\emph{(A)}: SE exponents computed from power law fits to high T and low T data. The high T exponent systematically  increases with increasing $k$ using {$\tau_A(k)$} but no such systematic trend is seen with {$\tau'_A(k)$}. Low $T$ exponents show a clear $k$ dependence using either definitions. This indicates that in the KA model, the conventional timescale ({$\tau_A$}) is not long enough to recover the normal SE relation at high T. Hence for subsequent analyses in this model, we consider only the longer timescale {$\tau'_A$}.
\emph{(B)}: Temperature dependence of different dynamical quantities. The behaviour of the shear viscosity $\eta$ is approximately proportional to that of {$\tau'_A(k=7.25)$} obtained from self intermediate scattering function using the definition {$F^A_s(k,\tau_A)=0.01$}.}
\label{fig:etaKA}
\end{figure}

\begin{figure}[h!]
  \centering
   \includegraphics[scale=0.3]{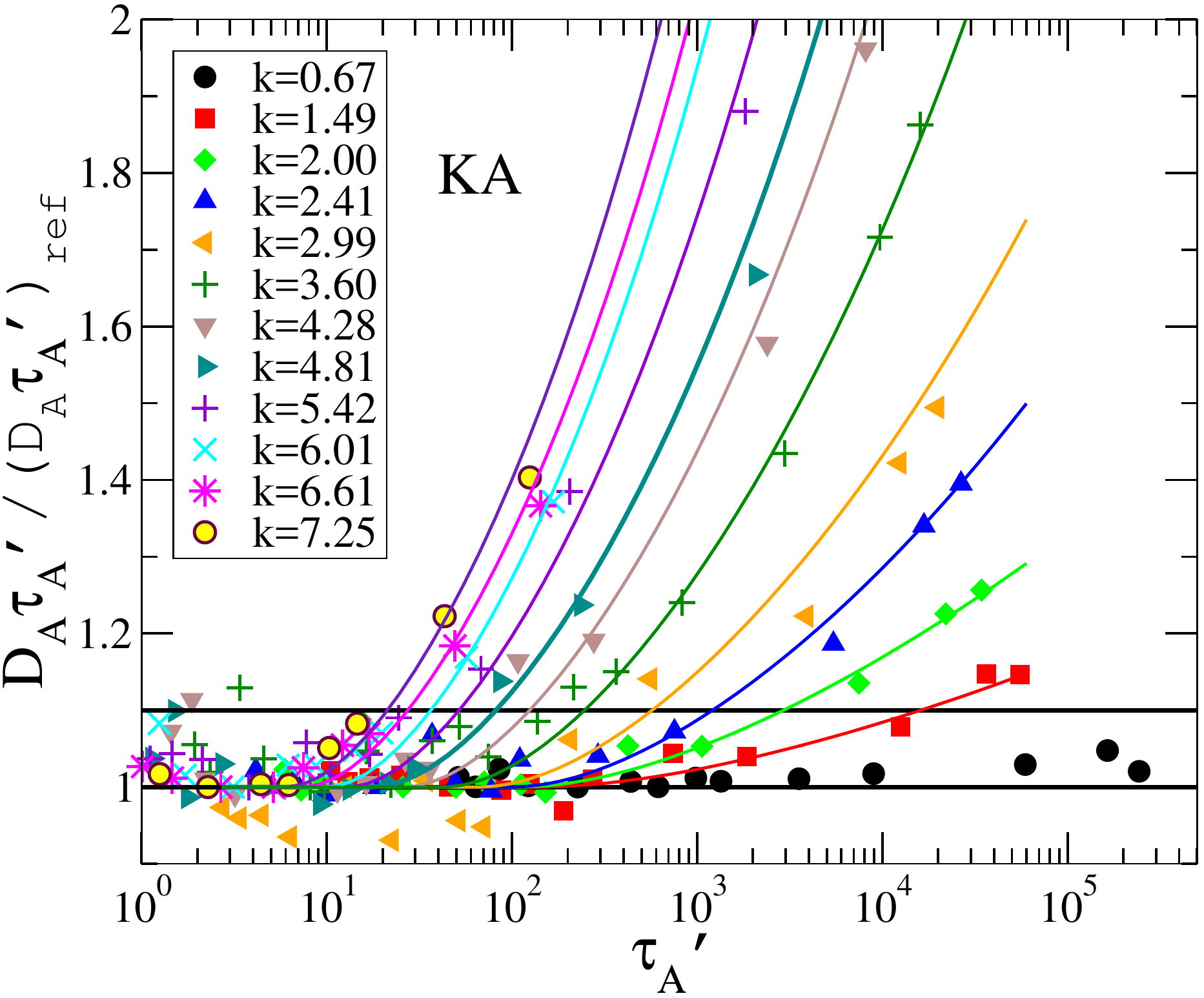}
  \caption{Wave vector ($k$) dependence of the product {$D_A \tau_A'(k)$} scaled by its value at a reference high temperature $T = 2.0$. For simple diffusive dynamics, this ratio should be 1. At a given $k$, the temperature where the ratio deviates from 1 (a threshold value = 1.1)\cite{Chong_s}, is taken to be the SE breakdown temperature $T_{SEB}$ for that $k$. Lines are fits which asymptotically go to {$y(\tau_A')=1$} in the normal SE regime and to {$y(\tau_A') = \tau_A'^{1-\xi}$} in the fractional SE regime. The fitting form is  $\ln y = (1-\xi) \left[ x + c_0 ({x_0 \over x})^n + c_1 ({x_0 \over x})^{2n} )\right]$ where {$y={D_A\tau_A' \over (D_A\tau_A')_{ref}}$} and {$x= ln \tau_A'(k)$}.}
\label{}
\end{figure}

\section{SqW model}
\subsection{Runlength and finite size effect}
Various system sizes ($N=$ $864$, $5324$ and $16384$) have been studied for temperature across the  Stokes Einstein breakdown (SEB) temperature $T=0.33$ (defined form $1/e$ decay of self-intermediate scattering functions at wave vector $k^*=2\pi$). The SEB temperature and exponent do not change much across the considered system size ($N=5324$).
\begin{figure}[h]
\centering
\includegraphics[scale=0.45]{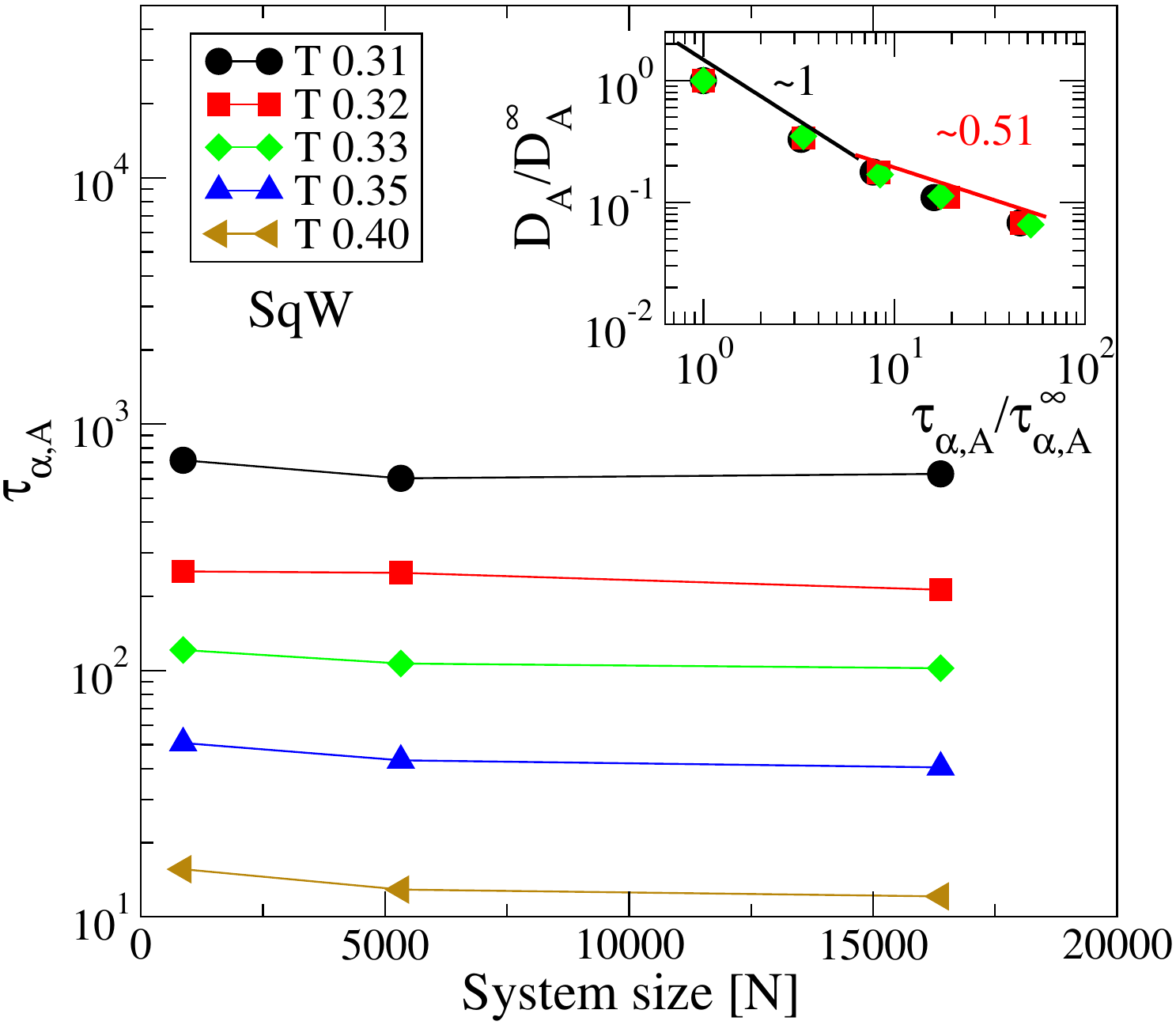}\vspace{-3mm}
\caption{The relaxation time estimated from $1/e$ value of the self-intermediate scattering function at wave vector $k=2\pi$, does not change much in the considered window of system size. The inset shows scaled SE plot, where, {the dynamical quantities for estimated for particles of type A, $D_A^\infty$ or $\tau_A^\infty$} stands for the high temperature values.}
\label{}
\end{figure}
\begin{figure*}[]
\centering
\includegraphics[scale=0.245]{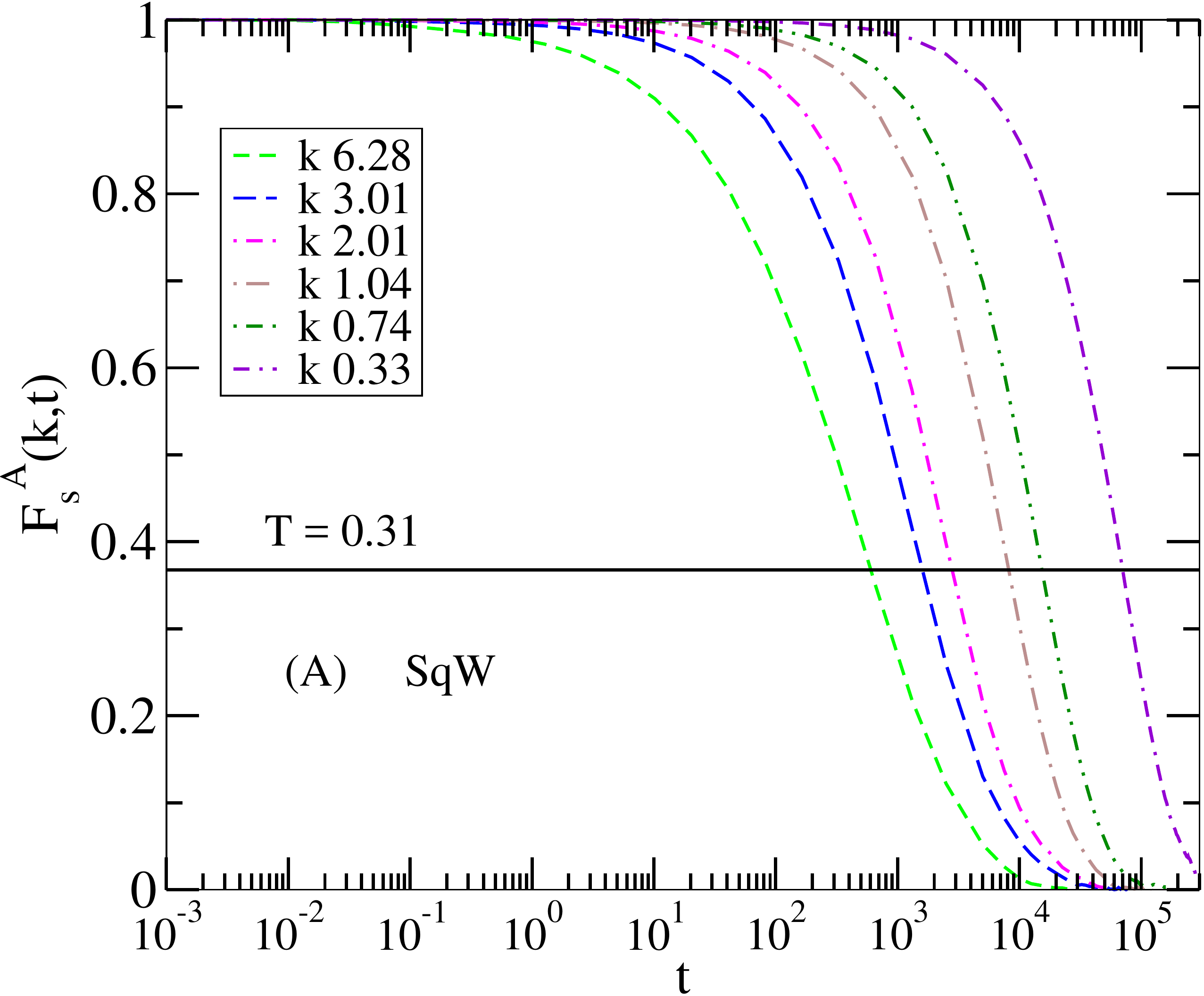}
\includegraphics[scale=0.245]{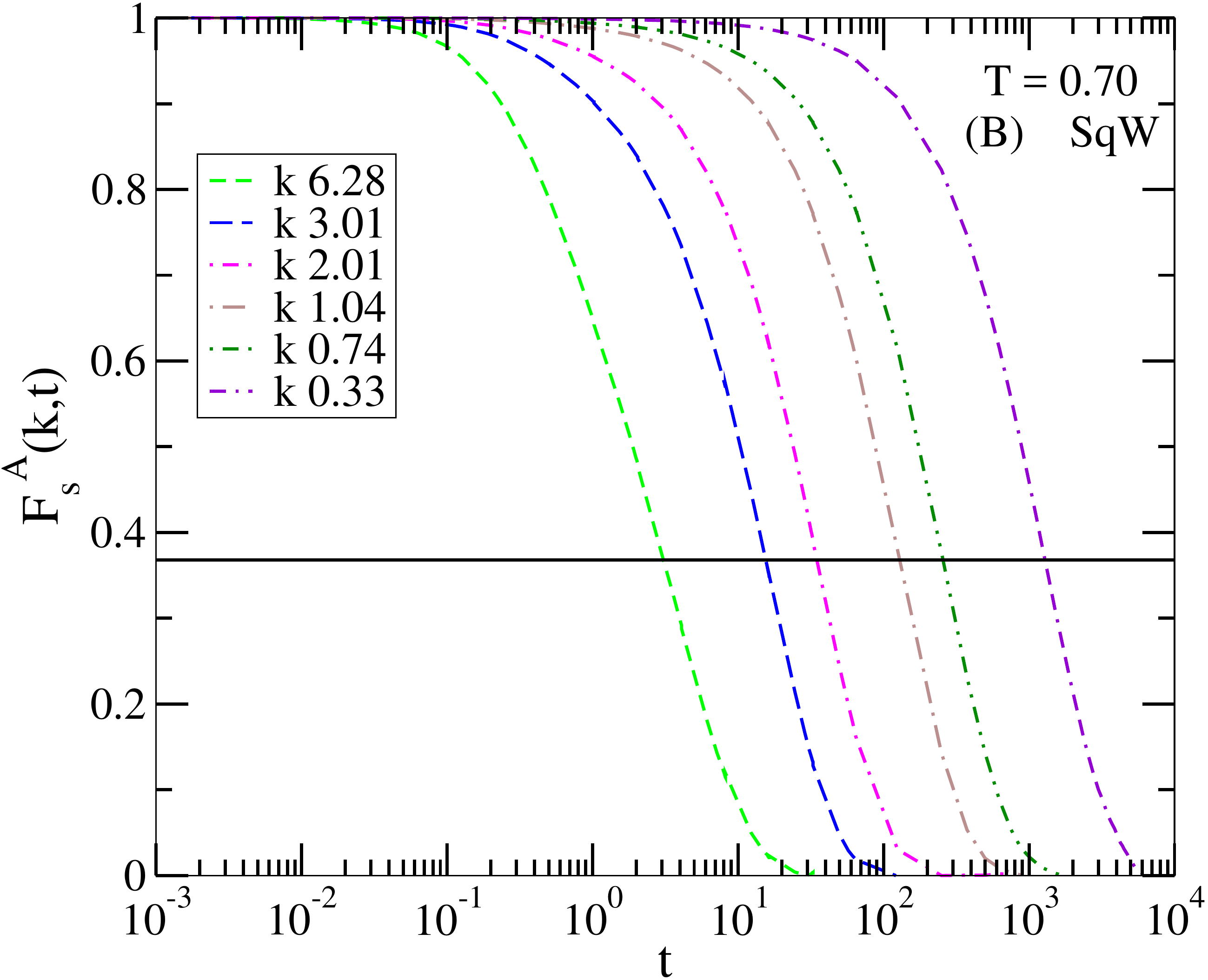}
\includegraphics[scale=0.245]{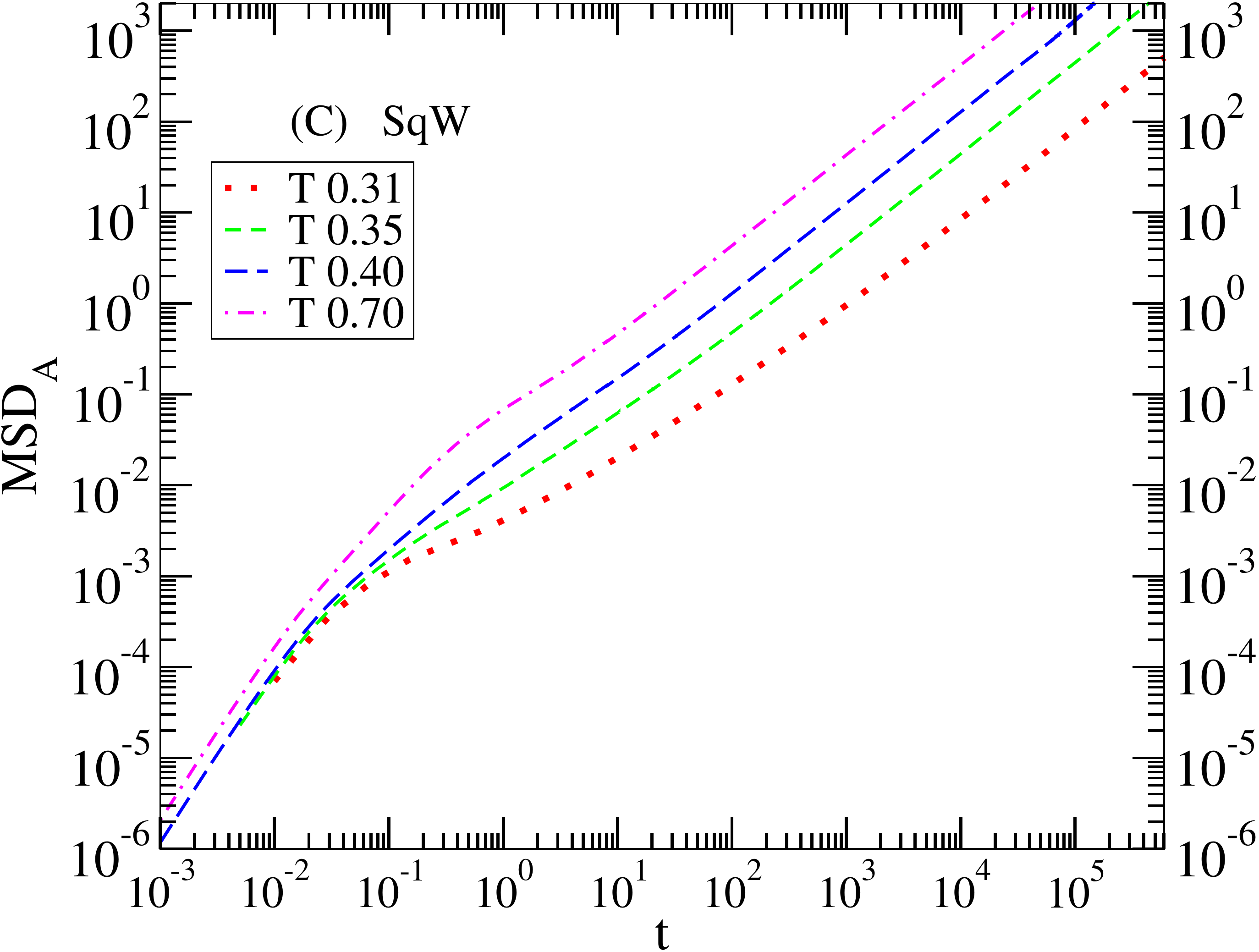}
\includegraphics[scale=0.245]{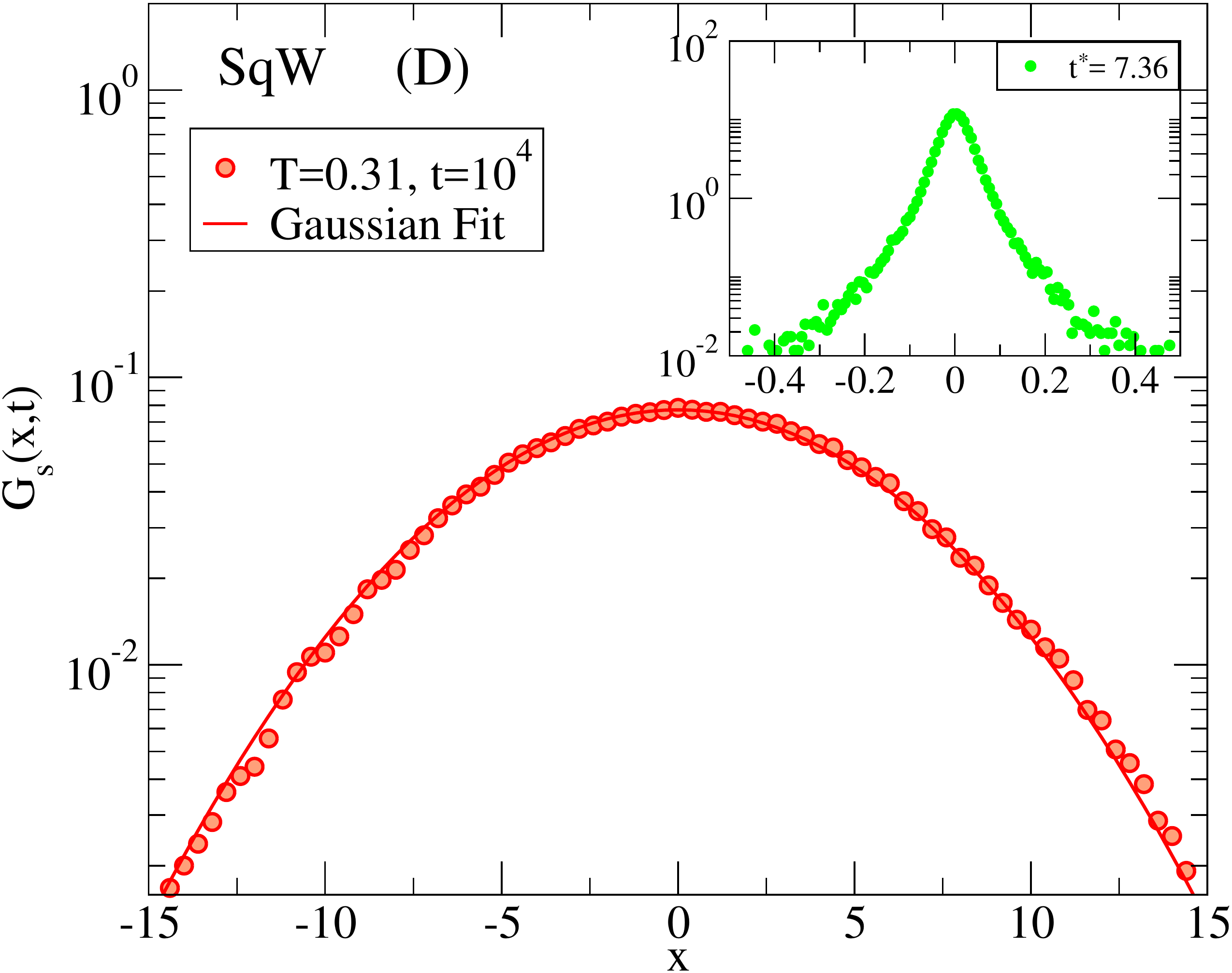}
\caption{ The self-intermediate scattering functions covering the full $k$ range studied at the lowest \emph{(A)} and the highest \emph{(B)} reported temperatures, \emph{(C)} the mean squared displacements for the full temperature range, and \emph{(D)} the self van Hove function, {for A type of particles}, at a long time-interval at the lowest temperature. The inset shows the self van Hove function at time $t^*$ (peak time of non Gaussian parameter, {for particle type A}), due to dynamical heterogeneous dynamics system exhibits strong deviation from the Gaussian (normal liquid) behavior. {The self part of intermediate scattering function $F^A_s(k,t)$} clearly decay to zero at all temperatures and for all $k$, thus proving that the system is well equilibrated at all temperature and $k$ points. This is further supported by the MSD which goes to a well-defined ballistic regime at all temperatures studied, and the van Hove function being Gaussian at the lowest temperature after a long time.}
\label{}
\end{figure*}
\begin{figure*}[h!]
\centering
\includegraphics[scale=0.325]{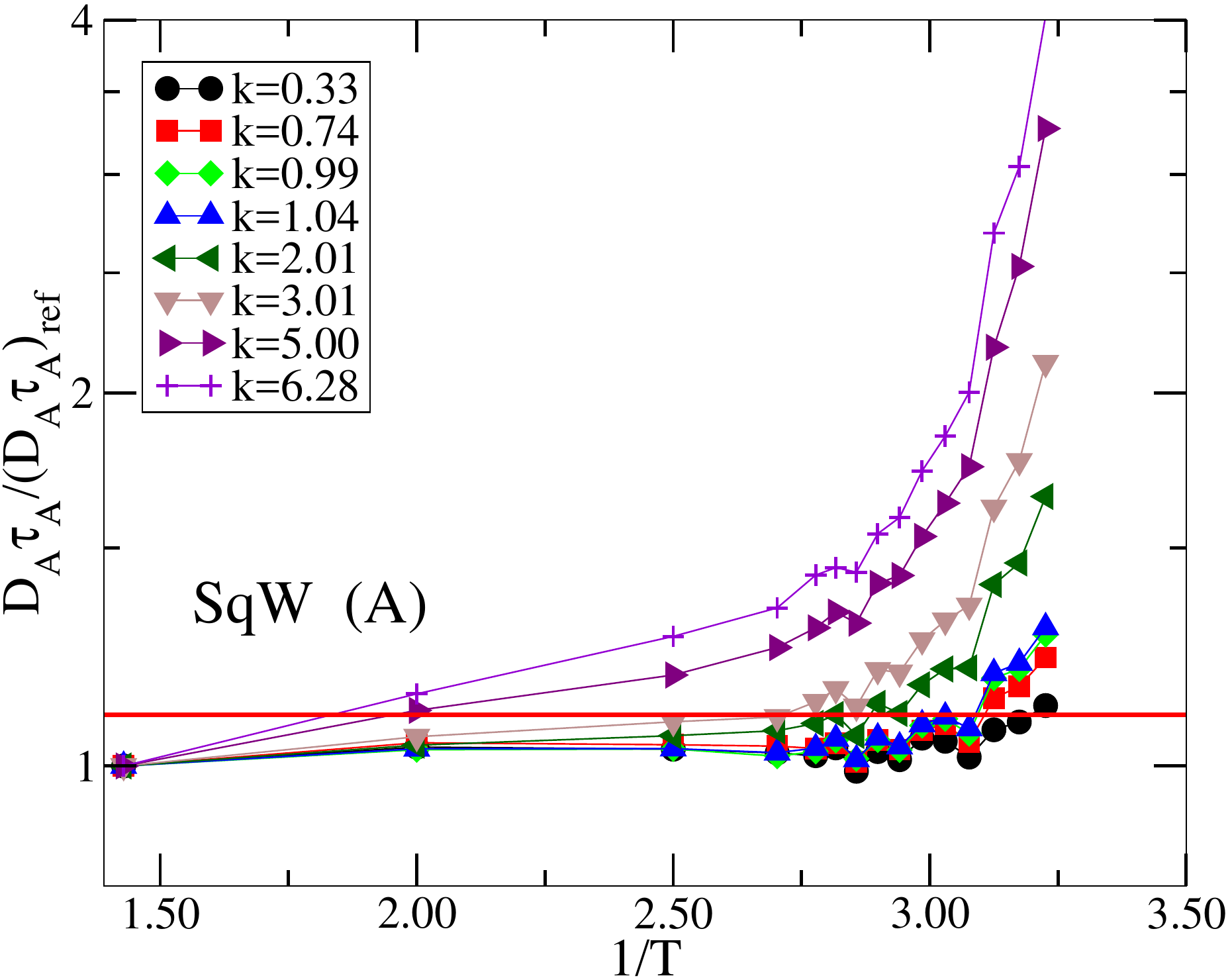}
\includegraphics[scale=0.325]{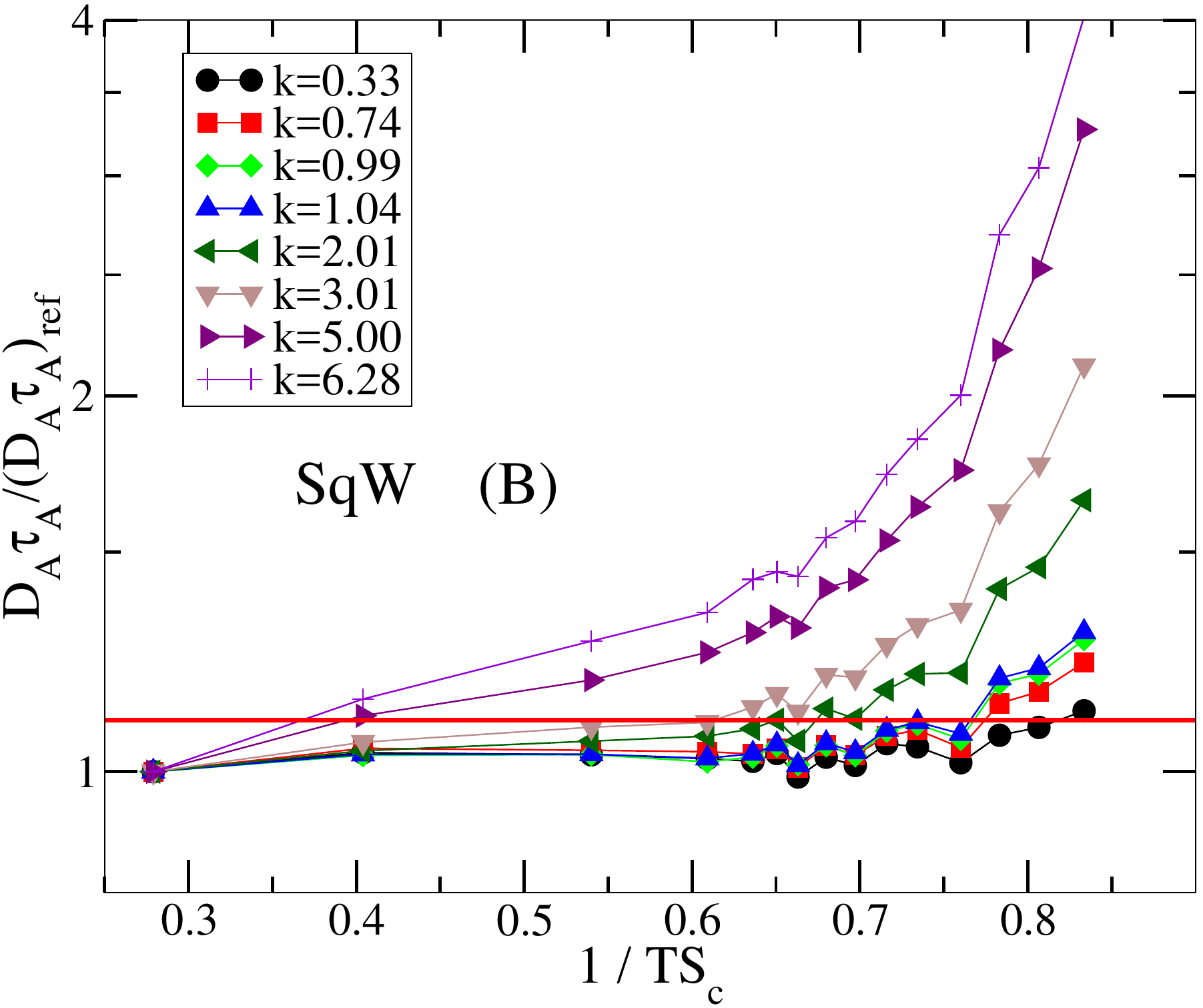}
\caption{Wave vector ($k$) dependence of the product {$D_A \tau_A(k)$} scaled by its value at a reference high temperature $T = 0.70$. For simple diffusive dynamics, this ratio should be 1. At a given $k$, the temperature where the ratio deviates from 1 (a threshold value = 1.1)\cite{Chong_s}, is taken to be the SE breakdown temperature $T_{SEB}$ for that $k$. Fig \emph{(A)} and \emph{(B)} shows the SE plot $vs.$ temperature ($T$) and  $1/TS_c$.}
\label{}
\end{figure*}

\clearpage

\subsection{Shear Viscosity}
The shear viscosity $\eta$ can be computed from the Green-Kubo and Einstein relation.  The Green-Kubo relation for the shear viscosity is defined from the integral of the  autocorrelation function of the stress  tensor ($J_{\alpha\beta}(t)$) -
\begin{equation}
\eta_{\alpha \beta} = \frac{V}{k_BT}\int_{0}^{\infty}dt \left< J_{\alpha  \beta}(t) J_{\alpha \beta}(0) \right>. 
\end{equation}
Further,  the stress tensor $J_{\alpha  \beta}$ is defined as 
\begin{equation}
J_{\alpha\beta}(t) = \frac{1}{V} \left( \sum_{i}^N \frac{p_{i\alpha}  
p_{i\beta}}{m} + \sum_{i}^N \sum_{j>i}^N r_{ij\alpha}f_{ij\beta} \right)
\label{stre_tensor}
\end{equation}
where, $[\alpha, \beta]$ are x, y and z components, $V$ is volume of the system of N particles, $m$ is mass of a particle, $p_i$ is the momentum of particle $i$, distance $r_{ij}=|\vec{r}_i -\vec{r}_j|$ and for the pair interaction $U(r_{ij})$ the force can be defined as $f_{ij}=-\frac{\partial U(r_{ij})}{\partial r}$.

The shear viscosity  also can be evaluated using Einstein relation 
\begin{equation}
\eta_{\alpha \beta} = \frac{1}{Vk_BT}\lim_{t\rightarrow \infty} \frac{\left < (A_{\alpha 
\beta}(t) - A_{\alpha \beta}(0))^2 \right >}{2t},
\label{Hel_num}
\end{equation}
where, the  Helfand moment $A_{\alpha \beta}(t)$  is described as 
\begin{equation}
\frac{dA_{\alpha \beta}(t)}{dt} = J_{\alpha \beta} V,
\end{equation}
also, 
\begin{equation}
A_{\alpha \beta}(t) - A_{\alpha \beta}(0) = V\int_{0}^{t}dt'J_{\alpha \beta}(t')
\label{int_num}
\end{equation}
The integral in the Eq. (\ref{int_num}) has been estimated numerically, which is not  affected by the periodic boundary conditions used in the MD simulations \cite{Leporini2001_s}. In the large time limit, the linear behavior of  $\left < (A_{\alpha\beta}(t) - A_{\alpha \beta}(0))^2 \right >$ (where, $\alpha, \beta \in [x,y,z], \alpha \ne \beta$) provides the estimate of the shear viscosity.

In the Kob-Andersen model particles are interacting $via$ continues model and  Helfand moment is computes from the stress tensor as defined in Eq. (\ref{stre_tensor}). In the Fig \ref{fig:etaKA}B, we show that computed viscosity is proportional to the $\tau'(k=7.25)$. 

 But, in the SqW model, the particles interaction is discrete and hence particles travel with constant velocity between two collisions ($\Delta t$, time between successive collisions). Hence, Helfand moment can be rewritten as \cite{haile_s,Alder_s,HM_s}

\begin{eqnarray}
\Delta A_{\alpha \beta}(t)  = m \sum_{\text{coll, s}}^t \Delta &t& {\sum_{i}^{N} \dot  r_{i,\alpha}(s) \dot r_{i,\beta}(s)} \nonumber \\ 
  &+& m \sum_{coll, s}^t  {\Delta \dot r_{i,\alpha}(s)\, r_{ij,\beta}(s)}
\label{eq10}
\end{eqnarray}

where, the first term is the kinetic part, evaluated at time $s$ and multiplied by the time interval between two successive collisions, the second term represents the change in the velocity between a colliding pair of particles $i$ and $j$. Further, the shear viscosity $\eta_{\alpha\beta}$ can be computed using Eq.  (\ref{Hel_num}). For the SqW model, we find that the shear viscosity is proportional to the relaxation time $\tau(k=2\pi)$ [see Fig \ref{eta_sqw}].

\begin{figure}
\centering
\includegraphics[scale=0.42]{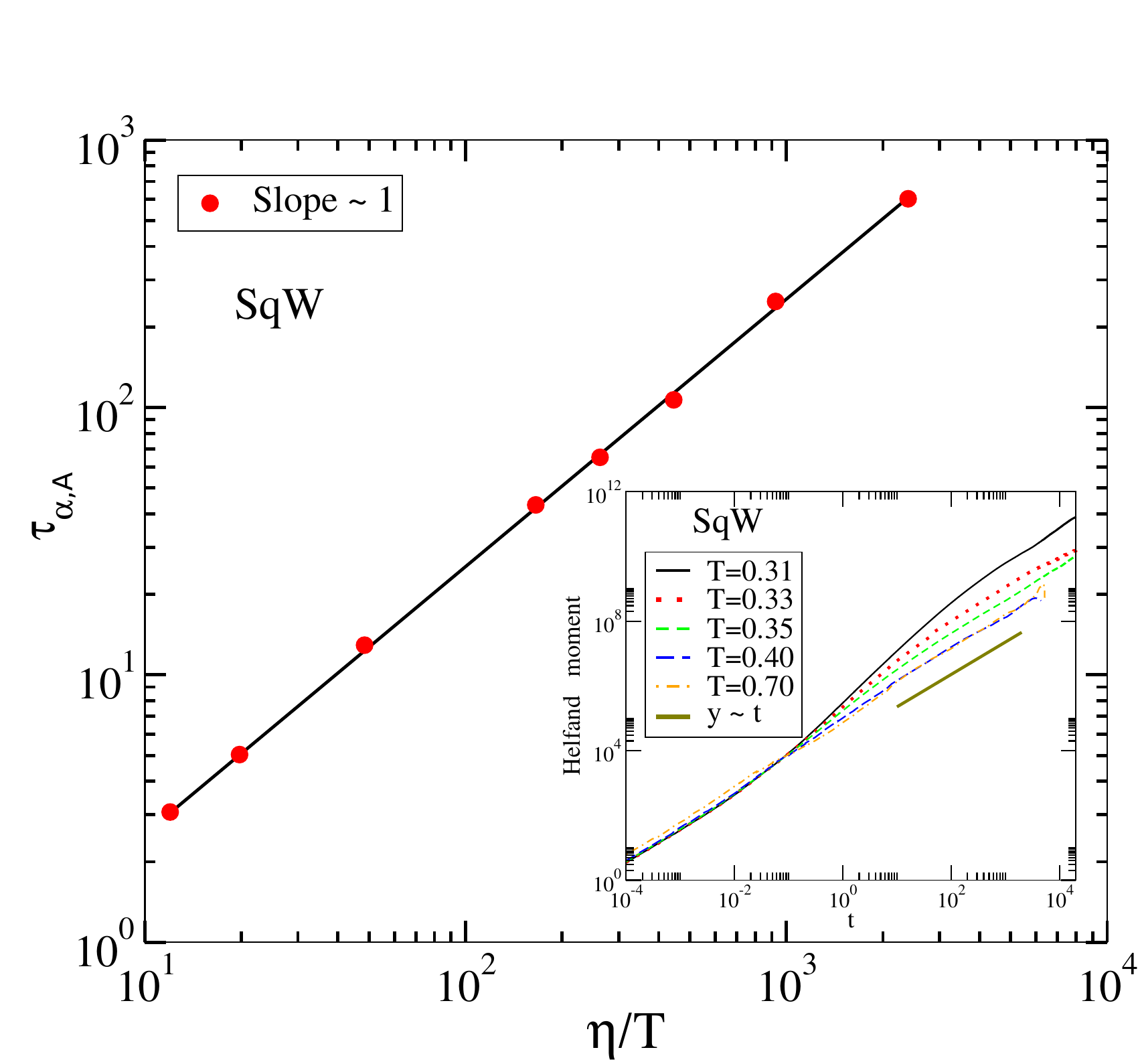} 
\includegraphics[scale=0.30]{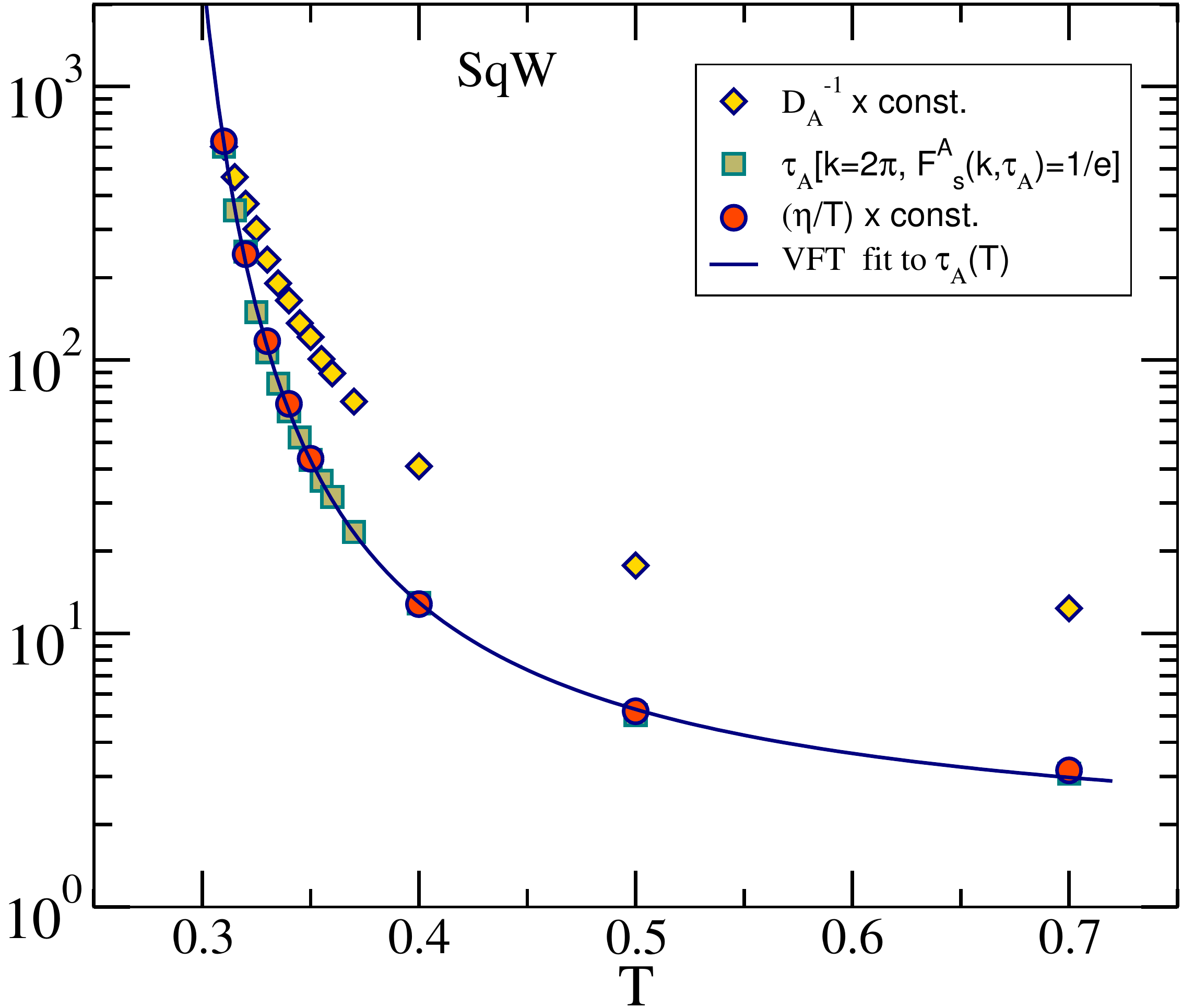}
\caption{The self-part ($\tau_A$) of the intermediate scattering function  {$F^A_{s}(k=2\pi,t)$} of type A particles plotted against viscosity showing that {$\tau_A(k=2\pi)$} $\propto \eta /T$ is a good description of data. In the presence of SE breakdown, the shear viscosity does not follow the diffusivity.}
\label{eta_sqw}
\end{figure}

\section{Dynamical heterogeneity}
\subsection{Four point correlation ($\chi_4(t)$) and Non-Gaussian parameter  {$(\alpha^A_2(t))$}}
The non-Gaussian parameter peaks at a time $t^*$, corresponding to the crossover between the so-called cage regime and the diffusive regime of the MSD. The  {maximum value of the parameter $\alpha_2^A (t^*)$}  is a measure of the heterogeneous dynamics, noticeably this value is quite high compared to well studied KA model \cite{kob1997_s}. Further to understand the DH, we study the particle displacement and the morphology of the mobile clusters.
\begin{figure}[h!]
\centering
\includegraphics[scale=0.235]{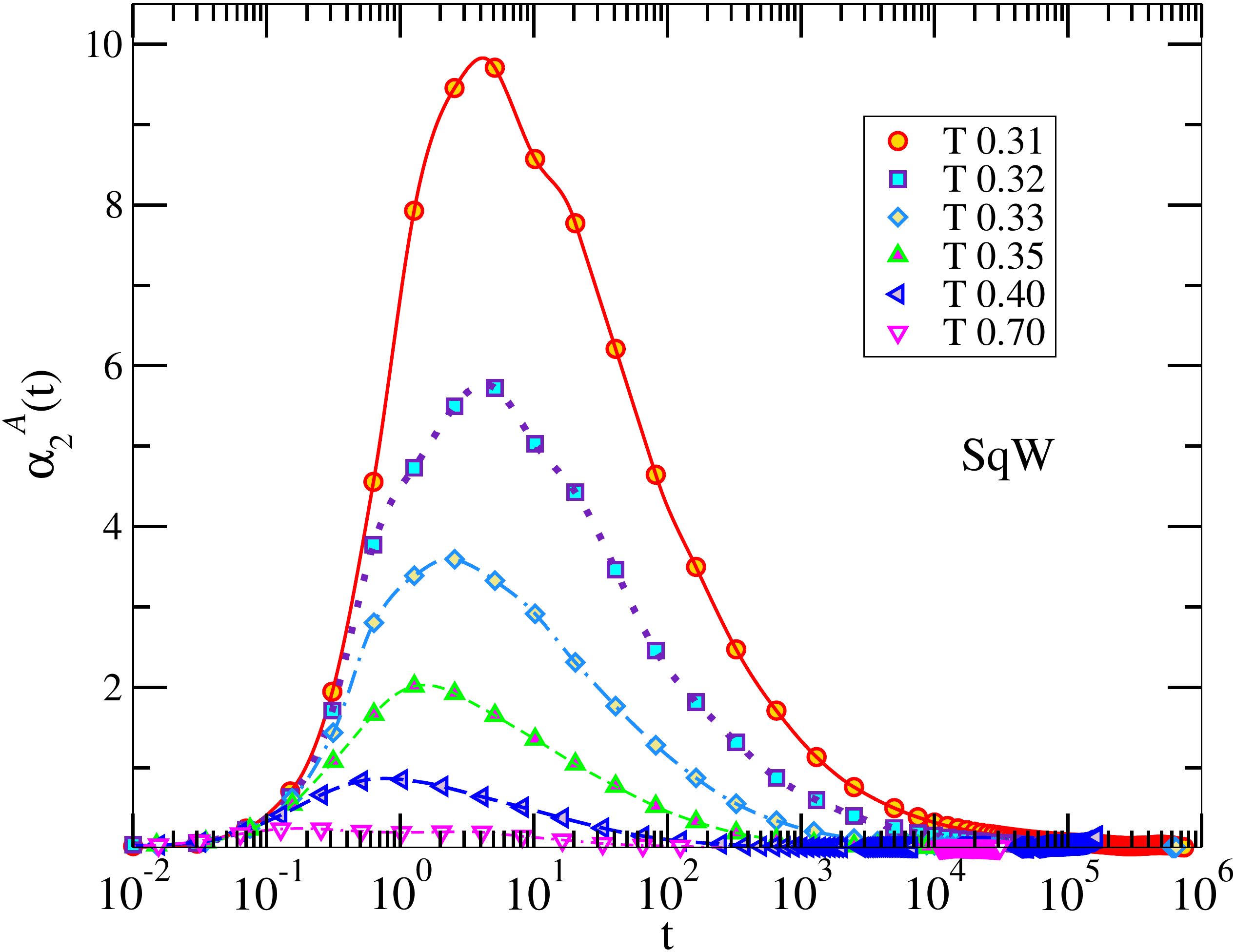}
\includegraphics[scale=0.235]{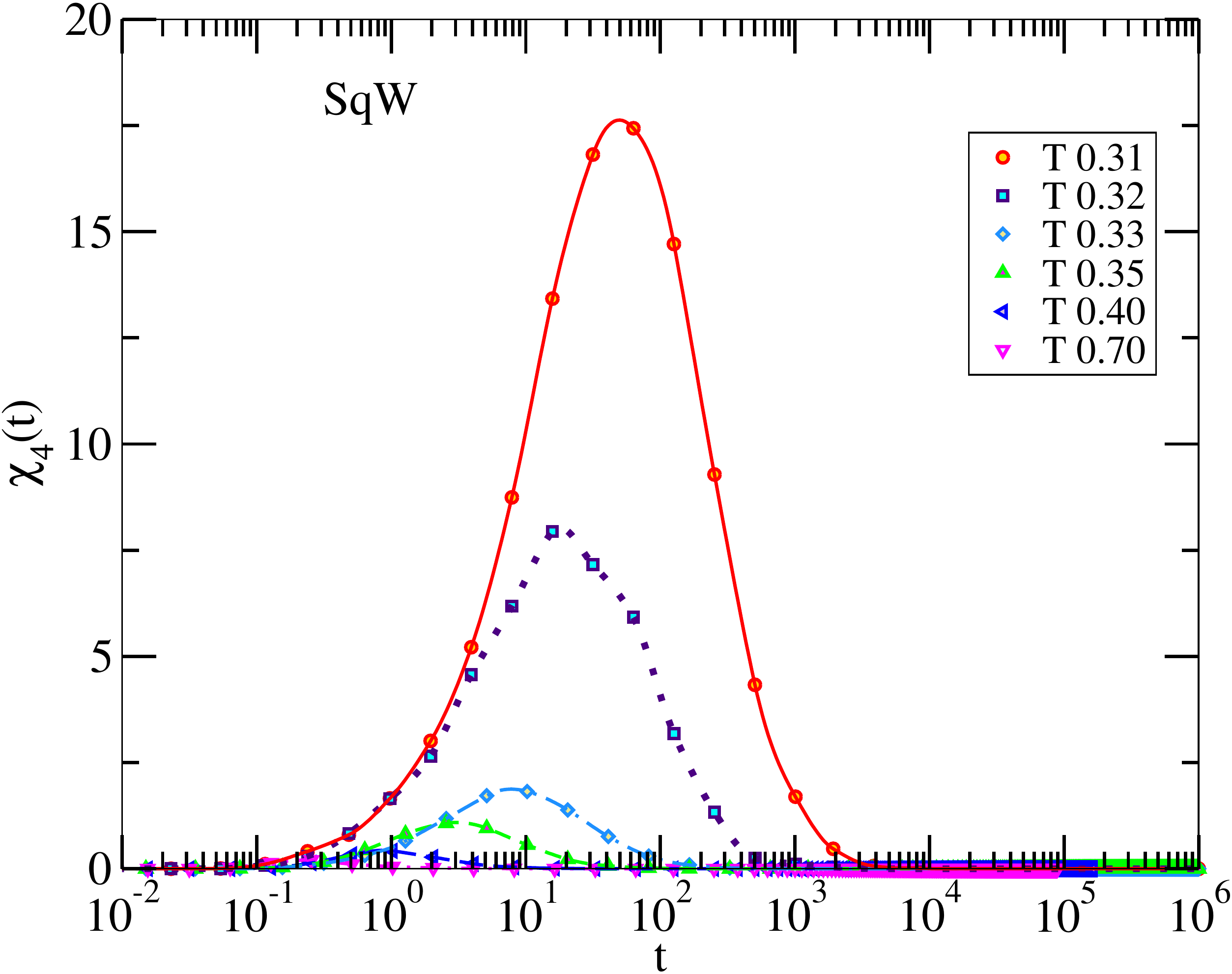}
\includegraphics[scale=0.32]{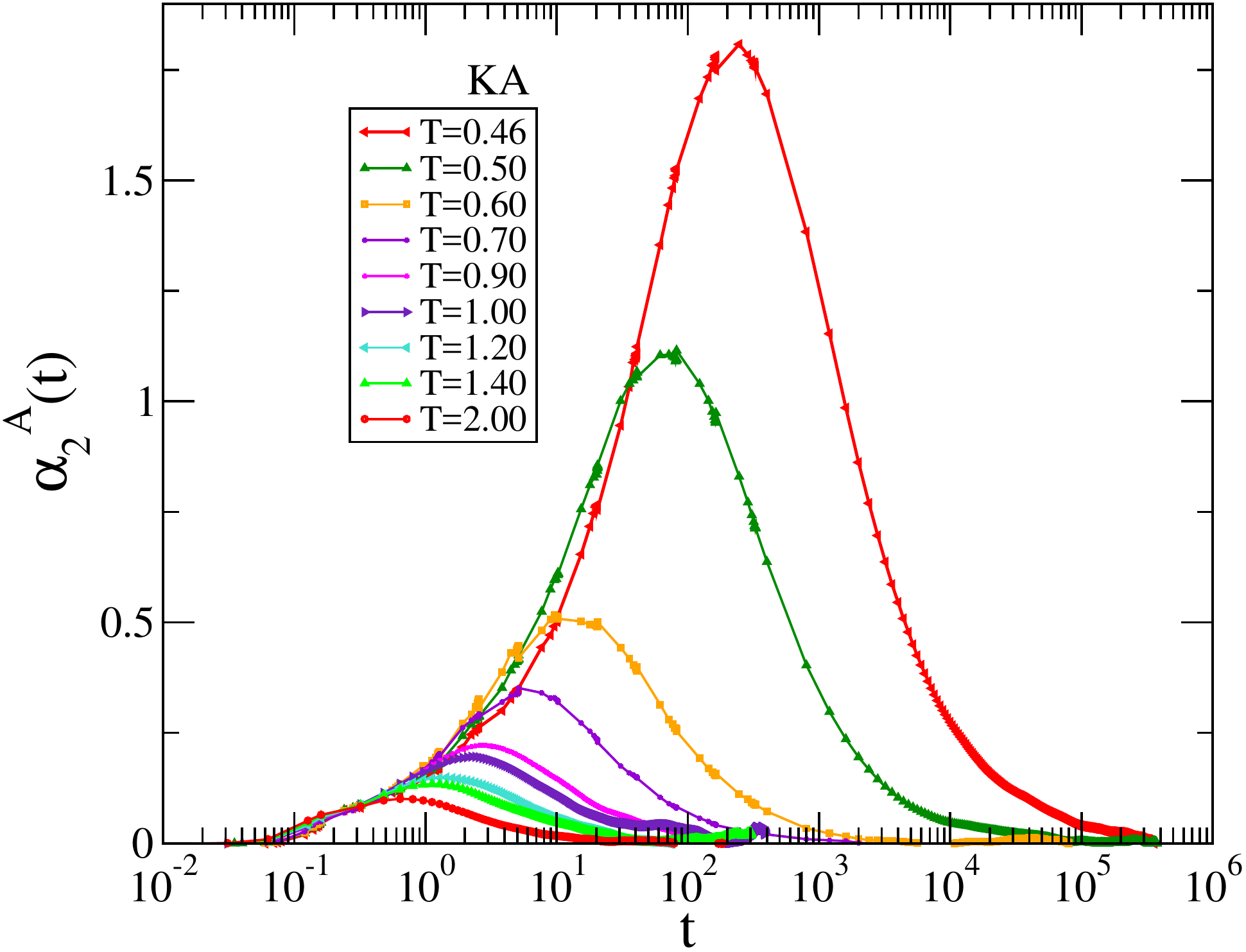}
\includegraphics[scale=0.32]{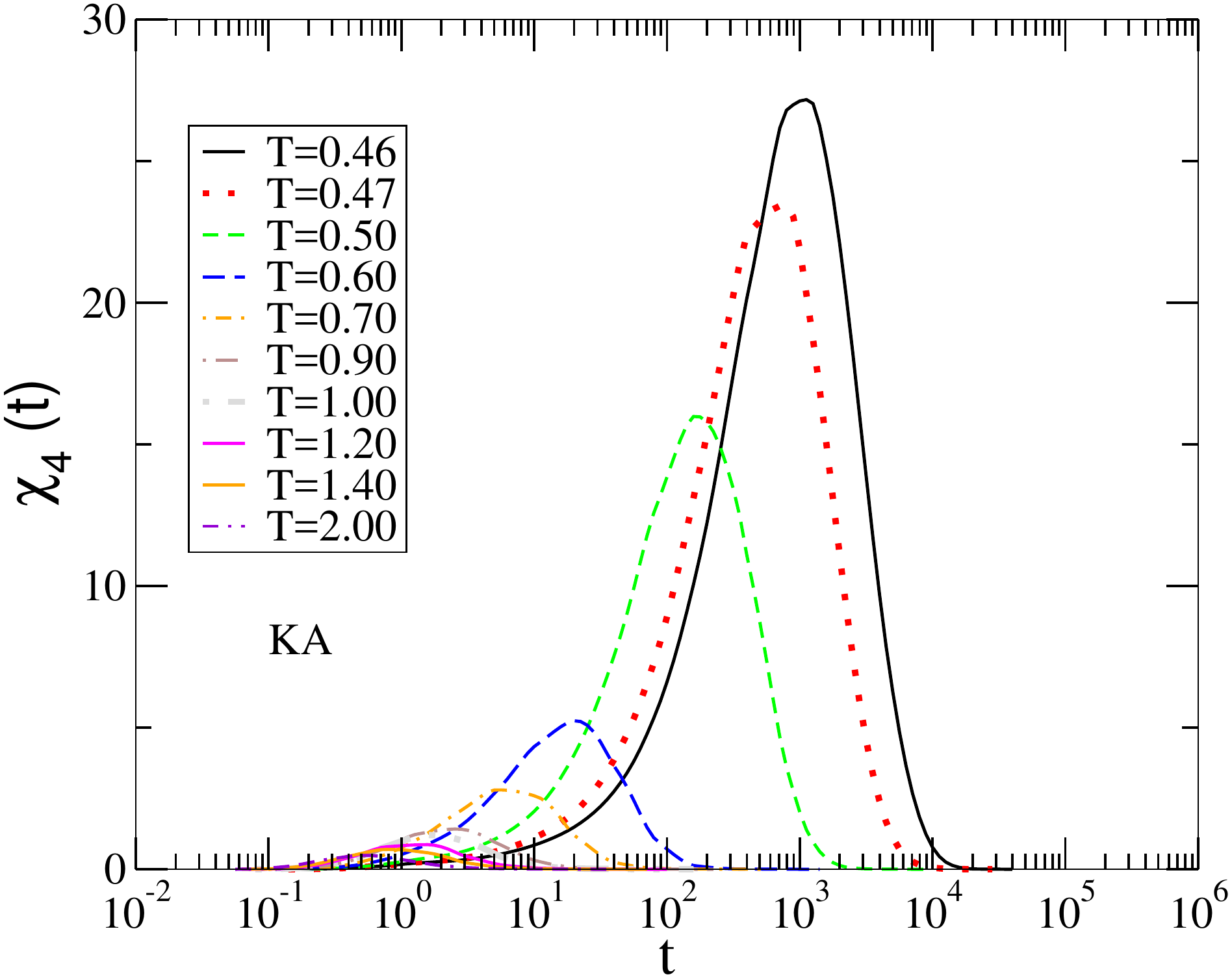} \vspace{-2mm}
\caption{The non-Gaussian parameter {for particles of type A ($\alpha^A_2(t)$)} and four-point correlation function ($\chi_4(t)$) shows a peak due to heterogeneous dynamics. The peak value of the $\alpha_2^{A}(t)$ and $\chi_4(t)$ for SqW model are quite high compare to KA model.}
\label{}
\end{figure} 

\subsection{The van-Hove function}
The van Hove distribution function is a dynamical correlation function which characterizes the spatial and temporal correlation of a pair, \emph{i.e.} the probability of finding a particle at time $t$ at distance $r$ from its position at time $t=0$. \\
  We define a Gaussian approximation to the self part of the van Hove function as
\begin{equation}
G_s^g(r,t) = \left (\frac{3}{2 \pi <r^2(t)>}\right )^\frac{3}{2} exp \left (-\frac{3 r^2}{2<r^2(t)>}\right )
\label{eq1}
\end{equation}
The deviation from the Gaussian approximation can be related to the presence of the  dynamical heterogeneity. {In this study van Hove function has been estimated for ``$A$" type of particles}.
\begin{figure}[htp]
\centering
\includegraphics[scale=0.35]{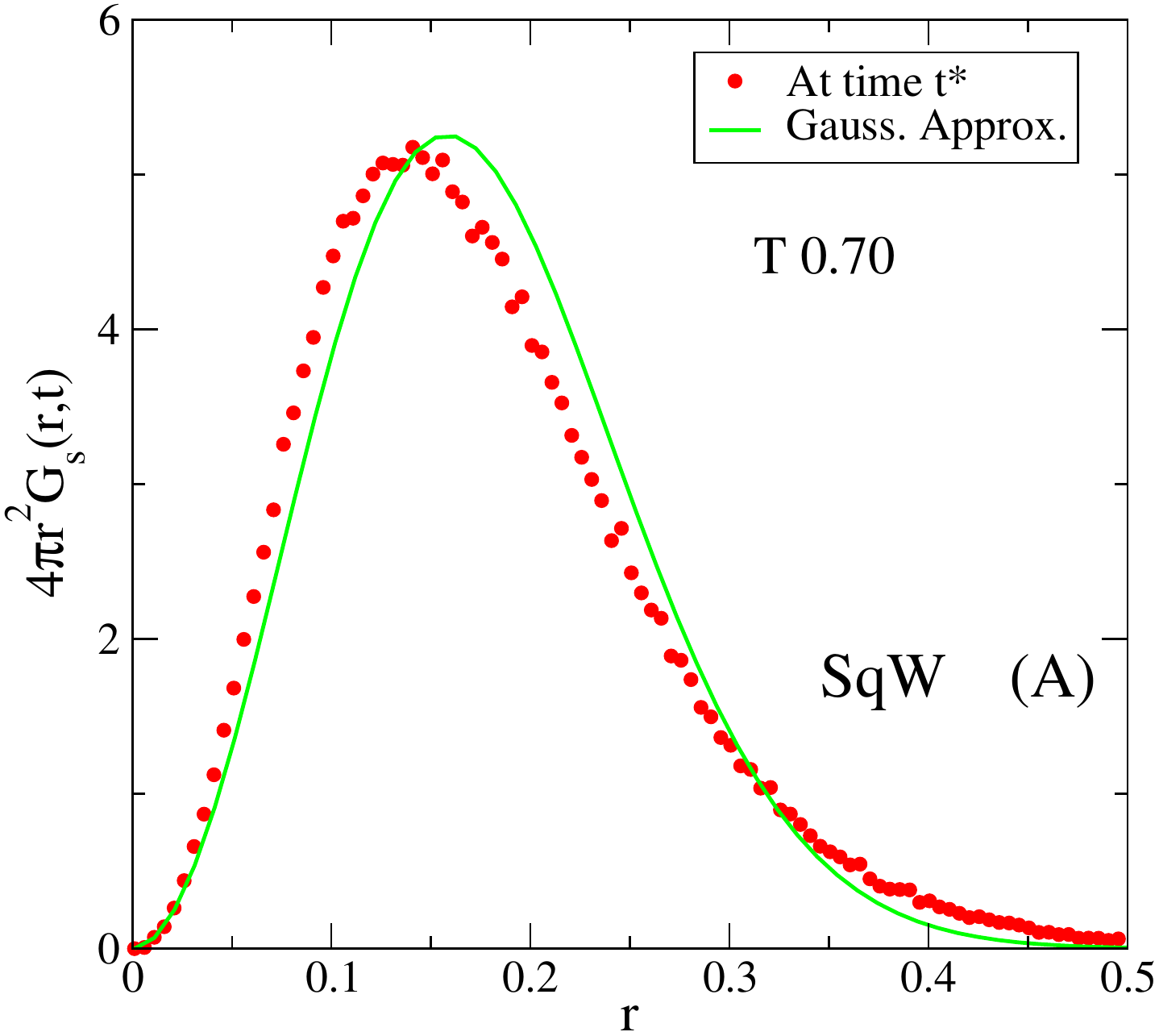} 
\includegraphics[scale=0.35]{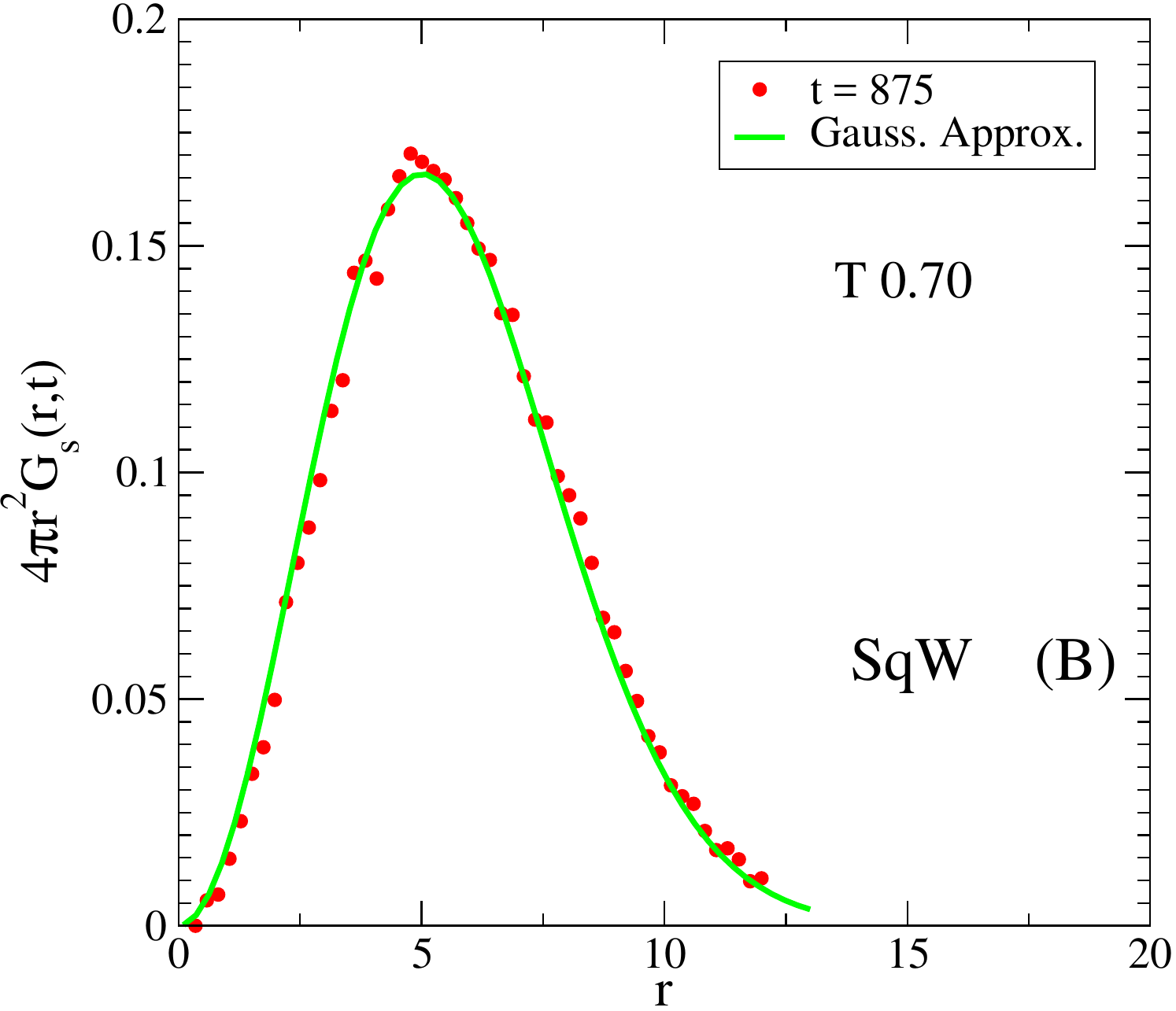} \vspace{-2mm}
\caption{A- At time $t^*$ the van Hove is quite different from the Gaussian approximation (Green line). Particles whose displacements are in the tail end, where probability of the displacement is higher than that given by Eq. (\ref{eq1}), are considered to be mobile particles. B- At large time, where $\alpha_2$ is quite small ($\approx 0$), $G_s(r,t)$ matches with Gaussian approximation of the probability density for diffusive particles.}
\label{}
\end{figure} 

\begin{figure*}[]
\centering
\includegraphics[scale=0.38]{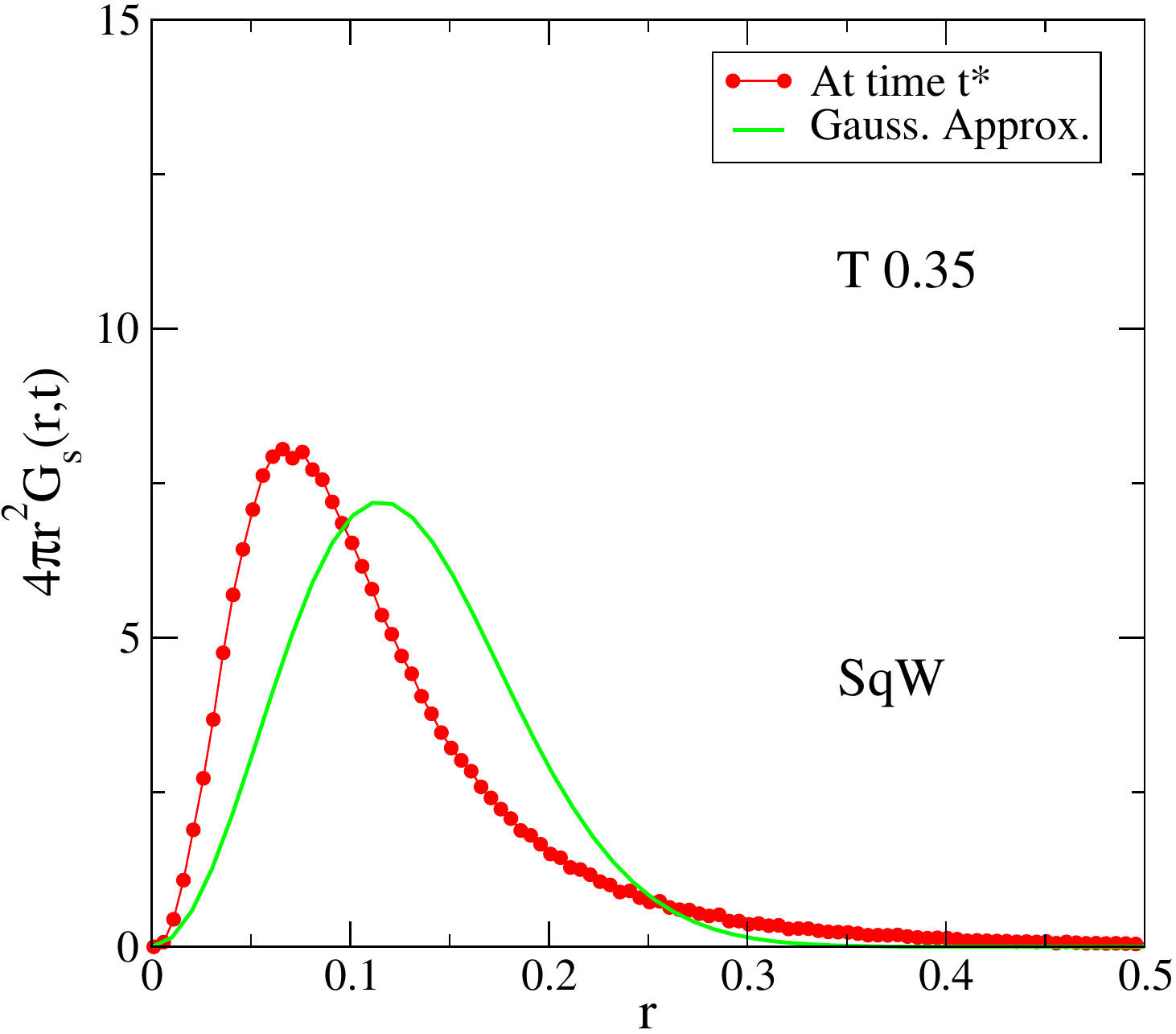}
\includegraphics[scale=0.38]{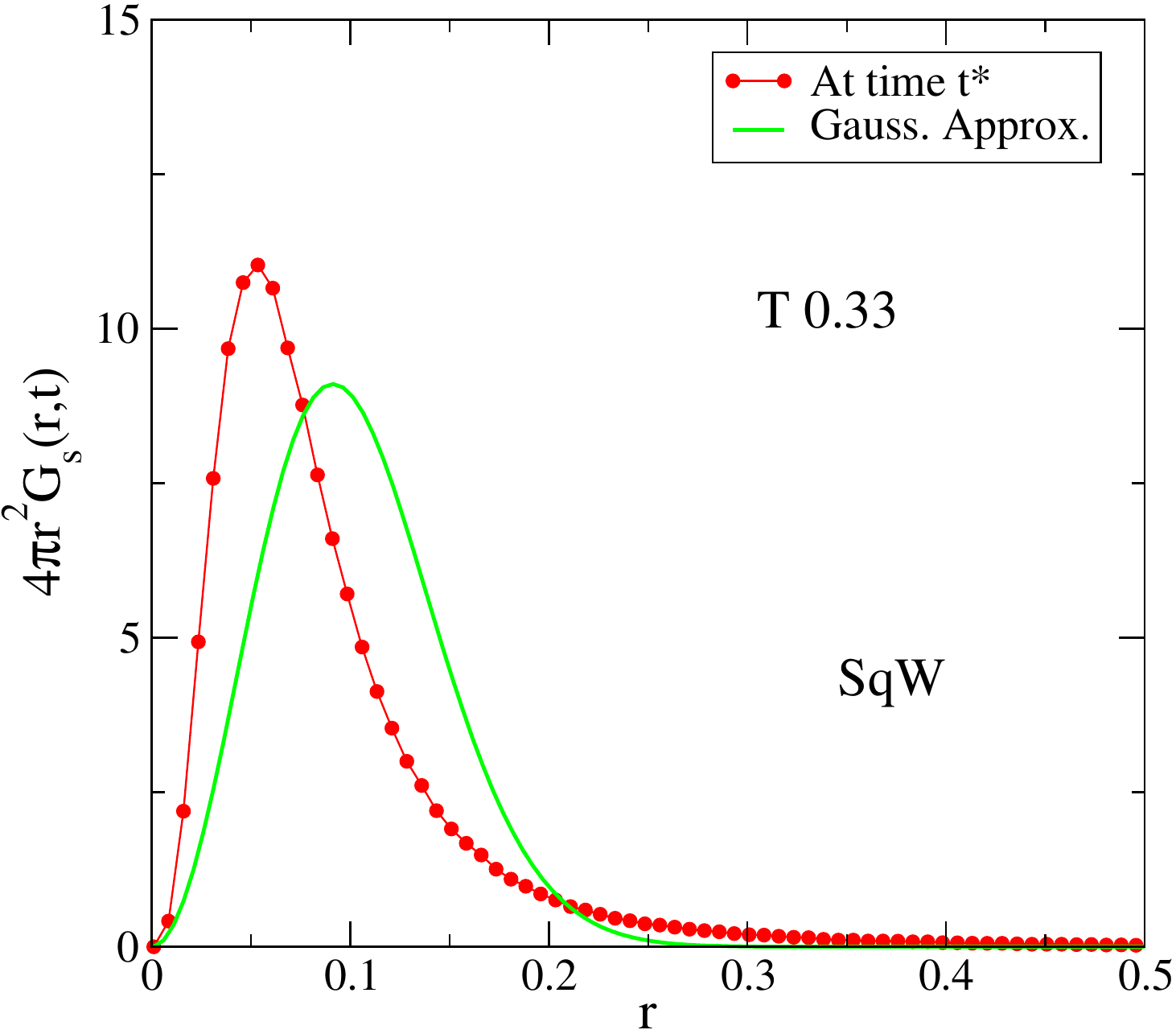} \\
\includegraphics[scale=0.38]{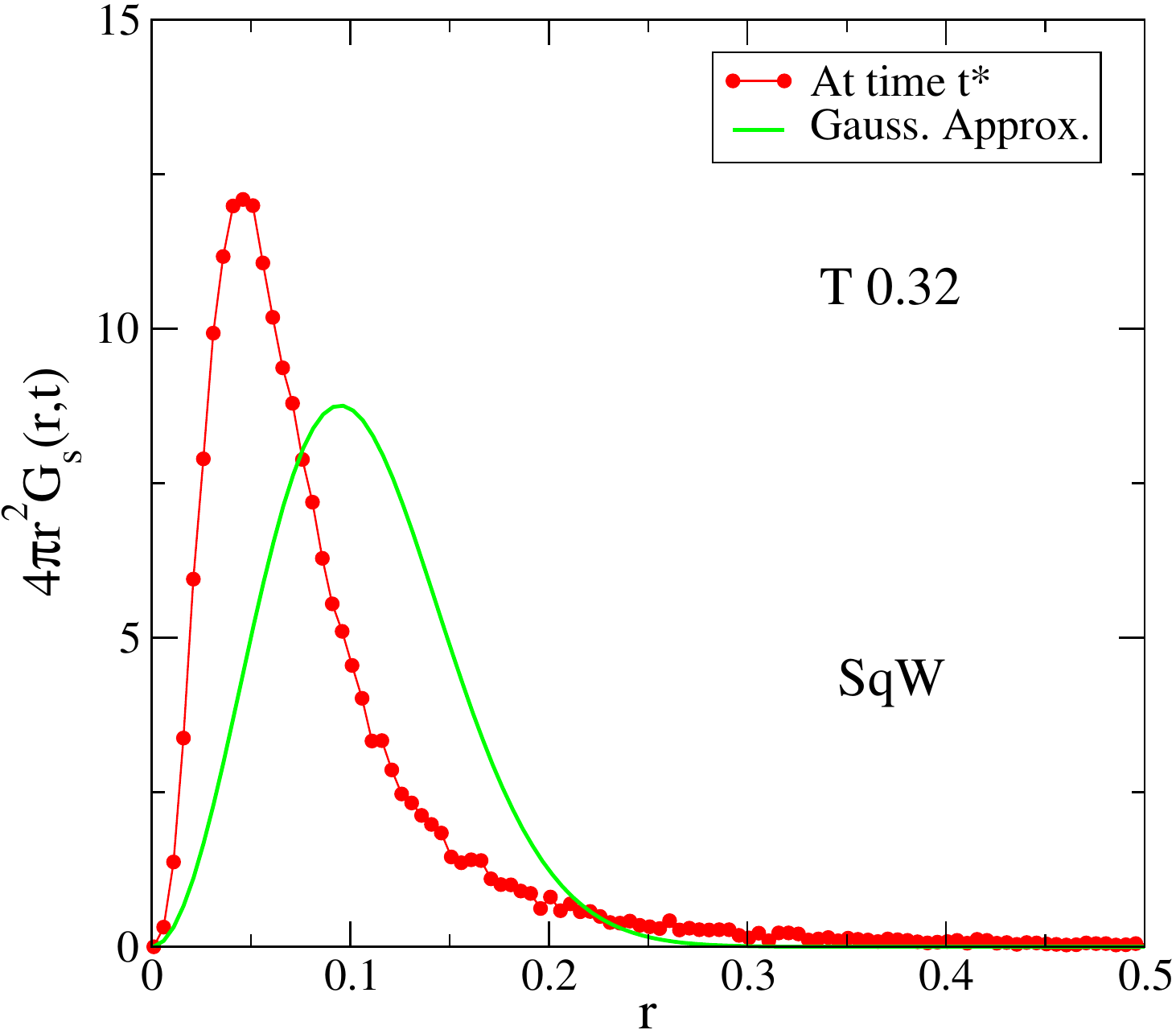}
\includegraphics[scale=0.38]{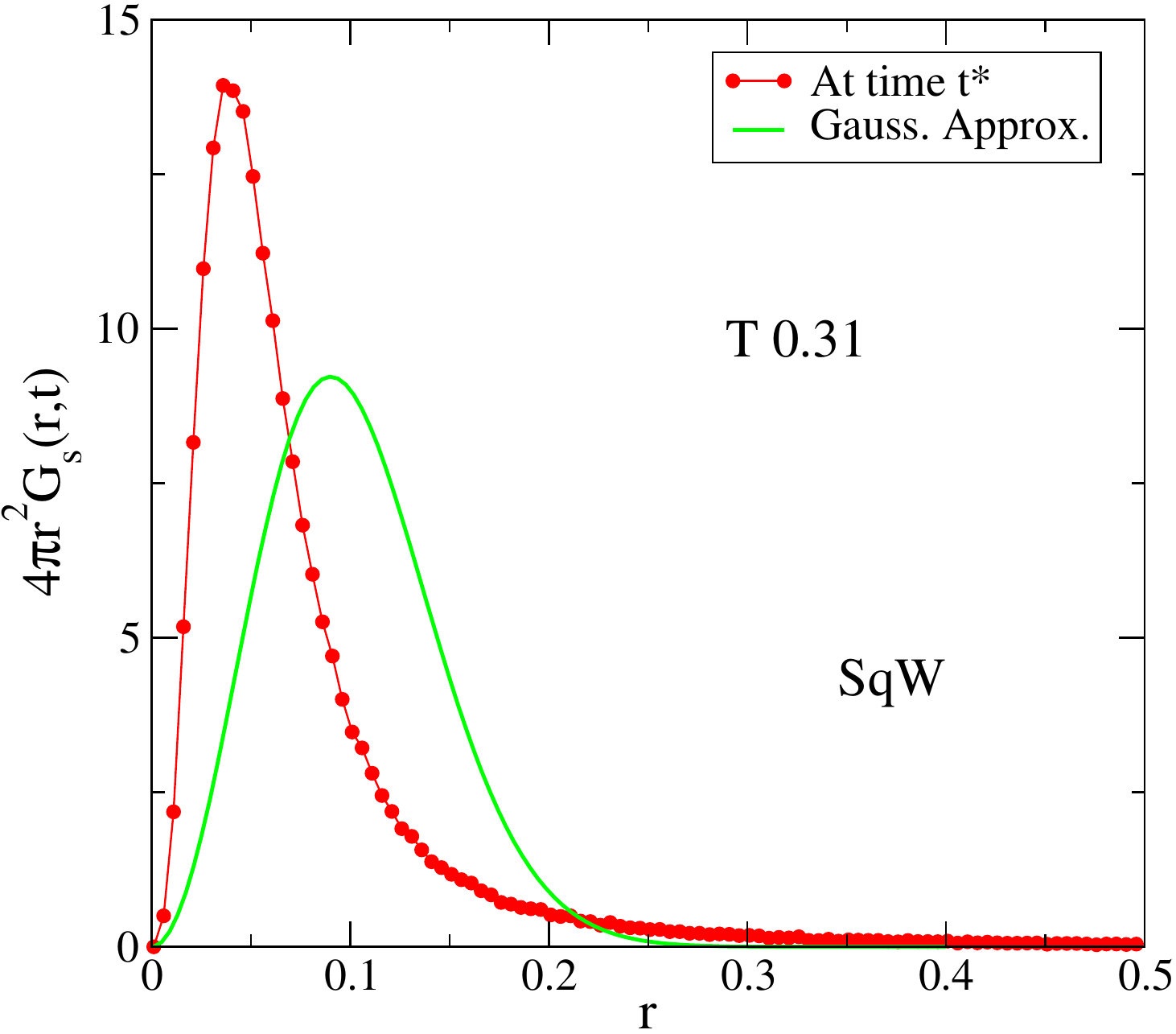}
\caption{The square well liquid manifests significant deviation from the normal liquid behavior at low temperatures.}
\label{}
\end{figure*} 

\subsubsection{ Time evolution of the self Van Hove function at different temperatures}
The van Hove function has been studied at various time, the range of the time is taken in the window of $F_s(k=2\pi,t)$. The distribution becomes bimodal for the low temperatures (especially below $T_{SEB}=0.33$).
\begin{figure*}[htp]
\centering
\includegraphics[scale=0.40]{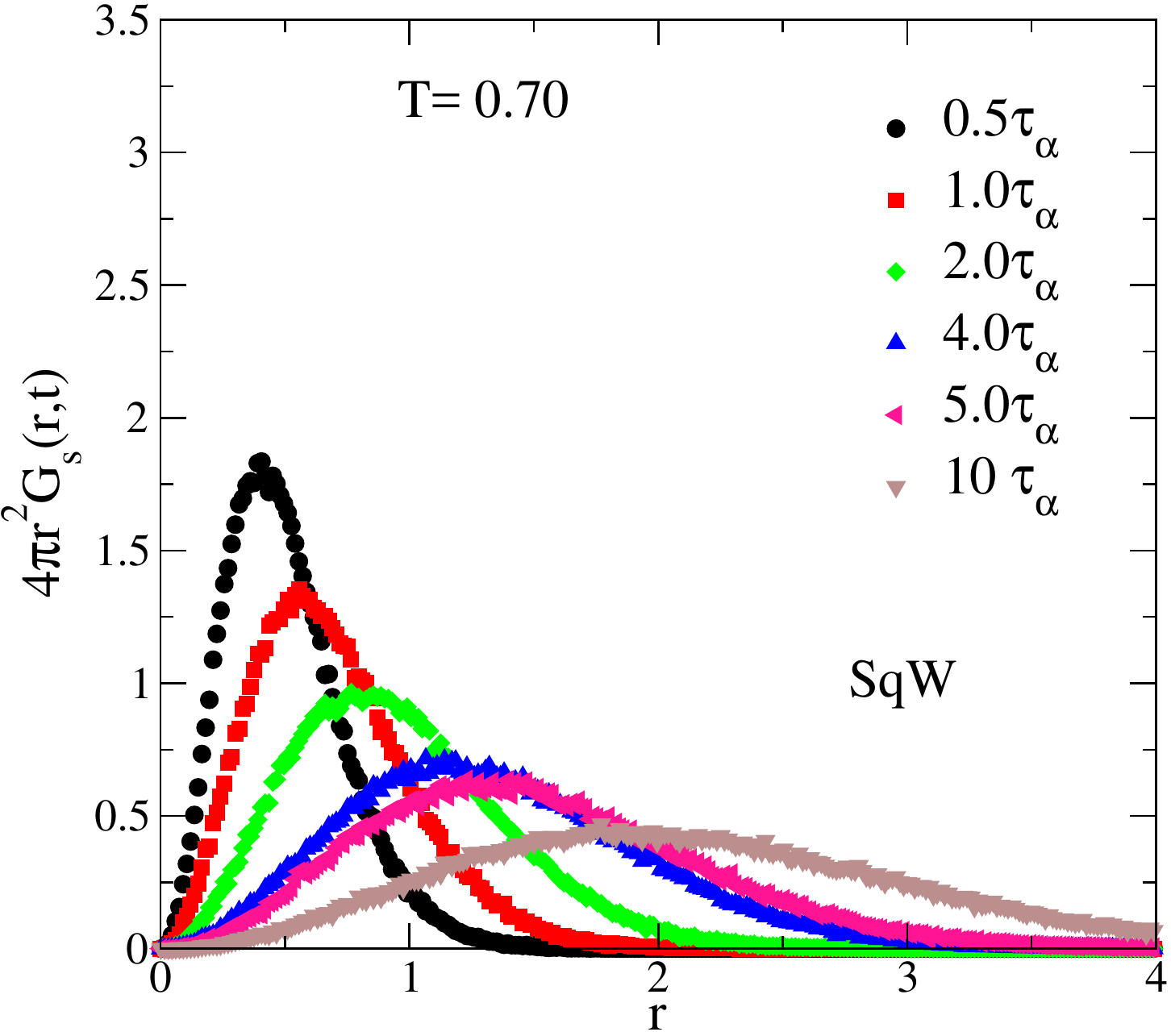}
\includegraphics[scale=0.40]{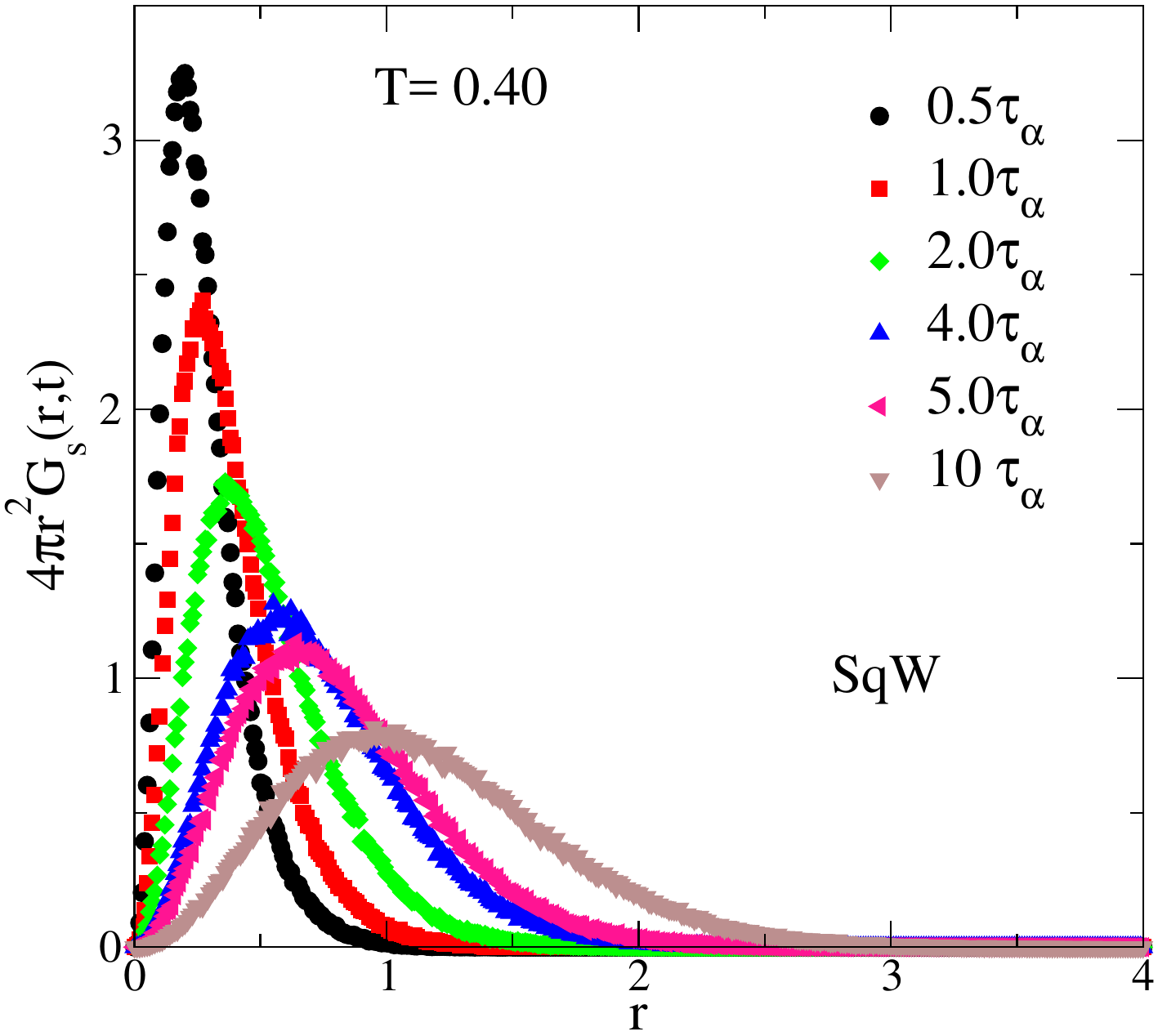}
\includegraphics[scale=0.40]{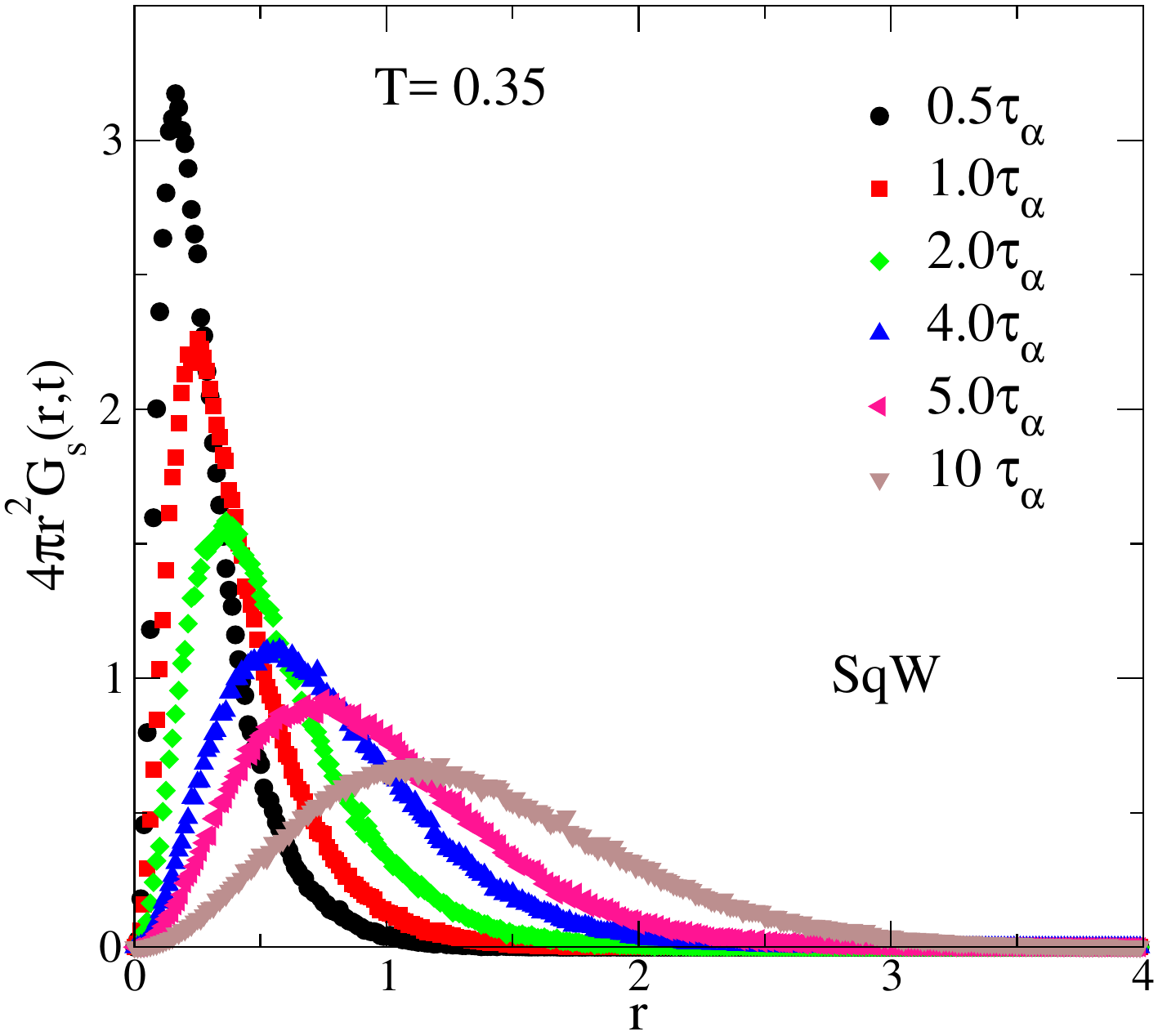}
\includegraphics[scale=0.40]{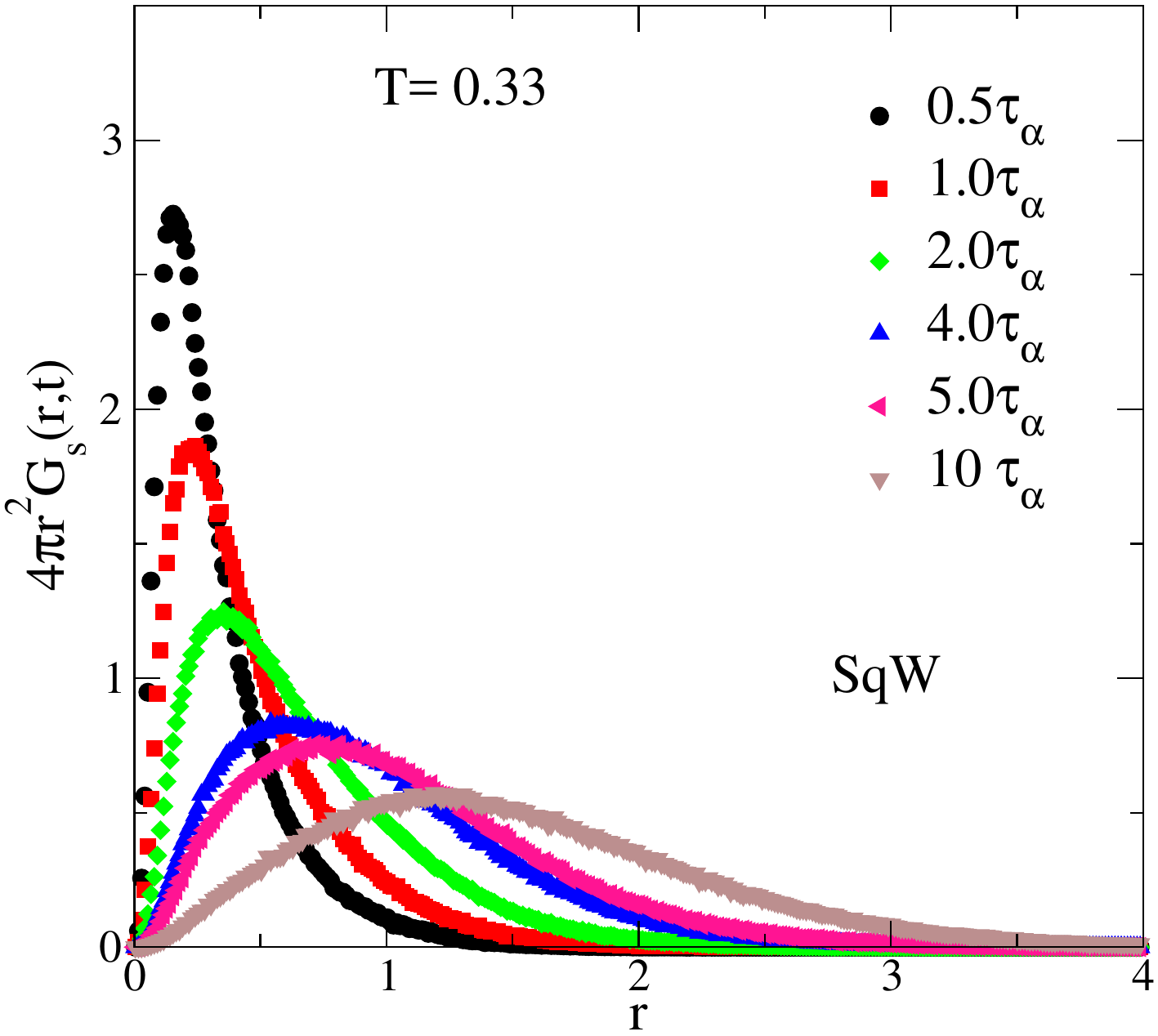}
\includegraphics[scale=0.40]{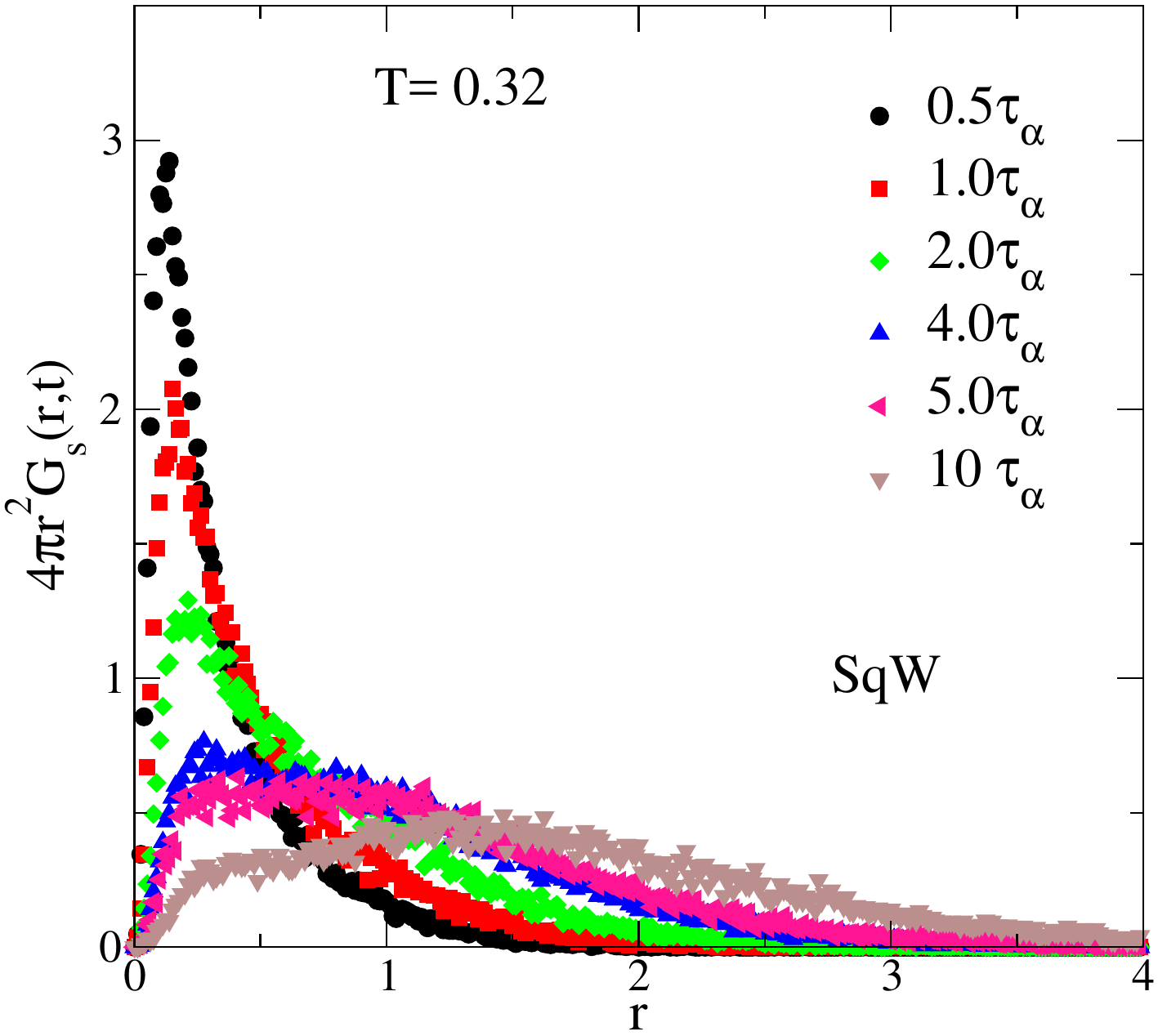}
\includegraphics[scale=0.40]{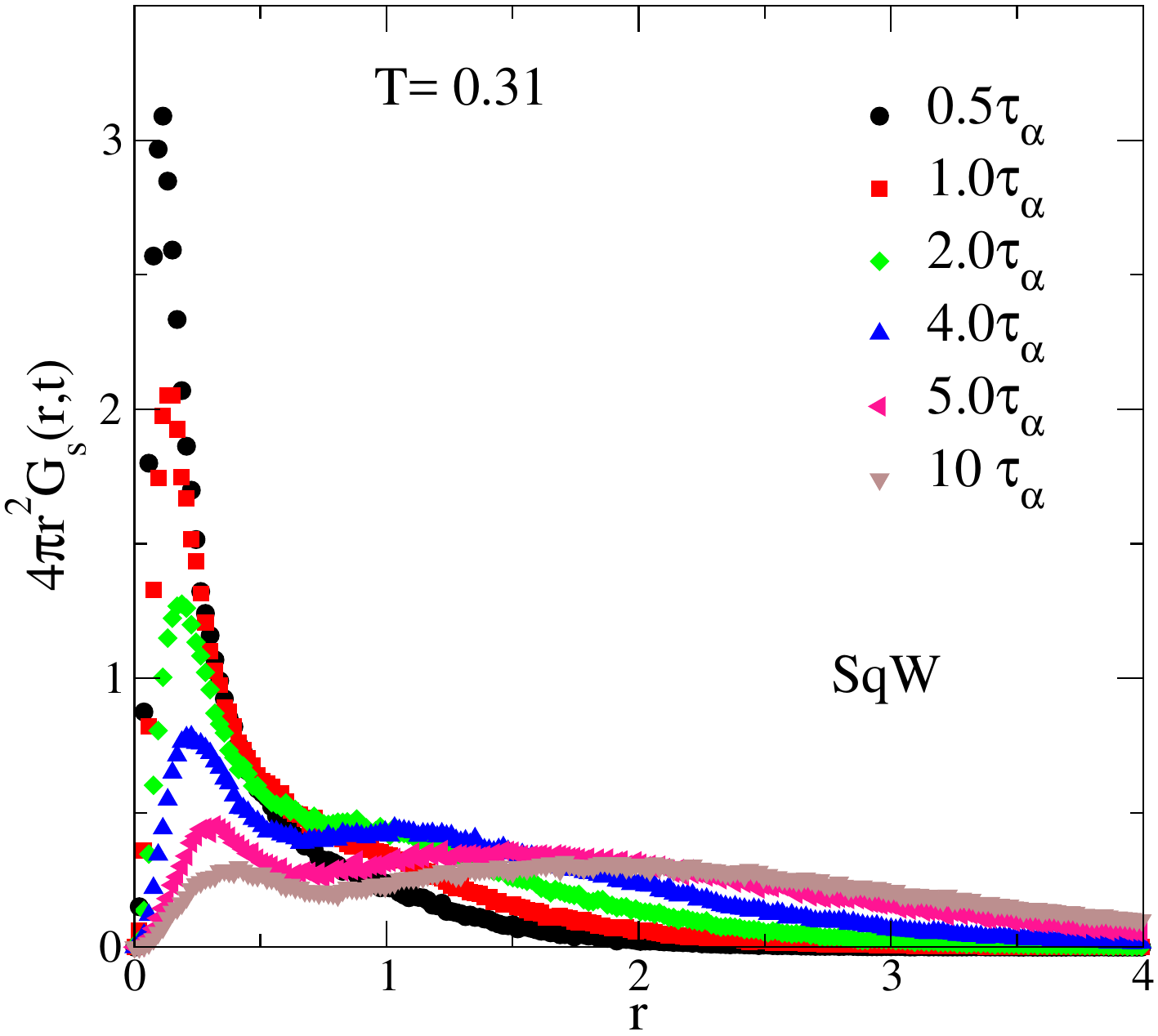}
\caption{At the lower temperatures the dynamics is quite different compared to high temperature liquid, in the intermediate time scales (before diffusion) system exhibits jump dynamics.}
\label{}
\end{figure*} 

\subsection{Morphology of the active particles (SqW model)}

At time $t^*$, top 10\% of the fast moving particles are considered as mobile particles ~\cite{AG-DH_s}. These particles are considered to be in the same cluster  if they are present in the first coordination shell ($\approx 1.7$).  The cluster size of these particles increases at lower temperatures. The morphology of the clusters of these mobile particles has been studied at time $t^*$.
\begin{figure}[htp]
\centering
\includegraphics[scale=0.40]{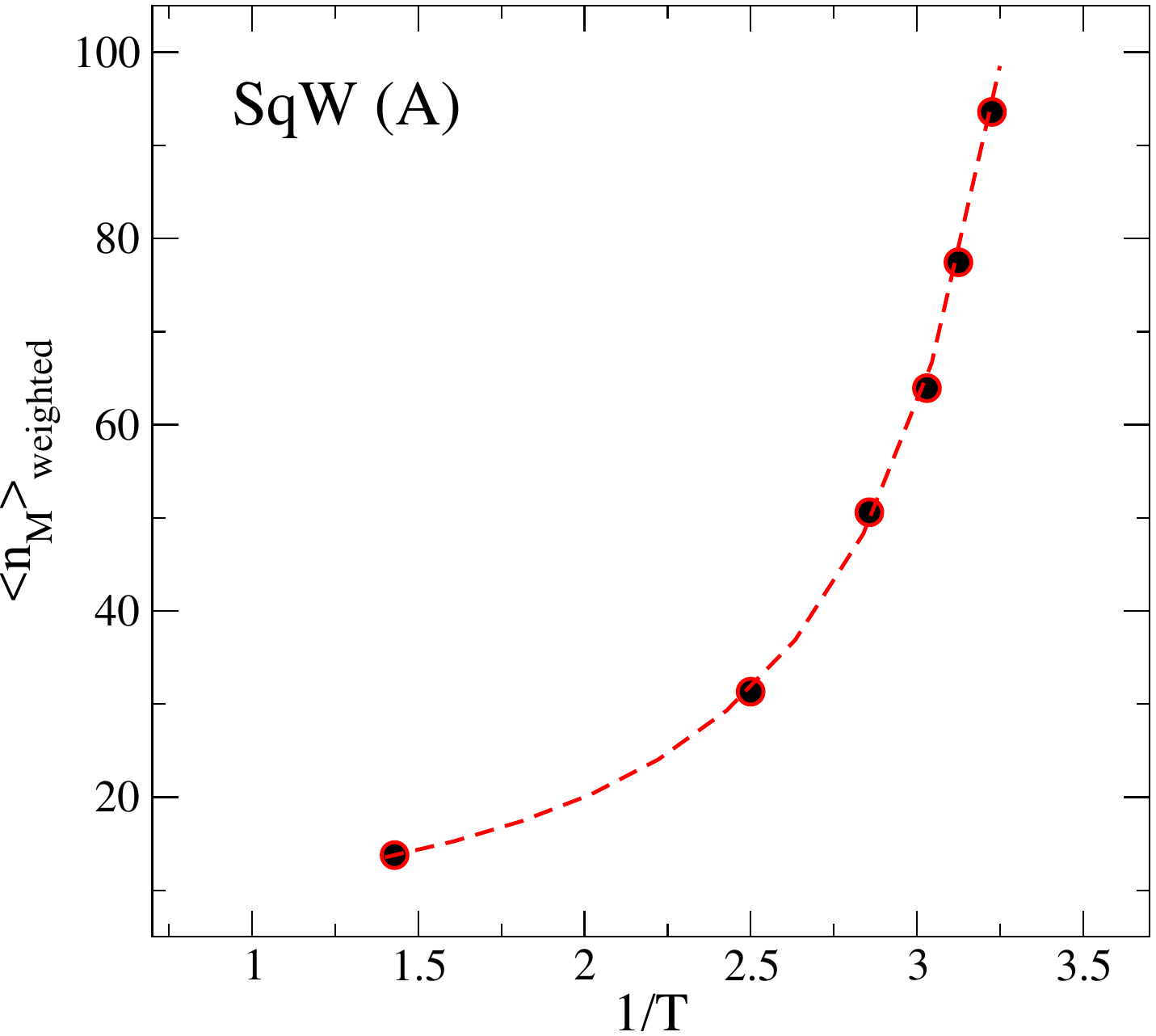} 
\includegraphics[scale=0.40]{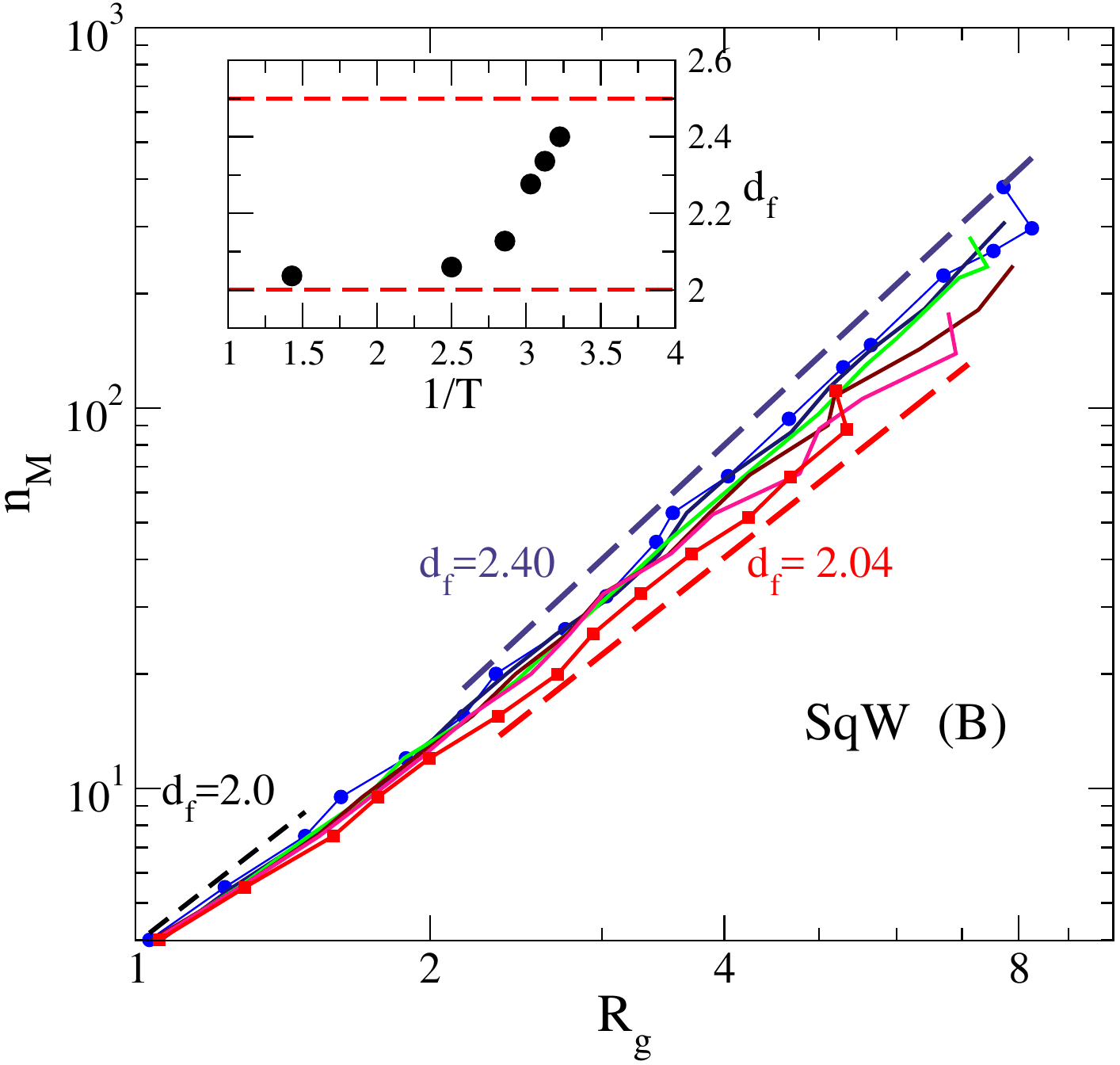}
\caption{\emph{(A)}: The mean cluster size at $t^*$ for various temperatures, the cluster size of these particles increases at lower temperatures. The ``weighted" mean cluster size  is defined as $<n_M> = \frac{\sum_{n_M} n_M^2 P(n_M)}{\sum_{n_M} n_M P(n_M)}$. \emph{(B)}: The cluster size of the active particles ($n_M$) is plotted against radius of gyration ($R_g$). The clusters at high temperatures are string like but at lower temperatures these clusters become compact.}
\label{}
\end{figure}

\section{Coupling of $t^*$ and diffusion time scales}
The present work provides the numerical evidence that the Adam Gibbs relation is valid for diffusion coefficient and not for relaxation times and viscosity but the precise, microscopic description of the mechanism(s) of structural relaxation in glass-forming liquids is currently lacking. In general, it is found that structural relaxation time {$\tau_{\alpha,A}$} is proportional to viscosity and is not proportional to the translational diffusion time scale {$D_A$}. This decoupling is related to the emergence of mobile and immobile clusters upon cooling (DH). The AG theory proposed a mechanism for structural relaxation via the concept of cooperatively rearranging regions (CRR), which have been identified as clusters of highly mobile particles, named ``strings"\cite{AG-DH_s,Freed_s,string2_s,string3_s}. In Ref. \cite{AG-DH_s} it is shown that the life time of these strings is proportional to $t^*$, which is proportional to $D^{-1}$. Hence, in the presence of SEB, the AG relation is described for $D$ and not for decoupled quantities $\tau_\alpha$ and $\eta$.  Fig. \ref{Tstar} shows that the diffusion coefficient and the $t^*$, {for particles type A}, are coupled in both the models considered in the present study.
\begin{figure}[htp]
\centering
\includegraphics[scale=0.40]{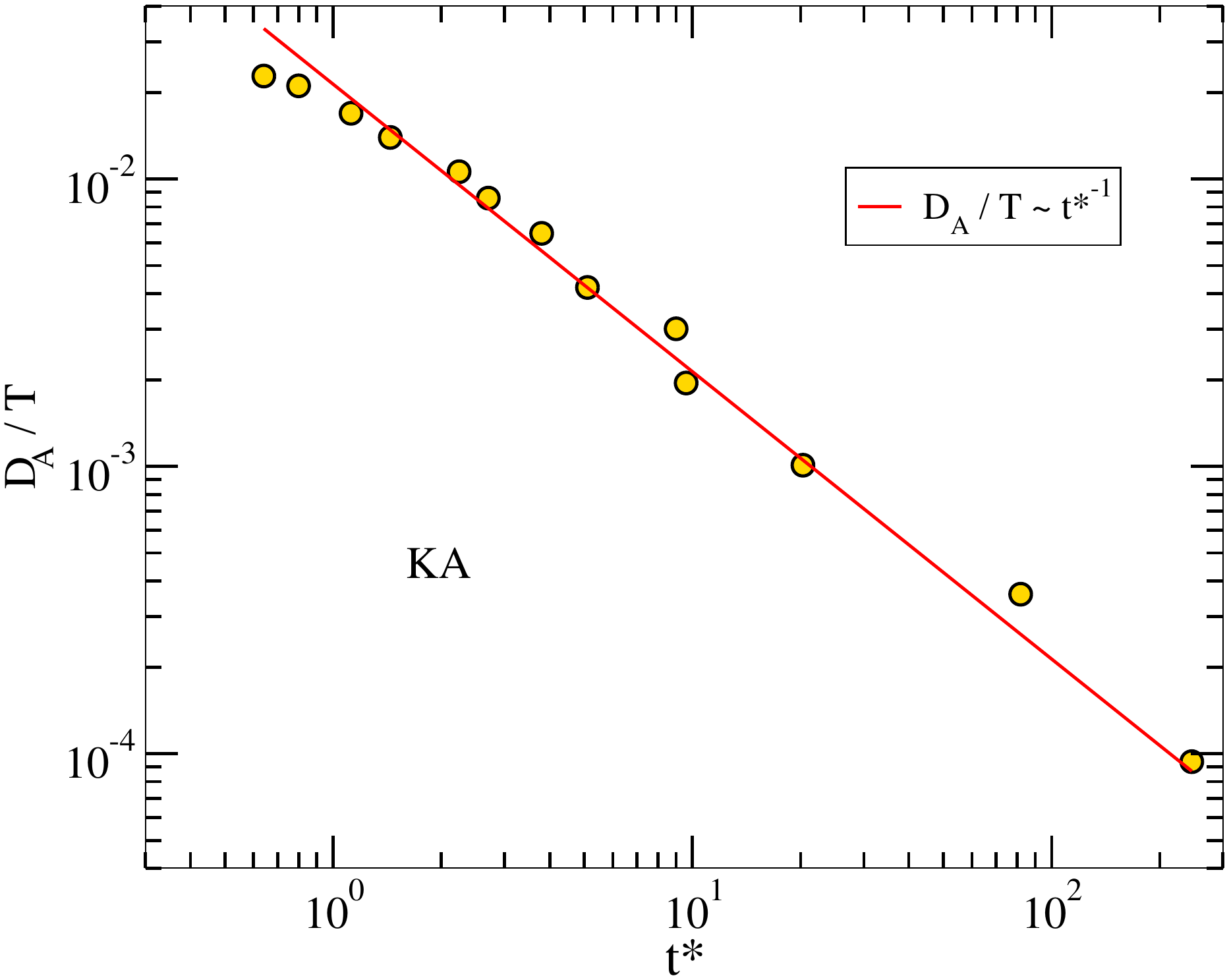} 
\includegraphics[scale=0.40]{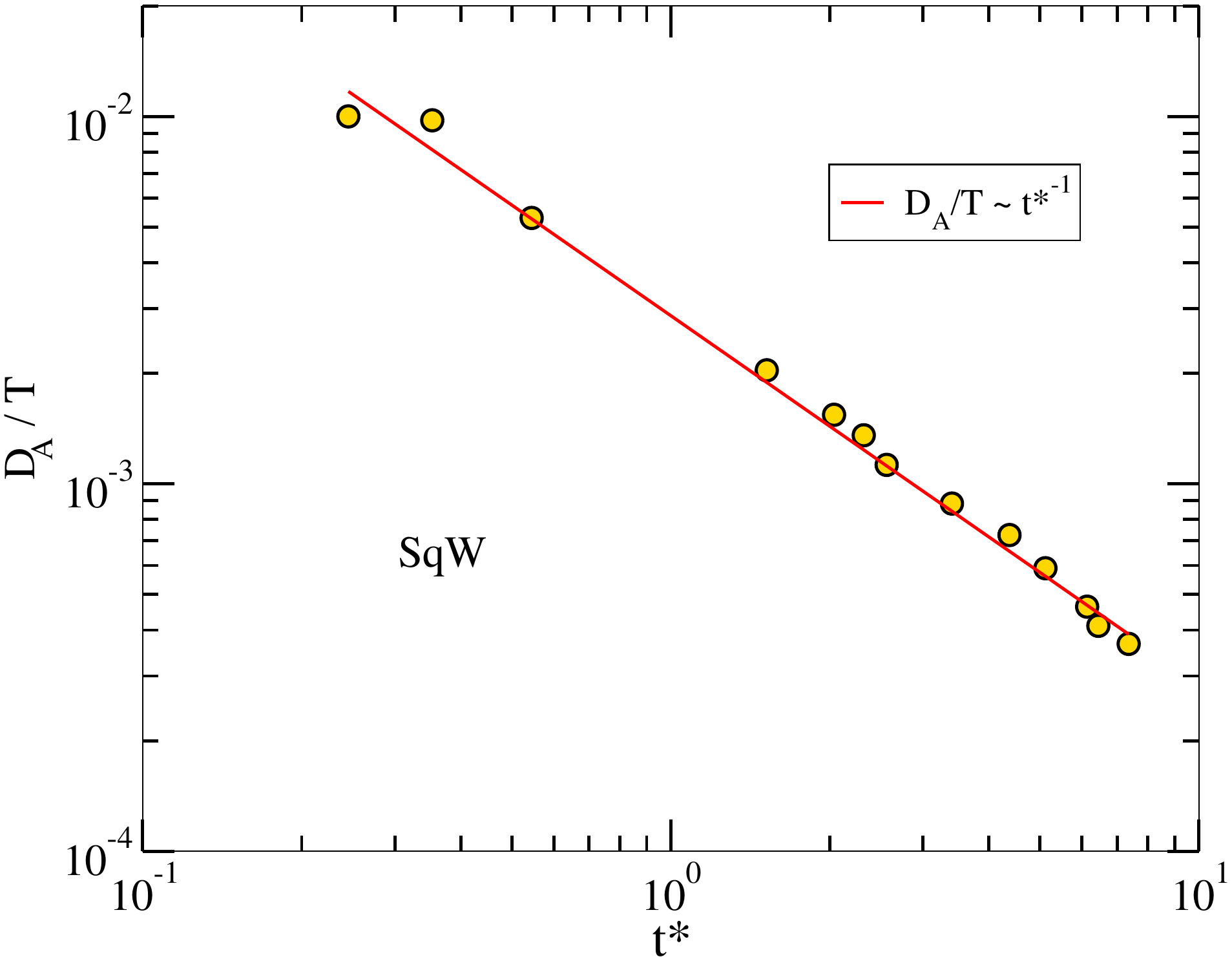}
\caption{The peak time $t^*$ of non-Gaussian parameter scales with the diffusion coefficient and consequently is decoupled from viscosity ($\eta$) and  relaxation times ($\tau_{\alpha,A}$) in both the KA and the SqW model.}
\label{Tstar}
\end{figure} 

\section{Stokes-Einstein and Adam-Gibbs relations for diffusion coefficients and relaxation times of  $A$ and $B$ components in the KA and SqW binary mixtures}
The studied model glass formers consist of two types of particles, with compositions of the KA and SqW models being $(80:20)$ and $(50:50)$ respectively. Changes in dynamics with temperature affect these components differently \cite{SEBKA-6_s}.  We study the relation of diffusion coefficients of the two components as a function of temperature for the two models, as well as relaxation times. Though the ratio of the diffusion coefficients (\emph{i.e.} $D_B/D_A$) is observed to be temperature dependent, as seen in Fig. \ref{DaDb},  we find that they have a fractional power law dependence 
\begin{figure}[h]
\centering
\includegraphics[scale=0.34]{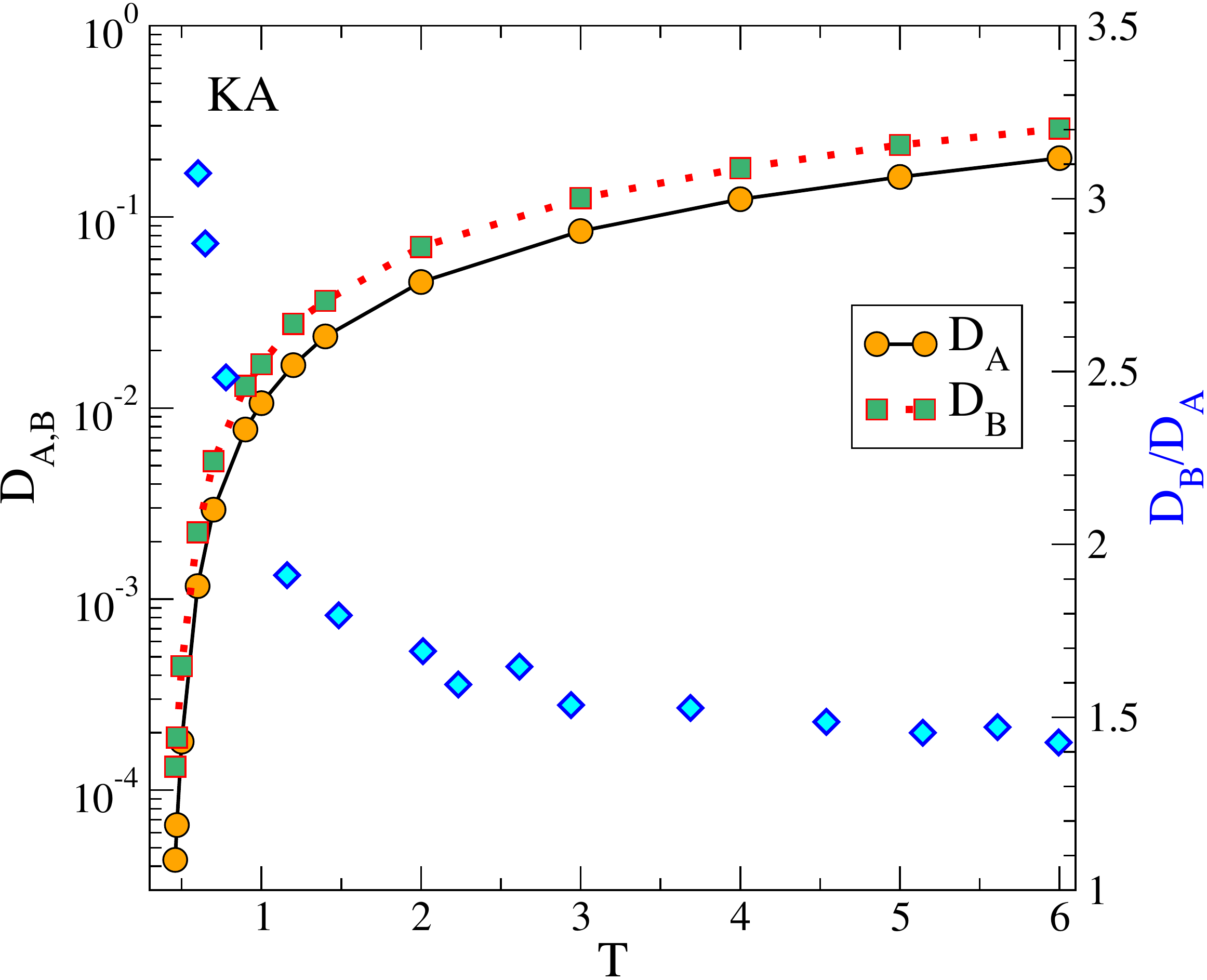}
\includegraphics[scale=0.50]{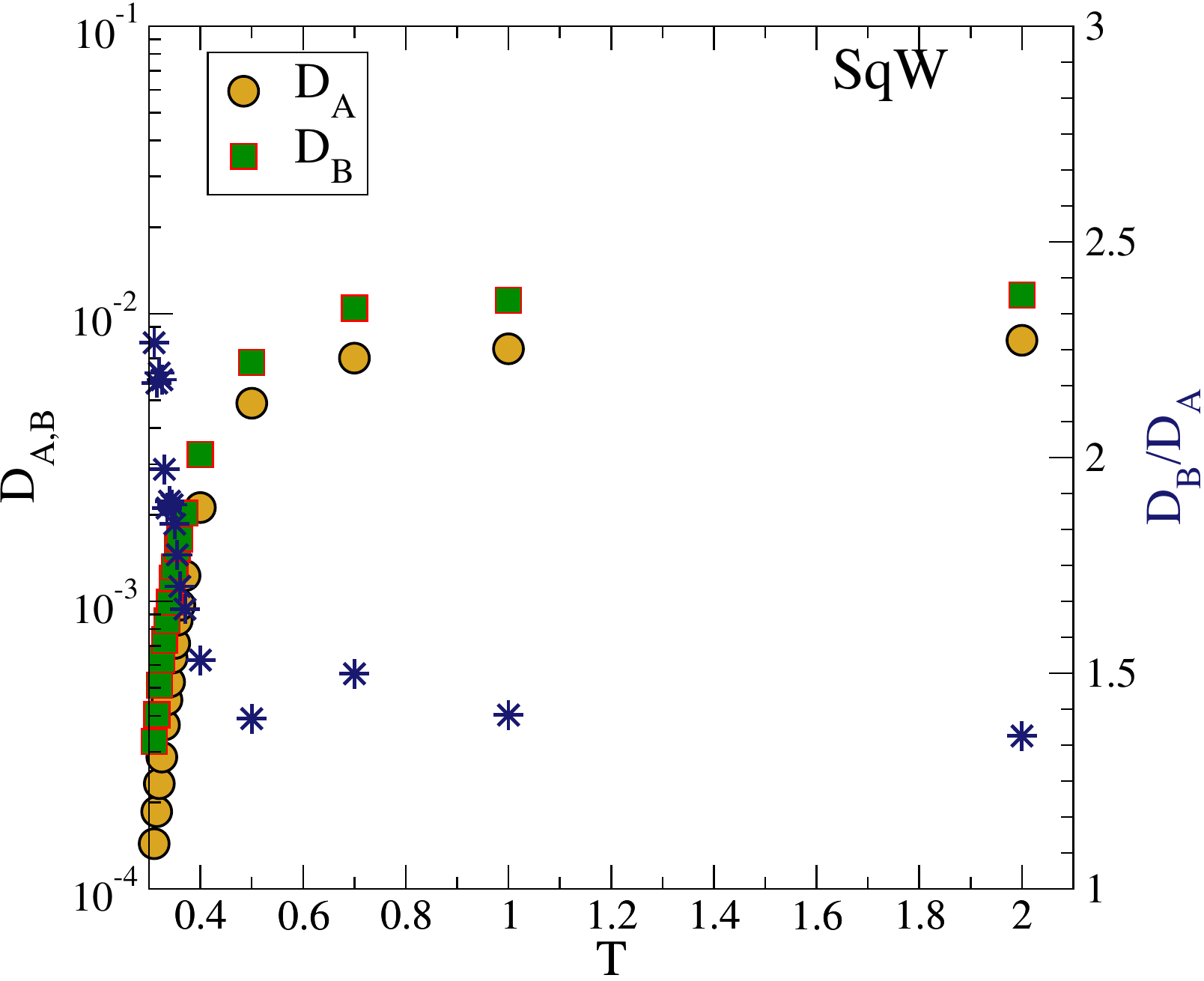}
\caption{Diffusion coefficients of A and B components, and their ratio, $vs.$ temperature for the KA and SqW models.}
\label{DaDb}
\end{figure} 
\pagebreak
on each other over a large temperature range (Fig. \ref{DaDbfrac}), extending well beyond the regime where the SEB is observed for any quantity we study. The Stokes-Einstein relation breaks down for particles of type $B$ as well. Fig. \ref{SEAB} shows that the fractional SE relation $D_{A,B} \propto \tau_{A,B}^{-\xi}$ is observed for both components (when we plot the diffusion coefficient of a given particle type against relaxation times computed for the same particle type) at low temperatures, with very similar characteristics. Because of their fractional power law dependence, the Adam-Gibbs relation holds for diffusion coefficients of both particle types in the considered temperature range, with different activation  energies. 
\begin{figure}
\centering
\includegraphics[scale=0.50]{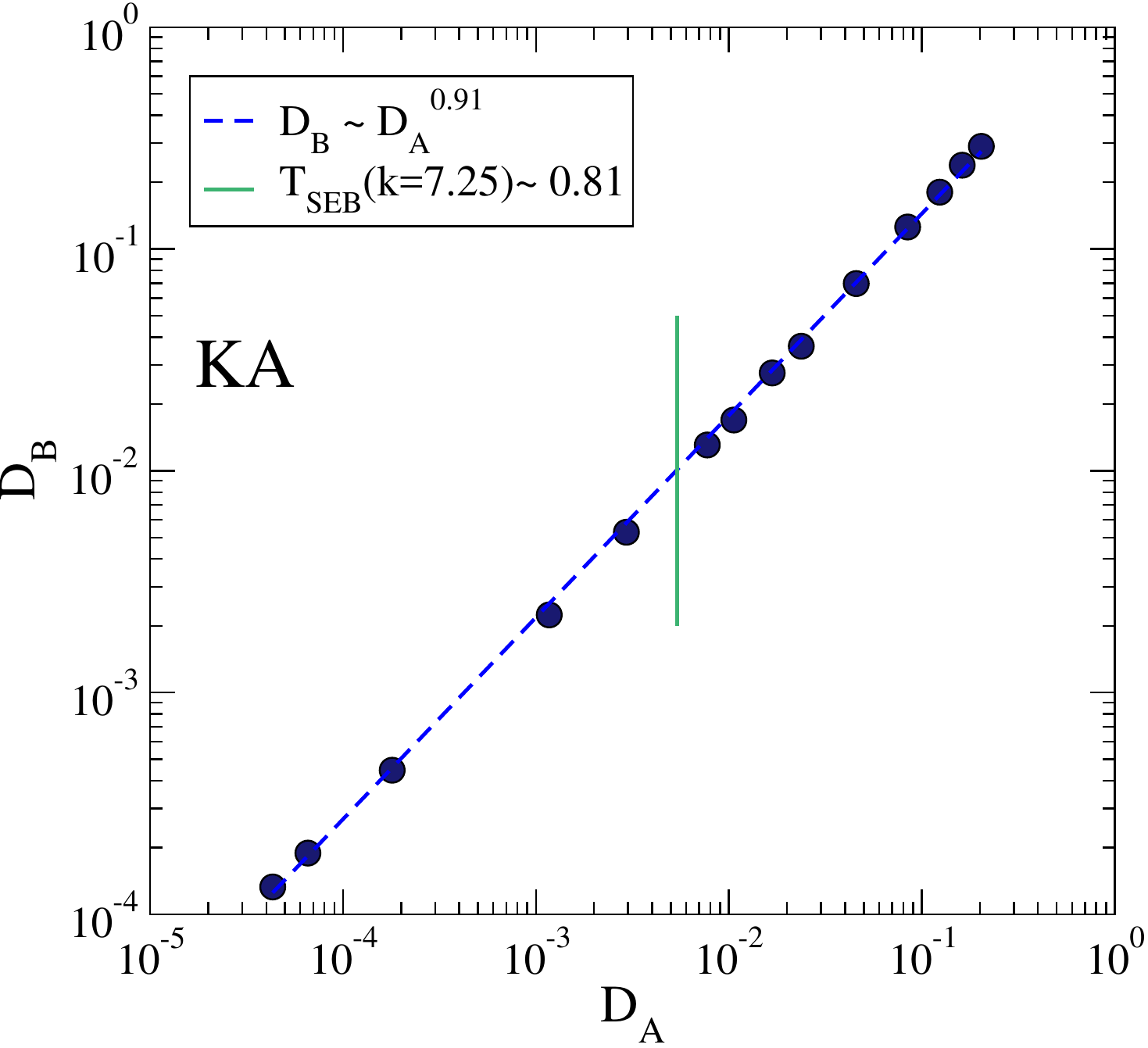}
\includegraphics[scale=0.50]{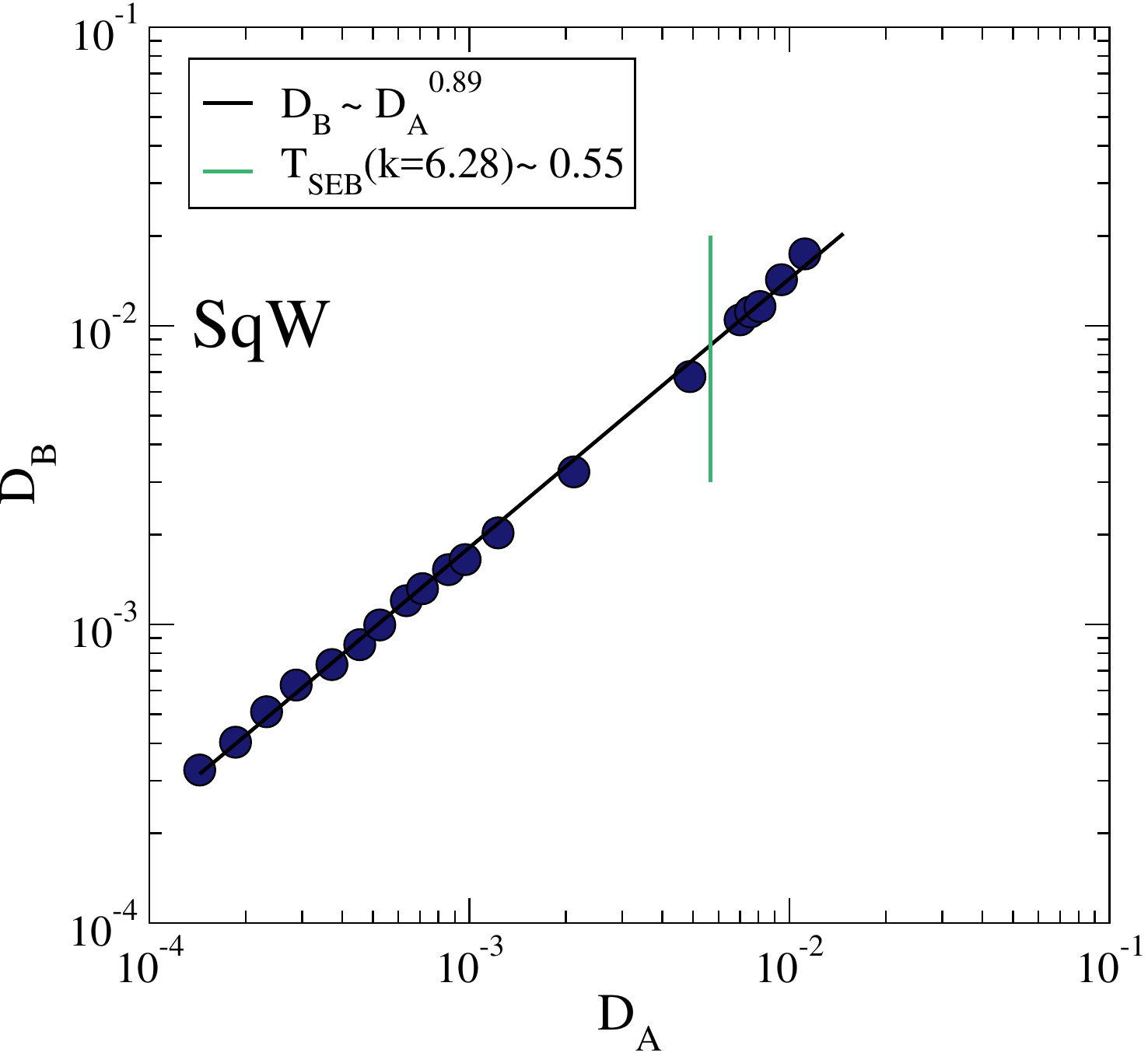}
\caption{The diffusion coefficient of the B species plotted against that of the A species, displaying a fractional power law dependence for a wide temperature range. (The temperature range is $[0.46,6.0]$ and $[0.31,10]$ respectively for the KA and SqW models). Vertical lines indicate the highest SEB temperatures we analyse.}
\label{DaDbfrac}
\end{figure} 

\begin{figure*}[]
\centering
\includegraphics[scale=0.44]{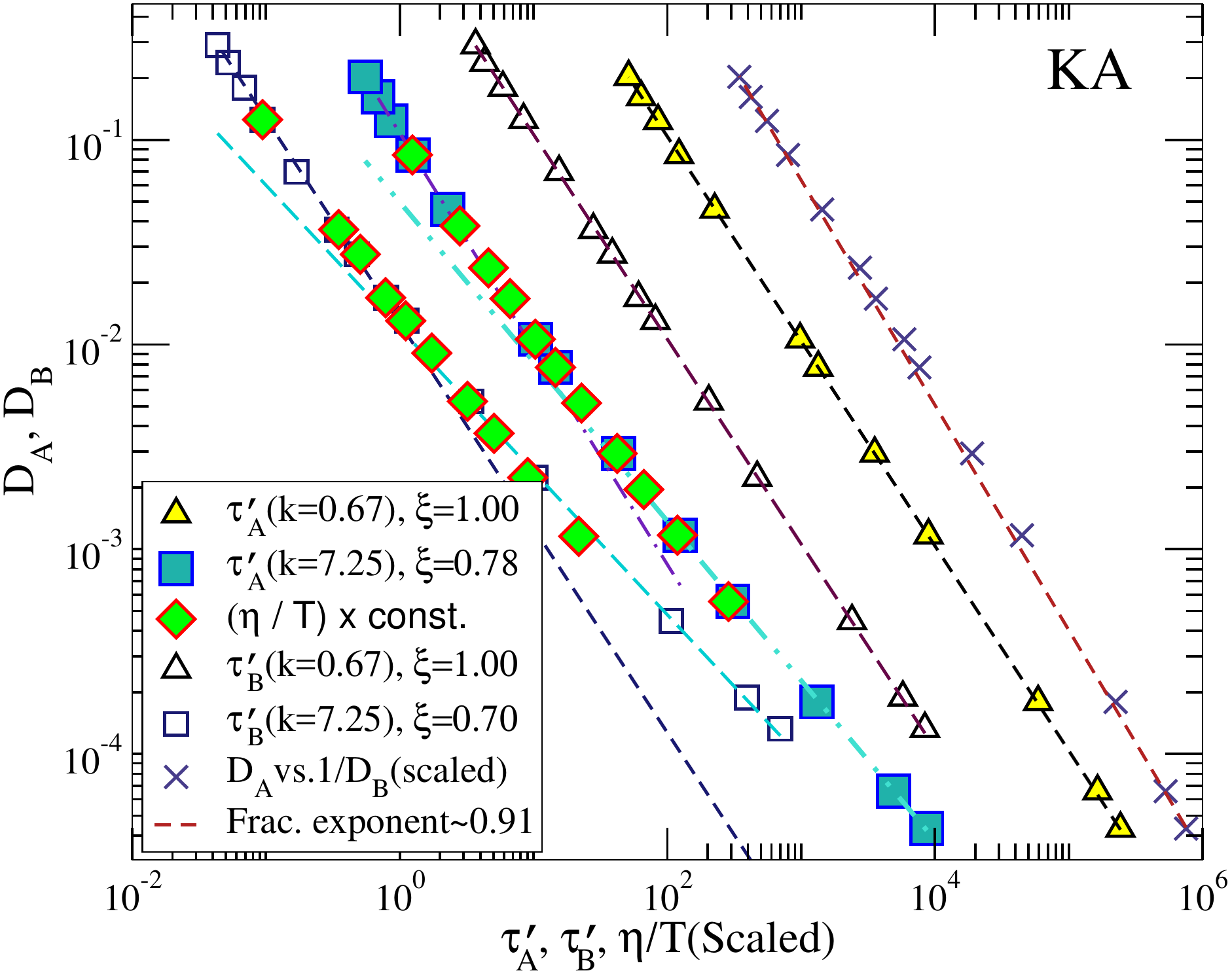}
\includegraphics[scale=0.44]{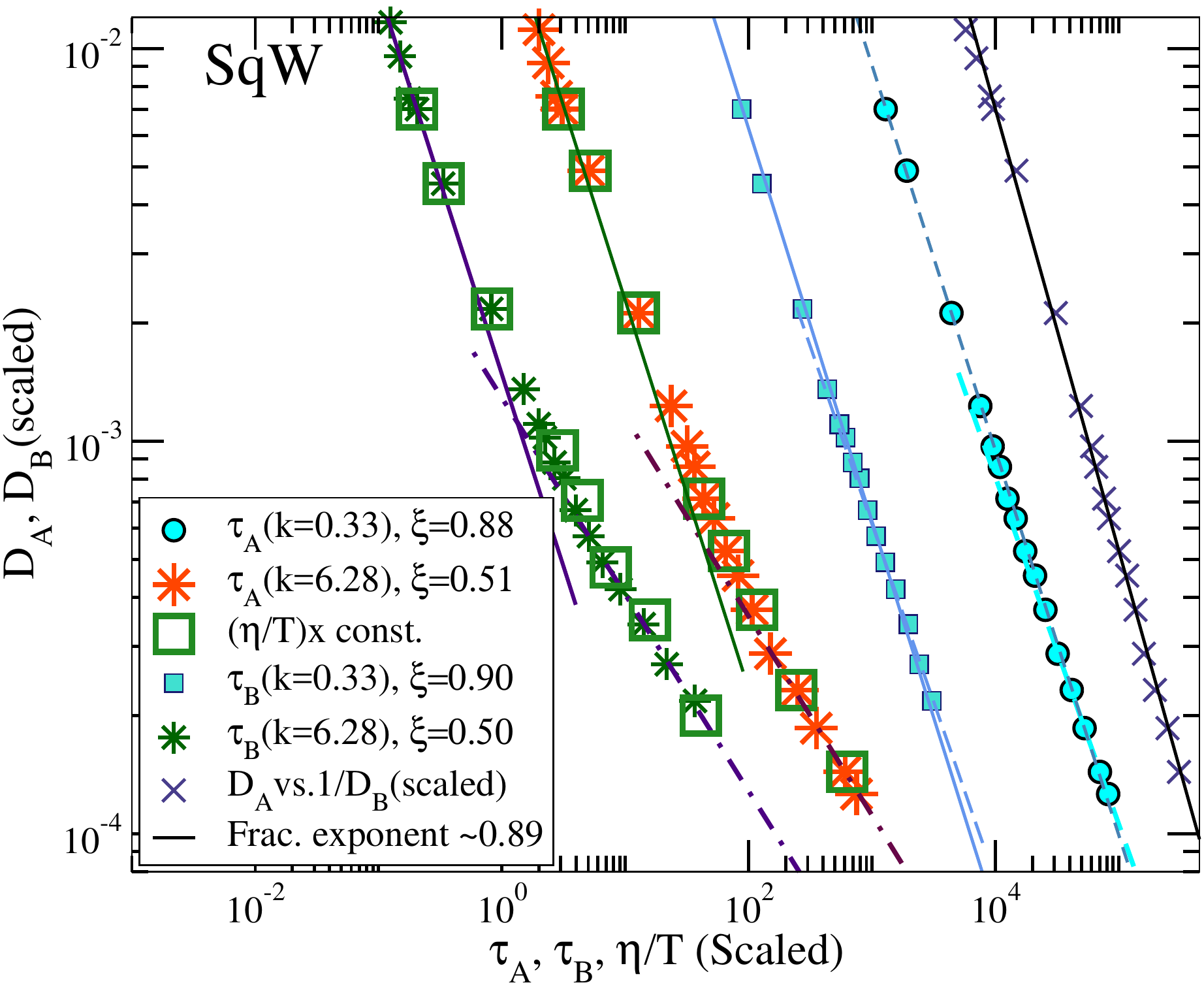}
\caption{The Stokes Einstein relation breakdowns in the KA and SqW models for both types of particles. The SEB exponents for the B type particles do not differ significantly from those for particles type $A$, which approach $1$ as the wave vector $k$ decreases. The lines are power law fits to $D_i = \mathcal{A} X_{i}^{-\xi}$, where $i$ is the particle type and $X$ is either $\tau_{A,B}$ or $\eta/T$.}
\label{SEAB}
\end{figure*}  
\begin{figure*}[]
\centering
\includegraphics[scale=0.44]{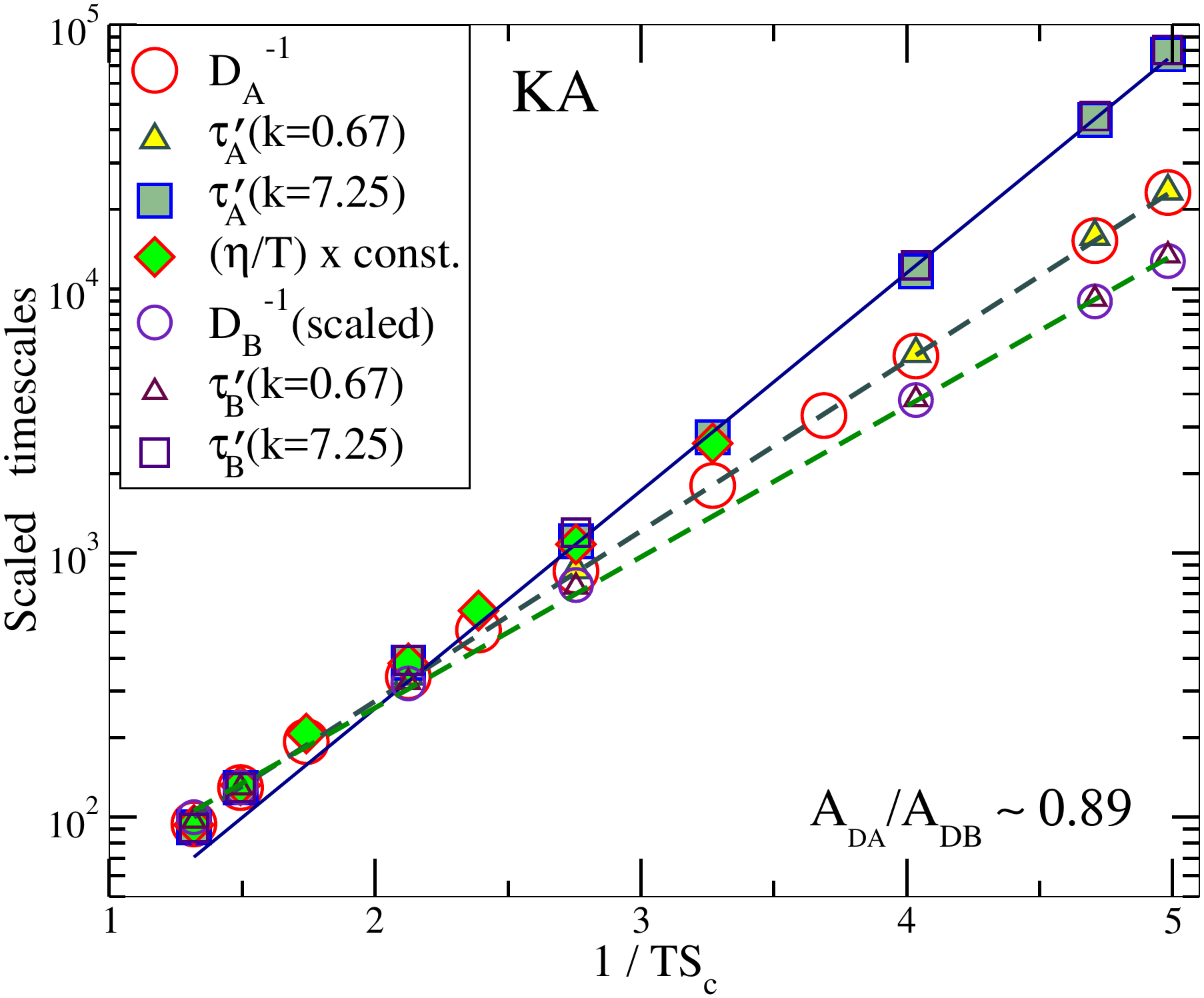}
\includegraphics[scale=0.42]{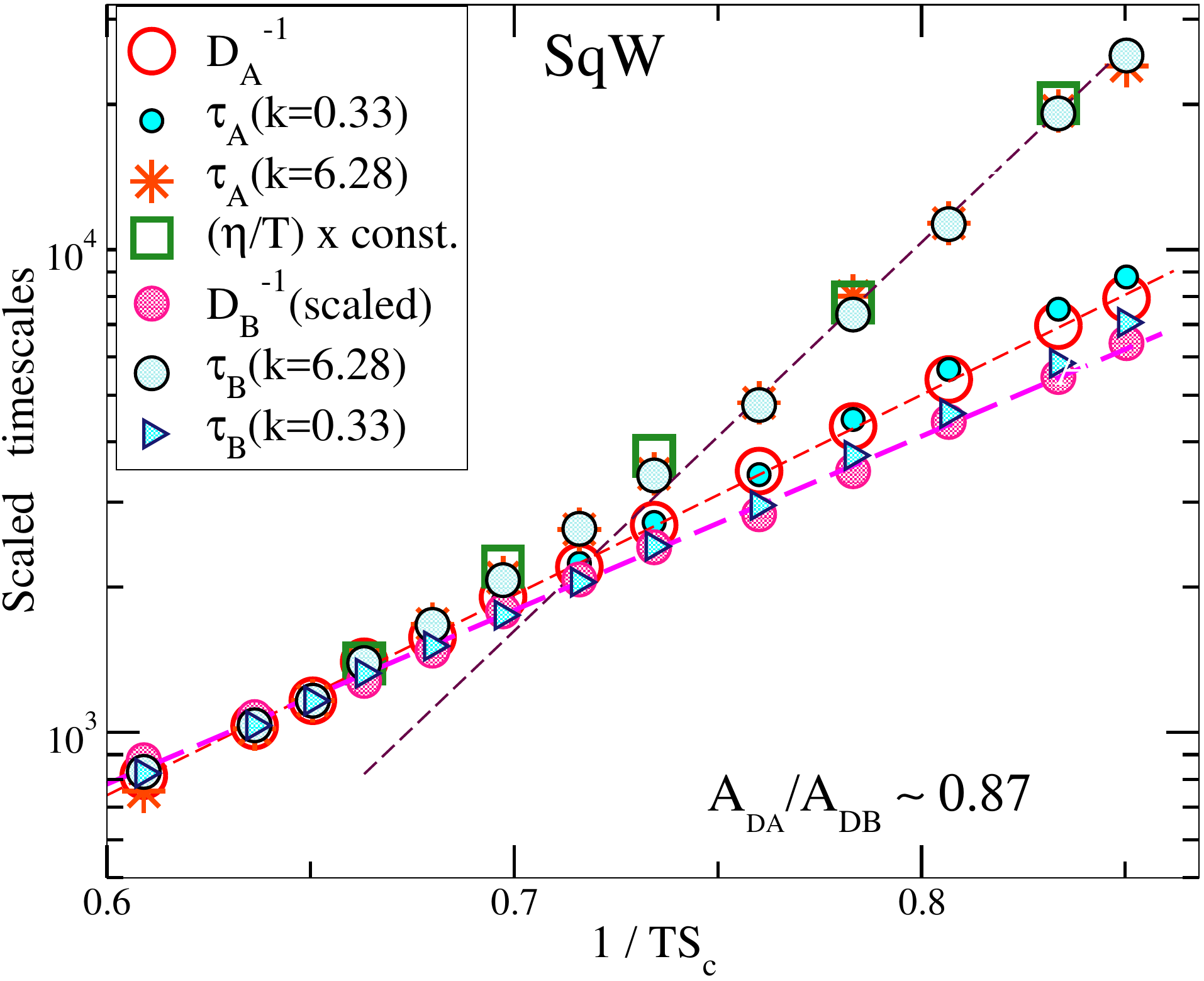}
\caption{The Adam Gibbs relation for $\eta/T$, $\tau_{A,B}$ and $D_{A,B}$ for the KA and SqW models. The $\tau_A(k)$, $\tau_B(k)$, $\eta/T$ and $D_B^{-1}$ have been scaled so that they coincide with $D_A^{-1}$ at one high temperature ($1/TS_c \approx 1.49~\&~0.65$, for KA and SqW, respectively).
The Adam Gibbs relation holds for the diffusivity of both types of the particles, with different activation energies that are  consistent with the fractional dependence of $D_A$ on  $D_B$. Structural relaxation times and viscosity show systematic deviation at low temperature from the AG relation due to the breakdown of the Stokes-Einstein relation.}
\label{}
\end{figure*}

\end{document}